\documentclass{aastex631}
\usepackage{graphicx}
\usepackage[T1]{fontenc}
\usepackage[utf8]{inputenc}
\begin{document}
\title{Multiplicity of Normal A-Type Field Stars of Low \\
Projected Rotational Velocity}
\author[0000-0002-4937-7015]{Daryl W. Willmarth}
\affiliation{NSF National Optical-Infrared Research Labratory, \\
950 North Cherry Avenue, Tucson, AZ 85719, USA; daryl.willmarth@noirlab.edu}
\author[0000-0002-9413-3896]{Francis C. Fekel} 
\affiliation{Center of Excellence in Information Systems, Tennessee State University, \\
3500 John A. Merritt Boulevard, Box 9501, Nashville, TN 37209, USA,}
\affiliation{Current Address: 8612 Sawyer Brown Rd, Nashville, TN 37221 USA}
\author[0000-0002-4531-7867]{Helmut A. Abt }
\affiliation{NSF National Optical-Infrared Research Labratory, \\
950 North Cherry Avenue, Tucson, AZ 85719, USA; daryl.willmarth@noirlab.edu}
\affiliation{deceased, November 23, 2024}


\begin{abstract}
While nearly all rapidly rotating ($v \sin i >$ 120 km s$^{-1}$) A stars
are observed to have normal spectra, a significant fraction ($\sim$43\%)
of slow rotators have peculiar spectra although many do not. For over
30 years, we conducted a radial velocity survey of 119 non-peculiar, slowly
rotating dwarf or subgiant A stars. We determined orbital elements and
related parameters for 43 systems and along with known visual binaries,
common proper motion companions and stars with proper motion anomalies,
find a multiplicity fraction of 0.64. We compared our results to recent
findings concerning similar stars. We also examined whether the slow 
rotation of our sample of could have been caused by tidal interactions in
short-period binaries. Of the 80 stars that we determined to be at
least binary in nature or to show other evidence of having companions,
only 6 have periods $<$ 10 days.  Thus, a large majority
have periods that are much too long for tidal forces to have reduced
their rotation rates while they are on the main-sequence. We also
observed a small sample of rapid rotators
($v \sin i = 125-230$ km s$^{-1}$) over a shorter time interval to
judge the accuracy of our measuring techniques on A stars with very
broad lines. Orbital elements for 5 non-survey A-type stars,
also with low $v \sin i$, are listed with our
results, but are not included in the statistical discussion.
\end{abstract}

\section{Introduction}
Compared to cooler F-, G-, and K-type stars, A-type stars in general have 
many fewer metal lines with much weaker line strengths. In addition, 
A-type stars are typically rotating much more rapidly than solar-type 
stars. In recent decades, ground-based binary star surveys have been 
carried out primarily with radial velocity spectrometers such as CORAVEL, 
ELODIE, and others \citep[e.g.][]{duq91, udry98}. 
Such spectrometers work well in obtaining radial velocities (RVs) for 
stars of F -- M spectral classes. However, the lack of numerous, 
strong, narrow metallic lines among the majority of A stars, makes 
such instruments much less useful for extensive binary star surveys. 
One exception is the survey of Carquillat and colleagues 
\citep[e.g.][]{car07} who carried out a search for spectroscopic 
binaries (SBs) among slowly rotating ($v \sin i <$ 40 km~s$^{-1}$), 
metallic-lined A stars (Am stars) primarily with the radial velocity 
spectrometer CORAVEL. Relative to their hydrogen line spectral classes,
the Am stars have enhanced line strengths of their heavy elements such 
as Fe, Sr, and Ba but diminished line strengths of their lighter 
elements like Ca and Sc.

From the third edition of The Bright Star Catalogue \citep{hof64},
\citet{abt95}, hereafter AM95, selected nearly 1700 stars with
spectral classes from A0 to F0 and determined their spectral types 
and rotational velocities. Among the 
1197 A0 -- A9 IV or V stars with measured $v \sin i$ in the AM95 
sample, 605 or 50.5\% have a value $\leq$ 100 km~s$^{-1}$. Of these with 
lower observed $v \sin i$, 261 or 43.1\% were classified as Am or Ap.  This 
leaves a substantial fraction of 56.9\% or approximately 344 stars with no 
detected peculiarities, but low $v \sin i$. \citet{roy02} 
recalibrated the rotational velocity scale used by AM95 so that 
the limiting $v \sin i$ for detecting the Am or Ap star peculiarities 
is increased to 120 km~s$^{-1}$. We adopt this value as the dividing 
line between low and high $v \sin i$ A stars. 

In a sample of 55 stars brighter than $V$ = 5.01 mag, \citet{abt65} 
examined the frequency of binaries of spectral types A4 -- F2~IV,V.
The RVs from that work and the literature resulted in the 
identification of 17 binaries. They determined orbits for 15
and found that none of the periods were less than 100 days. They also 
noted that their SBs had $v \sin i$ values that 
ranged from 50 to 240 km~s$^{-1}$. However, 24 stars of the sample were 
early-F stars reducing the number of A stars examined to just 31.
Of these 31, 10 were determined to be SBs and an additional 6 are 
known to be visual binaries for a binary fraction of 52\%. However, 
in a subsequent paper with additional new data \citep{abt74}, 3 of the 10 
suspected SBs were determined not to be binary, so the binary fraction was 
then reduced to 42\%.

We mention these earlier papers because their results have been cited
in subsequent papers, such as those of \citet{car07}, \citet{derosa14}, and
\citet{wais23a}, as representative of the binary fraction of A-type stars.
While there have been a number of studies of the binary statistics of Am
and Ap stars, no systematic spectroscopic surveys of normal A stars have
been done since the pioneering work of \citet{abt65} and \citet{ abt74}. 
The aforementioned studies by \citet{derosa14} and \citet{wais23a} 
explored the companion separation ranges of 30 to 45,000 au and $<$ 20 au, 
respectively, using adaptive optics, proper motions, and near-infrared 
interferometric techniques. These two papers also contain detailed 
histories of work on A-type companion searches as well as the 
astrophysical importance and implications of such statistics. \citet{defur25} 
noted the difficulties in conducting an RV search and performed a long-baseline 
interferometry (LBI) survey to search for companions of A-type star primaries
with separations <20 au.

As a follow-up to the work of \citet{abt65}, for more than three decades, 
we have conducted an RV search for SBs over the complete range of A star 
spectral classes that have dwarf or subgiant luminosities. Our 
$v \sin i$ upper limit is 120 km~s$^{-1}$. Whenever possible, orbital 
solutions have been obtained and a search at red wavelengths was made for 
the secondary of each binary. More recently, we observed a small sample 
(23 stars) of high $v \sin i$ A0 to F0 stars (140 -- 230 km~s$^{-1}$) to 
determine if our velocity measurement techniques could also be useful 
in detecting binaries among such broad-lined stars.
 
Another aspect of our study concerns the relation between the value of
$v \sin i$ and orbital period among the shorter period systems. One of 
the major drivers of this survey at the time it commenced (1992)
was to investigate whether low $v \sin i$ stars are short period 
binaries. Since the seminal work of \citet{zah77} and
references therein, tidal friction has been considered as a likely 
explanation of slow rotation and circular orbits among close binary stars.
While the tidal process can be very effective for binaries containing 
stars with masses less than 1.5 $M_\odot$ that have convective
atmospheres, earlier type binaries with radiative atmospheres 
require very small separations in order for the tidal slowing not to 
require a longer time period than their main-sequence (MS) lifetimes. In 
Table 1 of \citet{hur02} the ratio of separation to stellar radius 
($a/R$) is given for circularization and synchronization
of the orbits of binaries of various equal mass stars in a time 
period equal to 1/4 of their MS lifetimes. A pair of 1.2 $M_\odot$ 
stars would need an ($a/R$) ratio of 124.58 for synchronization 
which implies an orbital period of approximately 302 days. But for 
a pair of 2.0 $M_\odot$ stars, ($a/R$) would need to be  
no more than 6.75, yielding a period of approximately 2.08 days.
Equation 5.7 of \citet{zah77} indicates that the synchronization time
is dependent on $(a/R)^{17/2}$, so that small increases in the 
ratio result in very long time intervals that can exceed the A star MS 
lifetime.  In more recent work, \citet{smi24} studied the spin-orbit 
alignment of early-type astrometric binaries where the component 
stars are too far apart to have any significant degree of tidal 
interactions. They found that the primaries rotate more slowly 
than their single-star or wider binary counterparts and therefore, must
have acquired these properties during their formative evolution.

We describe our data sample in Section 2 with Table~\ref{prop_lo_vsini} 
listing the basic characteristics of the low $v \sin i$ sample. 
Table~\ref{prop_hi_vsini} provides the same information for the high 
$v \sin i$ stars. In Section 3 we describe our observations
and data reduction methods. In Section 4 we discuss the results of
our RV measurements with Tables~\ref{RVs_lo_vsini}, 
~\ref{RVs_lo_vsini_var},
and ~\ref{RVs_hi_vsini} listing the individual RVs for the low $v \sin i$
binaries, the low vsini stars that may have variable velocities, and
all the high $v \sin i$ stars, respectively.
Tables~\ref{RVsum_lo_vsini} and ~\ref{RVsum_hi_vsini}
contain summaries of the RV measurements for these groups,
as well as the status relating to whether a star was found to have one
or more companions. Table~\ref{Orbel_sb1} lists the orbital
elements of the single-lined binaries while Table~\ref{Orbel_sb2}
contains those of double-lined binaries or higher order multiple 
stars. Section 5 is a discussion of our derived multiplicity statistics
and how they compare with previous studies. Since this work commenced with
the hypothesis that the low $v \sin i$ stars might be members of binary systems,
we also discuss our findings regarding that relationship. Section 6
summarizes our conclusions. In the Appendix we provide notes on selected
low $v \sin i$ stars and all of the high $v \sin i$ stars.

\section{Stellar Sample}
From the list of AM95, our initial sample consisted of 69 stars with 
normal spectral classes from A0 -- A9, luminosity classes of V or IV, 
and $v \sin i$ values less than 50 km~s$^{-1}$. We soon expanded 
our sample by adding 50 more normal A-type dwarfs or subgiants from 
AM95 for which they determined $v \sin i $ values up to 100 km s$^{-1}$.
All those stars are listed in Table~\ref{prop_lo_vsini} with the columns 
of their properties described in the notes following the table.
The shortcomings of a magnitude limited survey such as this one are well
known and discussed in Section 5.1. At the time this survey commenced, a
large fraction of the sample stars either had no parallax values or they
differed significantly from those of the subsequent surveys of Hipparcos
\citep[herafter HIP2]{vanl07} and Gaia \citep[hereafter GEDR3]{gaia21}.
Since this work 
commenced, \citet{roy02} extended and recalibrated the $v \sin i$ 
measures in AM95, generally increasing them by the amount 
$1.05 (v \sin i_{\rm {AM95}}) + 7.5$. These revised values of \citet{roy02} 
are listed in column 6 of Table~\ref{prop_lo_vsini} except for the binaries
where our $v \sin i$ values are given, as noted.

More recently, we observed a small sample of 23 stars with spectral 
types ranging from A0 V or IV to F0 V or IV and high $v \sin i$ 
values. Spectra were obtained in 2014 -- 2016 with the same KPNO 
telescopes, and additional spectra were acquired with the Fairborn 
Observatory 2~m Automatic Spectroscopic Telescope (AST) in early 2024. 
These stars and their properties are listed in Table~\ref{prop_hi_vsini} 
with the same sources as those in Table~\ref{prop_lo_vsini}.
  
We have also included orbits for three A-type stars that were observed 
as radial velocity (RV) standards \citep[e.g.,][]{fek99} (HRs 3383, 6041, 
and 8641) but over the years have been identified as long-period 
binaries. In addition, two short period double-lined systems (HR 317 
and HD 144359), observed in a parallel program, also have their orbits 
listed here. While the properties of these 5 stars are given in 
Table~\ref{prop_lo_vsini}, these binaries are not included in the 
multiplicity statistics of our survey.

Since approximately half of the sample is more than 100 pc distant 
(up to 275 pc), the question arose as to how to correct for reddening 
considering the lower extinction in the local bubble. While the 
typical radius of the bubble is thought to be on the order of 100 pc, 
recent studies, for example \citet{zuc22}, have delineated its deviations
from a spherical volume. Nevertheless, we have assumed the reddening 
to be zero inside of the 100 pc radius and $(d - 100) * A_{v}$ for stars 
beyond 100 pc where $d$ is the distance in pc and $A_{v} = 0.0008$ 
mag~pc$^{-1}$. We then used the value of 3.1 for the ratio of total to 
selective absorption.

\vspace{1in}

\startlongtable
\begin{deluxetable}{lcDDccDD}
\tablecaption{Properties of Low $v \sin i$ Program Stars\label{prop_lo_vsini}}
\tabletypesize{\scriptsize}
\tablewidth{0pt}
\tablehead{\colhead{HR} & \colhead{HD} & \multicolumn2c{$V$} 
& \multicolumn2c{\bv} & \colhead{Spectral} & \colhead{$v \sin i$} 
& \multicolumn2c{$\varpi$} & \multicolumn2c{$M_{V}$} \\ 
\colhead{} & \colhead{} & \multicolumn2c{(mag)} 
& \multicolumn2c{(mag)} & \colhead{Type} & \colhead{(km s$^{-1}$)} 
& \multicolumn2c{(mas)} & \multicolumn2c{(mag)} }
\decimalcolnumbers
\decimals
\startdata
114 & 2628 & 5.214 & 0.253 & A9 IV & 21 & 16.0963 & 1.248   \\
196 & 4222 & 5.410 & 0.011 & A1 V & 36 & 9.2342 & 0.230   \\
317 \tablenotemark{a}& 6530 & 5.565 & -0.016 & A0 V & 52,49\tablenotemark{b} & 6.0959 & -0.561   \\
526 & 11031 & 6.460 & 0.090 & A2.5 V & 49 & 7.434\tablenotemark{c} & 0.788   \\
723 & 15385 & 6.198 & 0.187 & A9 IV & 29 & 11.6961 & 1.538   \\
892 & 18543 & 5.225 & 0.011 & A2 IVs & 49 & 8.9425 & -0.028   \\
971 & 20149 & 5.602 & 0.001 & A1 IV & 20,23\tablenotemark{b} & 6.7649 & -0.285   \\
1018 & 20980 & 6.351 & 0.015 & A1 V & 49 & 9.0074 & 1.115   \\
1041 & 21402 & 5.717 & 0.039 & A2 IV & 44 & 9.3697 & 0.570   \\
1086 & 22091 & 5.940 & 0.101 & A3 IVs & 33 & 5.4993 & -0.467   \\
1133 & 23193 & 5.588 & 0.069 & A2.5 V & 42 & 11.1661 & 0.828   \\
1181 & 23878 & 5.235 & 0.064 & A3 Vs & 24 & 12.5218 & 0.723   \\
1675 & 33266 & 6.167 & 0.009 & A2 IV & 15 & 6.7795 & 0.285   \\
1849 & 36473 & 5.525 & 0.004 & A0 V & 58 & 13.0736 & 1.107   \\
2103 & 40446 & 5.200 & -0.000 & A1 V & 68,26,9\tablenotemark{b} & 6.0104 & -0.959   \\
2163 & 41841 & 5.460 & 0.048 & A2 IV: & 57 & 6.5957 & -0.485   \\
2818 & 58142 & 4.610 & -0.016 & A0.5 V & 19 & 11.9021 & 0.043   \\
3383\tablenotemark{a} & 72660 & 5.792 & 0.005 & A1 II & 18 & 10.1869 & 0.832 \\
3657 & 79248 & 6.465 & 0.018 & A2 V & 42 & 6.8327 & 0.601   \\
3689 & 80064 & 6.390 & 0.046 & A3 V & 54 & 4.4734 & -0.456   \\
3702 & 80447 & 6.618 & 0.069 & A2 Vs & 15 & 6.1323 & 0.506   \\
3822 & 83104 & 6.333 & 0.061 & A1 V & 57 & 10.9433 & 1.528   \\
3861 & 84107 & 5.621 & 0.114 & A2 IV & 39 & 19.9483 & 2.121   \\
3894 & 85235 & 4.557 & 0.031 & A2 IV & 28 & 6.41\tablenotemark{d} & -1.454   \\
4033 & 89021 & 3.424 & 0.041 & A2 IV & 50 & 14.4594 & -0.776   \\
4331 & 96723 & 6.453 & 0.024 & A1 IVs & 23 & 7.2727 & 0.731   \\
4343 & 97277 & 4.449 & 0.026 & A2 IV & 49 & 10.3043 & -0.486   \\
4359 & 97633 & 3.350 & -0.020 & A1 IV & 23 & 20.3160 & -0.315   \\
4378 & 98280 & 6.649 & 0.055 & A2 Vs & 16 & 7.2098 & 0.908   \\
4585 & 104181 & 5.345 & 0.001 & A1 V & 44 & 10.8636 & 0.525   \\
4670 & 106819 & 6.045 & 0.090 & A3 V & 75 & 6.1840 & -0.048   \\
4717 & 107966 & 5.149 & 0.078 & A3 V & 51 & 11.1318 & 0.382   \\
4738 & 108382 & 4.959 & 0.086 & A4 V & 80 & 11.9394 & 0.344   \\
4760 & 108844 & 5.347 & 0.205 & A8 V & 89 & 11.7987 & 0.706   \\
4789 & 109485 & 4.773 & 0.000 & A0 V & 60 & 9.1148 & -0.117   \\
4805 & 109860 & 6.334 & -0.027 & A1 IVs & 73 & 4.6178 & -0.437   \\
4833 & 110462 & 6.013 & 0.021 & A2 IV & 50 & 5.5537 & -0.329   \\
4859 & 111270 & 5.885 & 0.190 & A7 V & 105 & 15.4538 & 1.830   \\
4866 & 111421 & 6.245 & 0.170 & A7 V & 50\tablenotemark{b} & 6.8208 & 0.377 \\
4921 & 112846 & 5.791 & 0.163 & A4 V & 114 & 13.3246 & 1.414   \\
4948 & 113865 & 6.510 & 0.076 & A3 V & 86 & 10.1298 & 1.538   \\
4963 & 114330 & 4.397 & -0.012 & A2 IVs & 18 & 11.1834 & -0.360   \\
4974 & 114504 & 6.510 & -0.035 & A0 IV & 91 & 4.8962 & -0.124   \\
5021 & 115709 & 6.620 & 0.032 & A2 IV & 64 & 7.5905 & 0.996   \\
5057 & 116706 & 5.745 & 0.084 & A3 Vs & 54 & 11.2467 & 1.000   \\
5076 & 117242 & 6.345 & 0.242 & A9 V & 107 & 9.9058 & 1.324   \\
5079 & 117281 & 6.960 & 0.235 & A8 V & 82 & 7.1801 & 1.209   \\
5129 & 118660 & 6.489 & 0.261 & A8 V & 94 & 14.2029 & 2.251   \\
5144 & 119055 & 5.756 & 0.020 & A1.5 V & 60,12\tablenotemark{b} & 10.3308 & 0.827   \\
5146 & 119086 & 6.605 & 0.023 & A1.5 V & 107 & 6.4935 & -0.165   \\
5163 & 119537 & 6.498 & 0.054 & A1 IVs & 13 & 7.8334 & 0.946   \\
5216 & 120874 & 6.447 & 0.087 & A2 V & 81 & 11.7452 & 1.796   \\
5220 & 120934 & 6.098 & 0.040 & A1.5 V & 81 & 9.8916 & 1.073   \\
5229 & 121164 & 5.909 & 0.193 & A8 IV & 75 & 13.5993 & 1.577   \\
5255 & 121996 & 5.748 & -0.023 & A0 V & 75 & 5.9260 & -0.443   \\
5280 & 122866 & 6.148 & 0.025 & A1 V & 91 & 10.3764 & 1.229   \\
5333 & 124713 & 6.377 & 0.193 & A7 V & 70\tablenotemark{b} & 15.3517 & 2.308   \\
5360 & 125349 & 6.195 & 0.050 & A1 IV & 86 & 10.5723 & 1.316   \\
5374 & 125658 & 6.445 & 0.153 & A5 IVs & 26 & 11.9629 & 1.834   \\
5397 & 126367 & 6.593 & 0.095 & A2 IV & 54 & 5.5630 & 0.256   \\
5406 & 126722 & 6.161 & 0.106 & A2 IV & 102 & 14.6280 & 1.987   \\
5415 & 127067 & 7.111 & -0.056 & A0 Vp & 65 & 3.7082 & -0.179   \\
5468 & 129002 & 5.387 & 0.025 & A0 V & 104,23\tablenotemark{b} & 17.1844 & 1.563   \\
5531 & 130841 & 2.725 & 0.160 & A3 V & 102 & 43.03\tablenotemark{d} & 0.894   \\
5569 & 132029 & 6.122 & 0.097 & A2 IV & 74\tablenotemark{b} & 9.4783 & 1.001   \\
5574 & 132145 & 6.504 & 0.012 & A1 IVs & 15 & 8.3221 & 1.089   \\
5665 & 135263 & 6.298 & 0.060 & A1 V & 84\tablenotemark{b} & 9.0480 & 1.072   \\
5676 & 135502 & 5.272 & 0.054 & A1 V & 86 & 13.3668 & 0.902   \\
5748 & 137928 & 6.436 & 0.061 & A2 IV & 34 & 8.1997 & 0.987   \\
5754 & 138245 & 6.532 & 0.132 & A3 V & 102 & 11.4097 & 1.818   \\
5759 & 138338 & 6.427 & 0.086 & A3 V & 65 & 10.6611 & 1.566   \\
5760 & 138341 & 6.452 & 0.167 & A3 V & 35,63\tablenotemark{b} & 5.3665 & 0.031   \\
5789 & 138918 & 4.146 & 0.235 & A9 IV & 84\tablenotemark{b} & 10.8056 & -0.686   \\
5791 & 138936 & 6.550 & 0.272 & A9 V & 69 & 17.5052 & 2.766   \\
5859 & 140775 & 5.580 & 0.039 & A1 Vs & 75 & 9.5642 & 0.480   \\
5875 & 141378 & 5.524 & 0.122 & A3 IV & 107 & 18.7970 & 1.894   \\
5886 & 141653 & 5.194 & 0.046 & A1 V & 108 & 11.3856 & 0.476   \\
\nodata & 144359\tablenotemark{a} &  6.770 & 0.0570 & A0 & 22,10\tablenotemark{b} & 6.7196 & 0.907 \\
6041\tablenotemark{a} & 145788 & 6.255 & 0.095 & A0 V & 16 & 6.0589 & 0.115   \\
6074 & 146738 & 5.779 & 0.034 & A3 IV & 117\tablenotemark{b} & 5.2987 & -0.671   \\
6170 & 149650 & 5.910 & 0.036 & A2 V & 102 & 8.6980 & 0.595   \\
6246 & 151862 & 5.928 & 0.014 & A1 IV & 81 & 10.8467 & 1.104   \\
6341 & 154228 & 5.908 & 0.005 & A1 IV & 42 & 12.2259 & 1.344   \\
6362 & 154713 & 6.441 & 0.081 & A2 IVs & 39 & 6.3445 & 0.407   \\
6367 & 154895 & 6.091 & 0.074 & A2 V & 76 & 11.9129 & 1.471   \\
6378 & 155125 & 2.420 & 0.050 & A2 IV & 23 & 36.91\tablenotemark{d} & 0.256   \\
6399 & 155860 & 6.166 & 0.143 & A3 V & 102 & 10.6511 & 1.303   \\
6412 & 156208 & 6.182 & 0.167 & A1 IV & 40,74\tablenotemark{b} & 4.6496 & -0.573   \\
6421 & 156295 & 5.540 & 0.197 & A7 V & 107 & 22.8472 & 2.334   \\
6432 & 156653 & 5.992 & -0.005 & A1 IV & 43 & 6.3943 & -0.024   \\
6455 & 157097 & 5.357 & 0.043 & A3 IVs & 15 & 7.2227 & -0.380   \\
6457 & 157198 & 5.120 & -0.039 & A1 IV & 107 & 7.4167 & -0.557   \\
6480 & 157728 & 5.712 & 0.197 & A7 V & 92 & 23.4066 & 2.559   \\
6499 & 158067 & 6.379 & 0.083 & A7: V & 44 & 5.3030 & -0.069   \\
6521 & 158716 & 6.462 & 0.049 & A2 IV & 15 & 10.6846 & 1.606   \\
6548 & 159480 & 5.810 & 0.016 & A2 IVs & 29 & 8.6746 & 0.489   \\
6559 & 159834 & 6.100 & 0.166 & A5: V & 19 & 6.4269 & 0.096   \\
6581 & 160613 & 4.228 & 0.082 & A3 IV & 107 & 20.4665 & 0.844   \\
6633 & 161941 & 6.225 & 0.072 & B9 V & 35 & 3.6491 & -1.103   \\
6681 & 163336 & 5.929 & 0.038 & A2 IV & 60 & 12.7459 & 1.456   \\
6753 & 165358 & 6.243 & 0.024 & A2 V & 119 & 6.4553 & 0.249   \\
6794 & 166230 & 5.099 & 0.159 & A7 V & 42 & 9.5493 & -0.005   \\
6883 & 169111 & 5.976 & 0.045 & A2 V & 107 & 7.0611 & 0.187   \\
7001 & 172167 & 0.030 & 0.000 & A0 V & 24 & 130.23\tablenotemark{d} & 0.604   \\
7003 & 172187 & 6.311 & 0.131 & A6 V & 109 & 7.9654 & 0.797   \\
7019 & 172741 & 6.489 & 0.211 & A6 V & 56\tablenotemark{b} & 6.9325 & 0.658   \\
7060 & 173664 & 6.187 & 0.073 & A3 IV & 71 & 6.5532 & 0.227   \\
7250 & 178187 & 5.776 & 0.103 & A4 IV & 44 & 9.9907 & 0.774   \\
7357 & 182239 & 6.644 & 0.251 & A9 V & 57 & 9.7064 & 1.577   \\
7362 & 182369 & 5.009 & 0.229 & A8 V & 54 & 12.8494 & 0.553   \\
7371 & 182564 & 4.576 & 0.042 & A2 IV & 25 & 13.8403 & 0.281   \\
7519 & 186689 & 5.890 & 0.166 & A7 V & 42 & 18.8840 & 2.271   \\
7695 & 191174 & 6.284 & 0.053 & A3 V & 41 & 12.1953 & 1.715   \\
7781 & 193592 & 6.004 & 0.050 & A2 IV & 23\tablenotemark{b} & 10.9904 & 1.209   \\
7818 & 194882 & 6.441 & 0.037 & A2 IV & 40,30\tablenotemark{b} & 3.6439 & -0.891   \\
7891 & 196724 & 4.800 & -0.020 & A0 IV & 52 & 18.2449 & 1.106   \\
7938 & 197734 & 6.160 & 0.001 & A1 IV & 15,40\tablenotemark{b} & 5.0079 & -0.421   \\
7981 & 198552 & 6.617 & 0.040 & A1 IV & 52 & 8.4872 & 1.247   \\
8272 & 205939 & 6.194 & 0.176 & A9 IV & 18 & 5.5186 & -0.162   \\
8459 & 210516 & 6.438 & 0.061 & A5 IV & 49 & 4.4461 & -0.422   \\
8605 & 214203 & 6.419 & 0.014 & A2 IV & 33 & 7.6520 & 0.813   \\
8624 & 214698 & 6.324 & -0.002 & A1 IV & 33 & 5.1385 & -0.198   \\
8641\tablenotemark{a} & 214994 & 4.79 & -0.01 & A1 III & 18 & 11.6546 & 0.122 \\
8837 & 219290 & 6.307 & -0.009 & A1 IV & 54 & 8.5110 & 0.943   \\
\enddata
\tablecomments{The table columns list the following parameters:
(1) HR number \citep{hof82}; 
(2) Henry Draper number (HD); 
(3) visual ($V$) magnitude;  
(4) \bv value from the Tycho2 catalogue \citep{hog00}; 
(5) spectral type from AM95;
(6) $v \sin i$ measures from \citet{roy02} (see text); 
(7) parallax ($\varpi$) from GEDR3 as listed on the 
SIMBAD web portal except where noted; 
(8) $M_{V}$ corrected for reddening (see text).}
\tablenotetext{a}{not a survey star}
\tablenotetext{b}{Our measures; primary, secondary, tertiary}
\tablenotetext{c}{Gaia DR2 value; no GEDR3 $\varpi$}
\tablenotetext{d}{HIP2 value; no GEDR3 or GDR2 $\varpi$}
\end{deluxetable} 

\vspace{0.5in}
\nopagebreak

\startlongtable
\begin{deluxetable}{lcDDccDD}
\tablecaption{Properties of High $v \sin i$ Program Stars\label{prop_hi_vsini}}
\tabletypesize{\scriptsize}
\tablewidth{0pt}
\tablehead{\colhead{HR} & \colhead{HD} & \multicolumn2c{$V$} & \multicolumn2c{\bv} &
\colhead{Spectral} & \colhead{$v \sin i$} & \multicolumn2c{$\varpi$} & \multicolumn2c{$M_{V}$} \\
\colhead{} & \colhead{} & \multicolumn2c{(mag)} & \multicolumn2c{(mag)} &
\colhead{Type} & \colhead{(km s$^{-1}$)} & \multicolumn2c{(mas)} & \multicolumn2c{(mag)}}
\decimalcolnumbers
\decimals
\startdata
5085 & 117376 & 5.458 & 0.002 & A0V & 155\tablenotemark{a} & 13.8279 & 1.162  \\
5107 & 118098 & 3.367 & 0.121 & A2IVn & 222 & 43.7467 & 1.572  \\
5112 & 118232 & 4.671 & 0.132 & A5V & 143\tablenotemark{a} & 17.5952 & 0.898  \\
5116 & 118295 & 6.832 & 0.194 & A9V & 124\tablenotemark{a} & 8.3811 & 1.434  \\
5127 & 118623 & 4.959 & 0.200 & F0Vp & 207 & 17.7243 & 1.202  \\
5162 & 119476 & 5.840 & 0.075 & A1.5V & 140\tablenotemark{a} & 14.2016 & 1.602  \\
5244 & 121607 & 5.904 & 0.206 & A7V & 135\tablenotemark{a} & 11.3738 & 1.184  \\
5290 & 123255 & 5.462 & 0.323 & F0IV & 142\tablenotemark{a} & 17.5627 & 1.685  \\
5392 & 126248 & 5.099 & 0.125 & A3V & 201 & 21.0624 & 1.716  \\
5532 & 130917 & 5.784 & 0.064 & A1Vn & 222 & 9.7343 & 0.724  \\
5715 & 136729 & 5.644 & 0.112 & A2.5IV & 143\tablenotemark{a} & 15.2070 & 1.554  \\
5818 & 139493 & 5.765 & 0.040 & A0Vn & 200\tablenotemark{a} & 13.9077 & 1.481  \\
5867 & 141003 & 3.645 & 0.080 & A1V & 207 & 21.5924 & 0.316  \\
5895 & 141851 & 5.093 & 0.122 & A3Vp & 196,17\tablenotemark{a} & 17.3357 & 1.288  \\
5919 & 142500 & 6.272 & 0.163 & A5Vn & 207\tablenotemark{a} & 12.4962 & 1.756  \\
5960 & 143466 & 4.961 & 0.249 & F0V & 129\tablenotemark{a} & 29.6204 & 2.319  \\
6004 & 144874 & 5.641 & 0.188 & A6V & 142\tablenotemark{a} & 10.0488 & 0.652  \\
6033 & 145607 & 5.428 & 0.119 & A3Vn & 175 & 12.7258 & 0.952  \\
6277 & 152569 & 6.252 & 0.259 & F0Vn & 177\tablenotemark{a} & 13.2145 & 1.857  \\
6317 & 153653 & 6.572 & 0.214 & A5V & 170 & 15.0258 & 2.457  \\
6507 & 158352 & 5.419 & 0.217 & A8V & 180 & 15.6711 & 1.395  \\
6877 & 168914 & 5.121 & 0.195 & A5IV & 196 & 11.8758 & 0.495  \\
8264 & 205767 & 4.683 & 0.160 & A6V & 170 & 18.2324 & 0.987  \\
\enddata
\tablecomments{The table columns list the following parameters:
(1) HR number \citep{hof82}; 
(2) Henry Draper number (HD); 
(3) visual ($V$) magnitude; 
(4) \bv value from the Tycho2 catalogue \citep{hog00}; 
(5) spectral type by AM95;
(6) $v \sin i$ measures from \citet{roy02} (see text);
(7) parallax ($\varpi$) from GEDR3 as listed on the SIMBAD web portal
(8) $M_{V}$ corrected for reddening (see text).}
\tablenotetext{a}{Our measure.}
\end{deluxetable}

\section{Observations, Reductions, and Measurements}
\subsection{Observations}
This long-term survey was begun in 1992 December with data obtained 
in most years up to early 2024. The KPNO 0.9 m coud\'{e} auxillary feed 
telescope and 2.1 m coud\'{e} spectrograph (hereafter KPNO CF) were used for a large 
share of the observations through mid-2014. The spectrographic setup 
consisted of the ``A'' grating (302 mm x 361 mm, 632 grooves mm$^{-1}$) 
in second order and the folded Schmidt camera number 5. Various CCD 
detectors were used during the interval, and the resolution was 
either 0.11 \AA\,  pix$^{-1}$ with 0.015 mm pixels (TI5, F3KB, LB1A), 
or 0.18 \AA\, pix$^{-1}$ with 0.024 mm pixels (T1KA, T2KA, T2KB). The 
spectral region was generally centered near 4500 \AA\, and covered 
approximately 88 to 368 \AA\ depending on which CCD was used. After 2014,
a number of night's data were obtained with the WIYN 3.5 m telescope utilizing
the Hydra multifiber positioner in single fiber mode, feeding the bench
spectrograph. Lastly, the Mayall 4 m echelle and RC spectrographs were used
on several occasions.

We obtained additional spectra of many of the survey stars at 
Fairborn Observatory in southeast Arizona \citep{ew04} with the 
Tennessee State University 2~m AST, a fiber-fed echelle spectrograph 
\citep{ew07} and two different detectors. For the early observations 
from 2004 through the first half of 2011, we employed a 2048 
$\times$ 4096 SITe ST-002A CCD with 15 $\mu$m pixels. In the summer 
of 2011 that CCD was replaced by a Fairchild 486 CCD that has a 
4K $\times$ 4K array of 15 $\mu$m pixels \citep{fetal13}. While the 
early CCD spectra span the wavelength range 4920 to 7100~\AA, the 
new spectra range from 3800 to 8260~\AA. \citet{fetal15} have 
provided additional information about the spectra from these two 
detectors. In 2024 March operation of the AST was discontinued 
because of a lack of funding.

\subsection{Data Reductions}
\subsubsection{KPNO Data}
The spectra were processed and extracted with standard 
IRAF\footnote{IRAF is distributed by the Community Science and Data 
Center at NSF NOIRLab, which is managed by the Association of 
Universities for Research in Astronomy (AURA) under a cooperative 
agreement with the U.S. National Science Foundation.}
routines, after which RVs were measured with the IRAF 
task FXCOR \citep{fit93}. Templates for the cross correlation from 
the early-type star list of \citet{fek99} and Fekel (private 
communication) were observed along with the survey stars.
Long-term observations of several of these proposed constant velocity
standards eventually showed variable velocity, requiring some of the 
apparently constant survey stars to be used as secondary standards.

The final RVs were determined in a two-step process. 
All spectra were measured initially with FXCOR with the use of the 
standard stars as described above. These RVs include 
corrections for velocity shifts in the spectrograph as determined
by the hourly ThAr lamp spectra. The average velocity was then 
calculated for each star with all data. Next, nightly correction 
curves were formed with the use of data points for each star on a 
given night consisting of the difference between a star's average 
velocity and its nightly velocity as a function of the time of 
observation. With the IRAF task CURFIT, coefficients were determined 
for each night's difference values with outliers due to stars with 
variable velocity rejected. The correction coefficients were applied to 
each program star as a function of time (in HJD) for each night. This 
second RV correction step improved the velocity accuracy 
significantly. For 68 apparently constant velocity stars without 
corrections, the average $\sigma$ = 1.05 km s$^{-1}$, while after 
correction $\sigma$ = 0.50 km s$^{-1}$.  This RV 
correction technique is similar to that used by \citet{mar89} in a 
search for substellar companions to M dwarfs. Their average standard 
deviation for stars not obviously varying was 0.23 km s$^{-1}$.

As expected, the measuring accuracy is a function of $v \sin i$,
as is evident in Figure~\ref{SDvsVsini}, where each star's $\sigma$ 
value is plotted against $v \sin i$. A curve has been fit to the 
$\sigma$ values for the apparently non-binary stars (red filled diamonds) 
and has the relation 
$\sigma_{v \sin i} = 0.2202 + 0.0076 (v \sin i)$ km~s$^{-1}$. An iterative
$\sigma$ rejection was done using the IRAF task CURFIT with the high rejection
parameter set to $2 \sigma$.
For the range of $v \sin i$ included in the stars surveyed here 
(15--229 km~s$^{-1}$), the calculated values of $\sigma_{v \sin i}$ are 
0.33 km~s$^{-1}$ to 1.96 km~s$^{-1}$.

\subsubsection{Fairborn Observatory Data}
Spectra from the AST and bench-mounted echelle spectrograph at Fairborn 
Observatory were calibrated and extracted with a program based on one 
written by \citet{hal94} and \citet{hal95} and described in more detail 
in \citet{ew07}.

\citet{fetal09} have provided a general description of the typical
velocity reduction, which is generally done in the spectral region
4920 to 7100~\AA\ for solar-type stars.  Because of the hotter 
temperature range of the A-type stars as well as our search for 
cooler companions, we employed several different line lists. 
For the slowest rotating A-type stars, $v \sin i$ $\lesssim$40 
km~s$^{-1}$, we primarily used our usual A-star line list, which 
contains ionized metal lines and covers the above mentioned 
wavelength region. For the somewhat more rapidly rotating A-type 
systems, those with $v \sin i$ from $\sim$40 km~s$^{-1}$ to 
$\sim$80 km~s$^{-1}$, we generally used a list of ionized metal 
lines in the blue region of the spectrum. Those lines have 
wavelengths that range from 4400 to 4920~\AA~\citep{getal22} 
and are typically much stronger than the ionized metal lines in 
the 4920 to 7100~\AA\ region. For the most rapidly rotating 
A-type stars, $v \sin i$ $\gtrsim$80 km~s$^{-1}$, we used a subset 
consisting of the more isolated lines of the above mentioned line 
list in the blue region of the spectrum that ranges from 
4400 to 4920~\AA. Finally, several of the cooler secondaries
were detectable only with our solar-type star line list.

To fit the individual line profiles we used a rotational broadening 
function \citep{slf11, fg11}. In the cases where blended double lines
were present, we simultaneously fitted the blended profiles with two 
separate rotational broadening functions, and allowed the depth,
width, and velocity of the fits to vary. The Fairborn velocities 
determined with the solar-type star line list have been shifted to 
the zero point of \citet{sca10}. For the velocities determined with 
the blue wavelength line lists, the measured average Fairborn RVs of 
six apparently constant stars (HRs 675, 1389, 2489, 2818, 
4359, and 8404) were compared to their average RVs determined 
by \citet{fek99}.
The average of those 6 differences was $+$0.2 $\pm$ 0.3 km~s$^{-1}$, 
and so no zero point correction was made to the RVs measured in that 
wavelength region.

\section{Radial Velocity Results, Variability and Orbit Calculation}
Over the course of the 34 years of data acquisition, approximately 
6000 RVs have been measured, often with 2 or more RVs 
per spectrum for multiple systems. The individual RVs for the 
low $v \sin i$ binary stars are listed in Table~\ref{RVs_lo_vsini} 
with the values in the columns described in the note following 
the table. Table~\ref{RVs_lo_vsini_var} contains the individual RVs for
stars that may have variable RVs. Table~\ref{RVs_hi_vsini} lists
similar data for the stars with high $v \sin i$.
For the stars without orbits and no
suspected variability, the individual RVs are available upon request. 
A summary table listing the average RV for the stars without orbits
and additional statistics for each of the low $v \sin i$ stars 
is compiled in Table~\ref{RVsum_lo_vsini}. 
Table~\ref{RVsum_hi_vsini} provides similar data
for the 23 high $v \sin i$ stars. 

\begin{figure}
\begin{center}
\hspace{2em}
\includegraphics[width=5in,keepaspectratio]{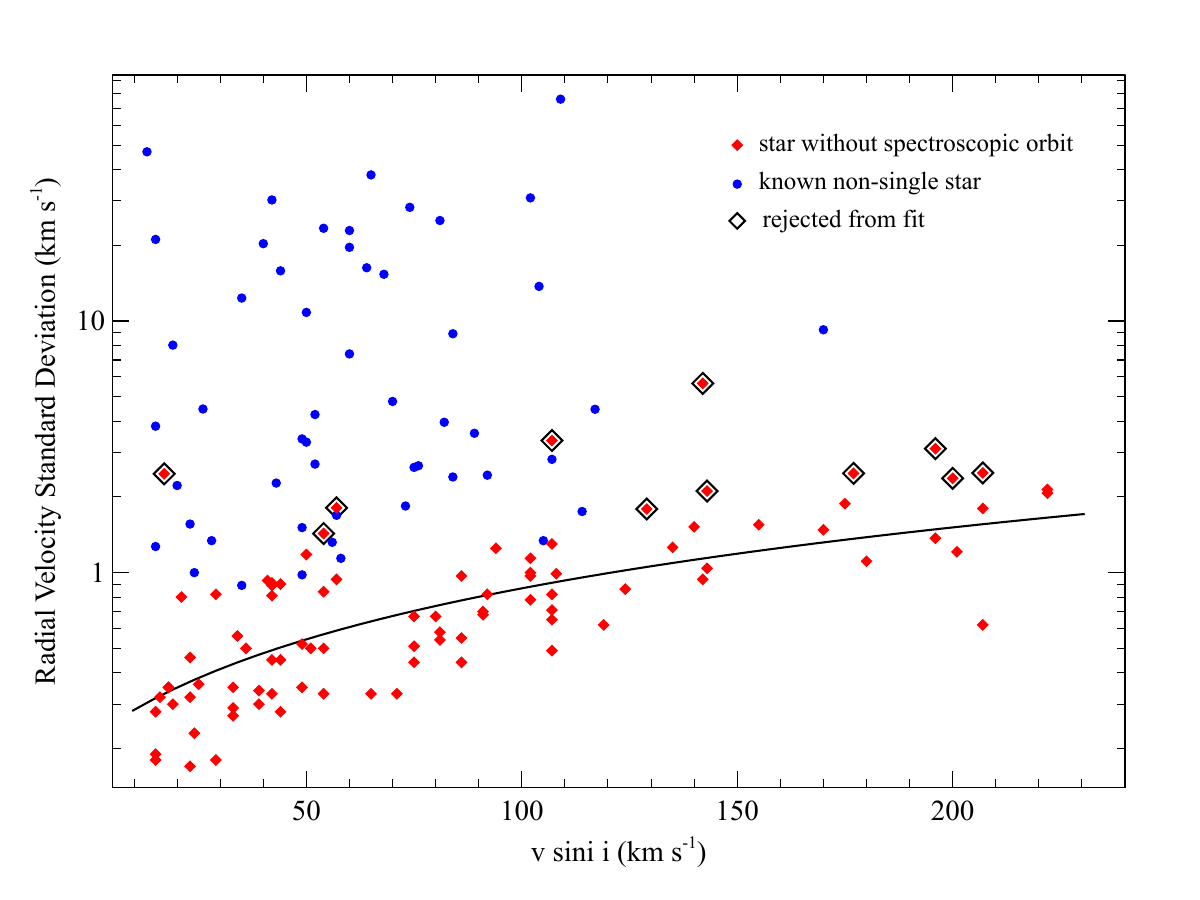}
\figcaption{Fit to the standard deviation (SD) versus $v \sin i$ 
(black curve) for apparently constant RV stars. A value of 2$\sigma$ 
was used to iteratively reject deviant points. The relatively large 
values of SD for a number of the stars without orbits are likely 
due to $\delta$ Sct pulsations and undetected binaries.\label{SDvsVsini}}
\end{center}
\end{figure}

At the time that our survey commenced, none of the stars had published
orbits although a few were suspected of having RV variations,
mainly from surveys conducted in the early 1900s. Two stars were known
as eclipsing binaries, but the relatively new field of speckle imaging had
not yet been employed for a sufficiently long interval to produce many orbits of
sub-arc second pairs. At present, 23 systems, including both the low and high 
$v \sin i$ samples have orbits or companions determined from speckle 
observations.
A number of the stars thought to be single have standard deviations larger 
than expected based on their $v \sin i$ values (Figure~\ref{SDvsVsini}
and show variable and asymmetric-line cross-correlation
profiles that may be caused by $\delta$ Sct or $\gamma$ Doradus
variations. We also detected several stars with composite spectra, which  
show both broad and narrow-lined components. All of these factors contribute to
uncertainties in the RV measurements. Additional details on 
selected individual stars with low $v \sin i$ can be found 
in Appendix A.1, while in Appendix A.2 we comment on all 23 
with high $v \sin i$.

Early in our project, we used a Fortran program \citep{mon79} to determine
orbital periods and then orbits with the Lehmann-Filhes method.  After
the computer system on which it resided failed, we determined preliminary
periods with the period-finding program PDFND, which uses the least-string
method \citep{bopp70}, or PGRAM, a program that fits data to a sine
curve and identifies the minimum sum of the residuals squared as the likely
period (G. Henry, private communication). Initial orbits were
determined with the program BISP \citep{wol67}, which uses the
Wilsing-Russell Fourier analysis method \citep{wil93, rus02}.
We refined the single-lined and double-lined orbits with the programs
SB1 and SB2, respectively, which are slightly modified programs from
\citet{bar67} that utilize differential corrections.

For the triple system HR~2103, we also used the general least 
squares program (GLS) of \citet{dan66}. That program simultaneously 
determines long-period and short-period elements for triple systems.

\begin{figure}
\begin{center}
\hspace{2em}
\includegraphics[width=6in,keepaspectratio]{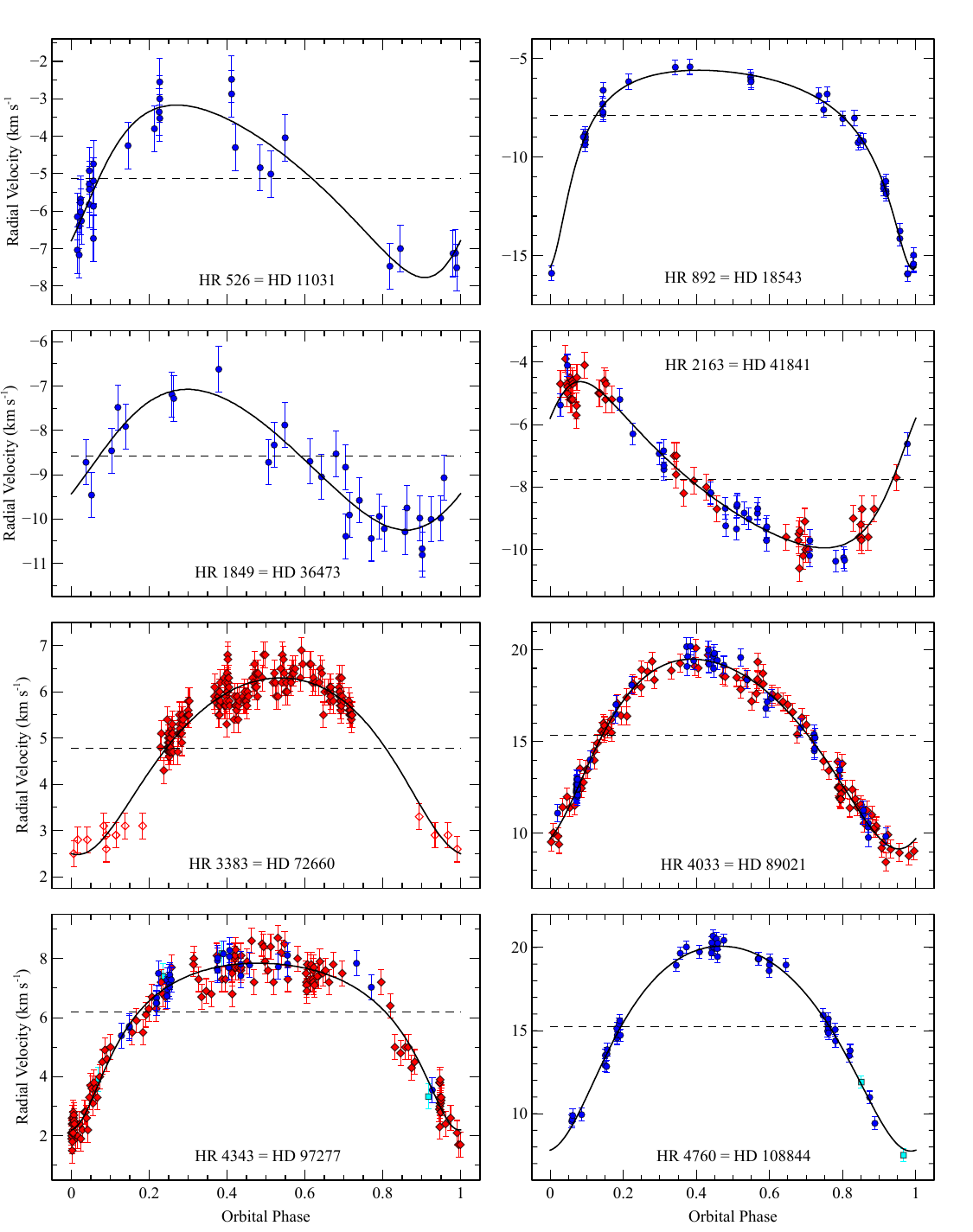}
\figcaption{Calculated orbital velocity curves and measured RVs. 
Error bars are the $\sigma_{\rm {RV}}$ of the ($O - C$) values for each 
star. The horizontal dashed line indicates the systemic velocity. 
The filled red diamonds are RVs from Fairborn Observatory; the 
filled blue circles are from Kitt Peak telescopes. For HR 3383, the open red
diamonds are from the KPNO CF telescope by FCF; for HR 4343, the 4 cyan squares are from 
\citet{gren99}; for HR 4760, the 2 cyan squares are from \citet{har37}.
\label{orbplots1}}
\end{center}
\end{figure}

\begin{figure}
\begin{center}
\hspace{2em}
\includegraphics[width=6in,keepaspectratio]{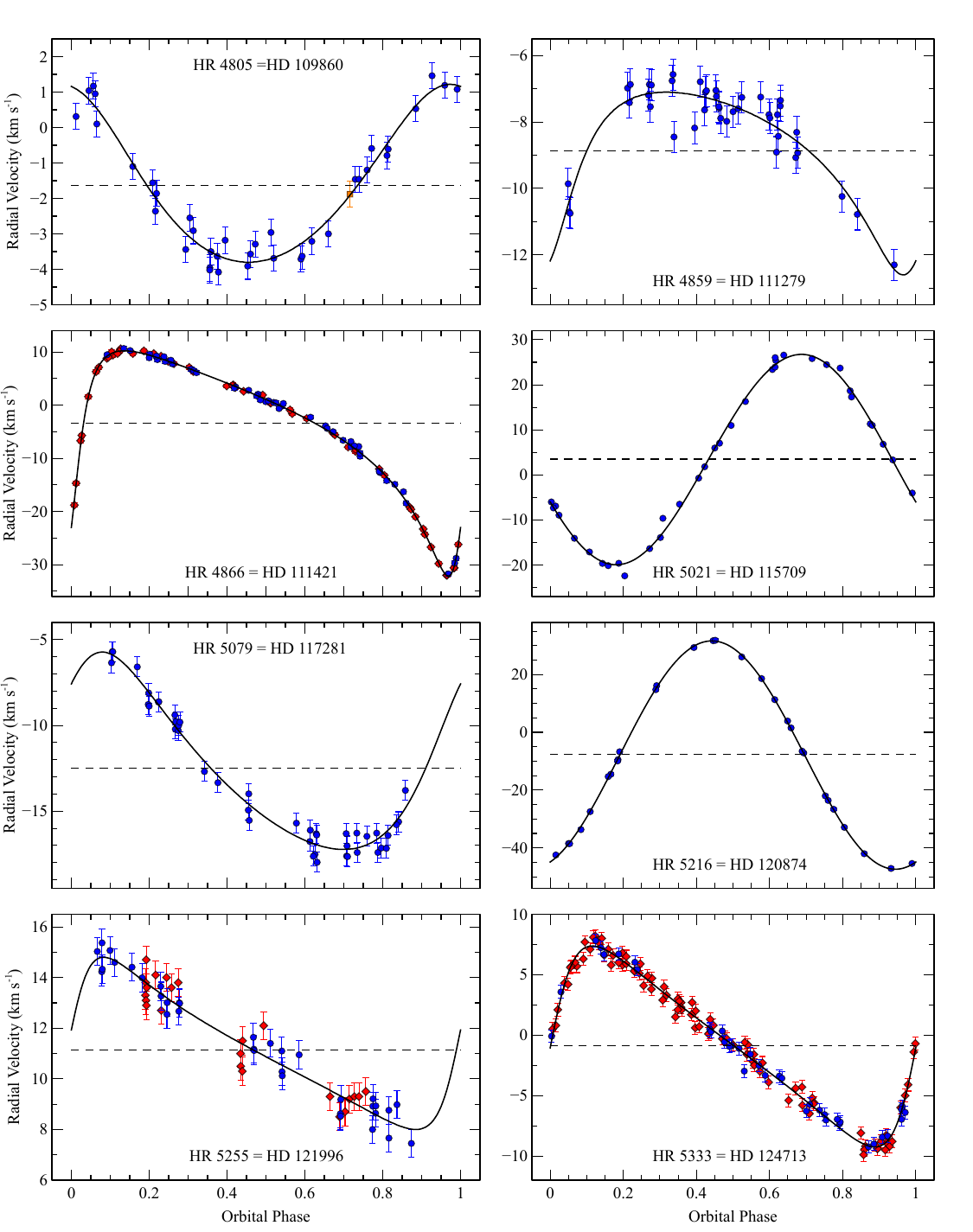}
\figcaption{Calculated orbital velocity curves and measured RVs. 
Error bars are the $\sigma_{\rm {RV}}$ of the ($O - C$) values for each 
star. The horizontal dashed line indicates the systemic velocity. 
The filled red diamonds are RVs from Fairborn Observatory; the 
filled blue circles are from Kitt Peak telescopes. For HR 4805, the orange
square is from \citet{gren99}.
\label{orbplots2}}
\end{center}
\end{figure}

\begin{figure}
\begin{center}
\hspace{2em}
\includegraphics[width=6in,keepaspectratio]{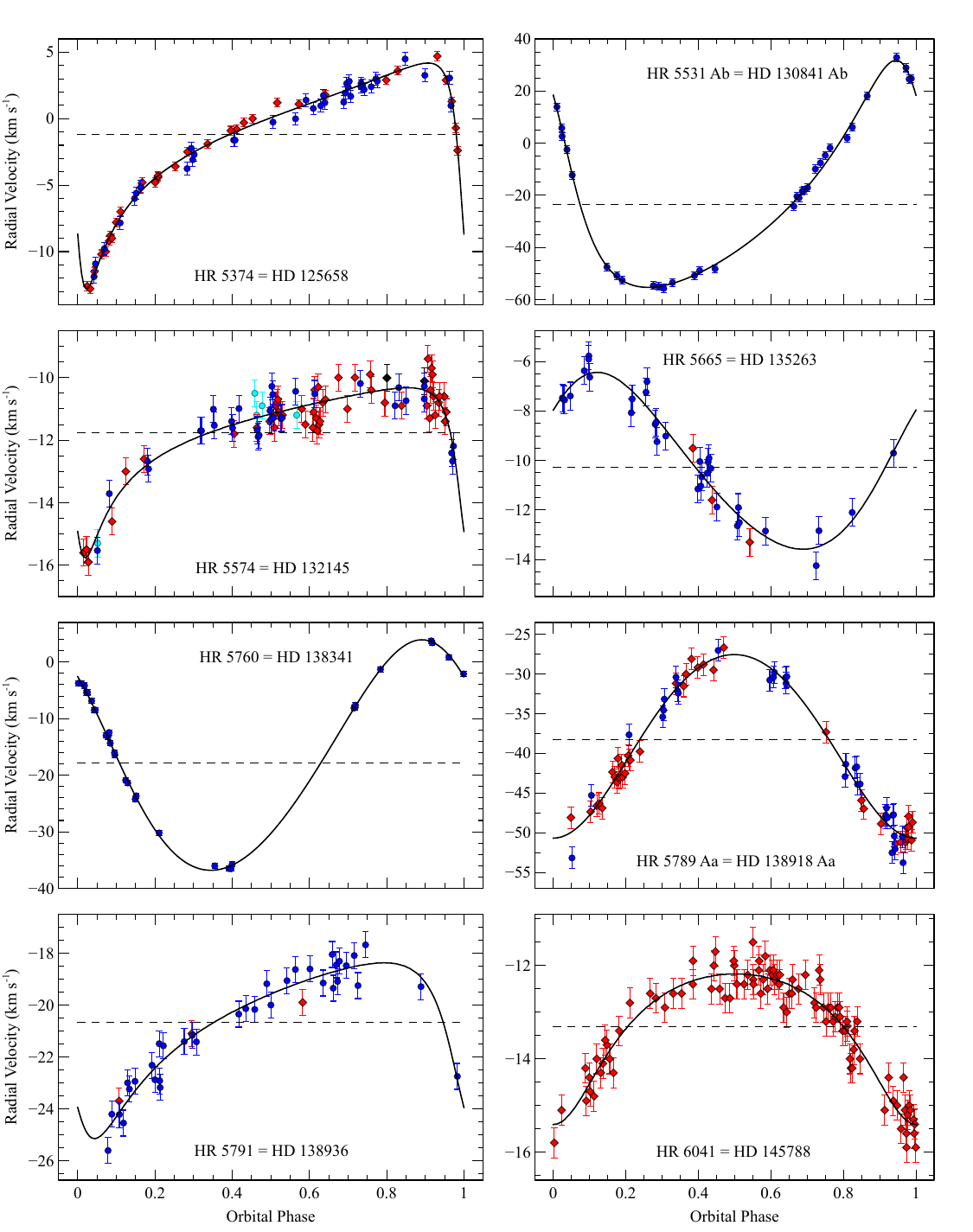}
\figcaption{Calculated orbital velocity curves and measured RVs. 
Error bars are the $\sigma_{\rm {RV}}$ of the ($O - C$) values for each star. 
The horizontal dashed line indicates the systemic velocity. The 
filled red diamonds are RVs from Fairborn Observatory; the filled 
blue circles are from Kitt Peak telescopes. For HR 5574, the 5 cyan
circles are from \citet{har37}, the 2 black diamonds are from 
\citet{roy14}.
\label{orbplots3}}
\end{center}
\end{figure}

\begin{figure}
\begin{center}
\hspace{2em}
\includegraphics[width=6in,keepaspectratio]{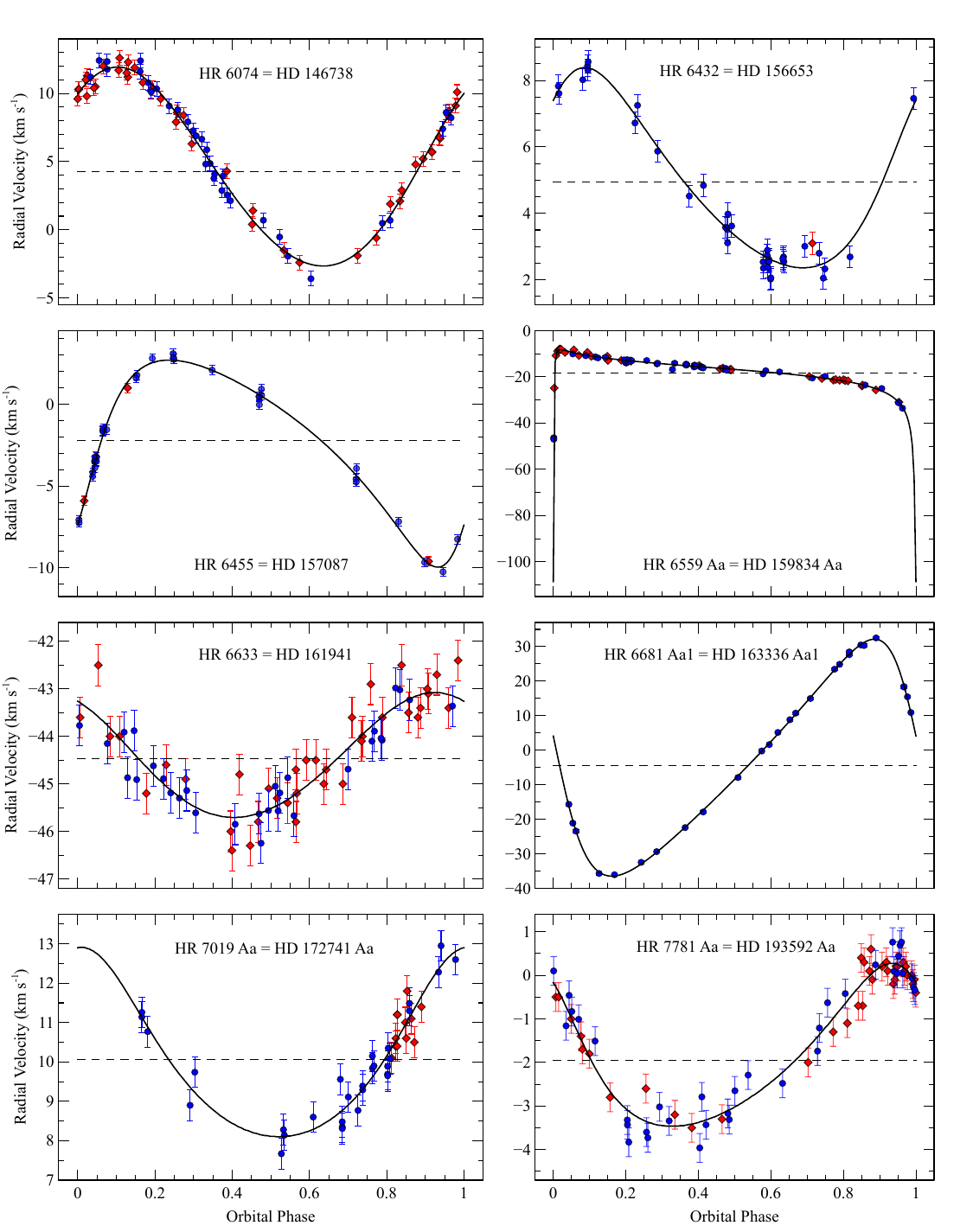}
\figcaption{Calculated orbital velocity curves and measured RVs. 
Error bars are the $\sigma_{\rm {RV}}$ of the ($O - C$) values for each 
star. The horizontal dashed line indicates the systemic velocity. 
The filled red diamonds are RVs from Fairborn Observatory; the 
filled blue circles are from Kitt Peak telescopes.
\label{orbplots4}}
\end{center}
\end{figure}

\begin{figure}
\begin{center}
\hspace{2em}
\includegraphics[width=6in,keepaspectratio]{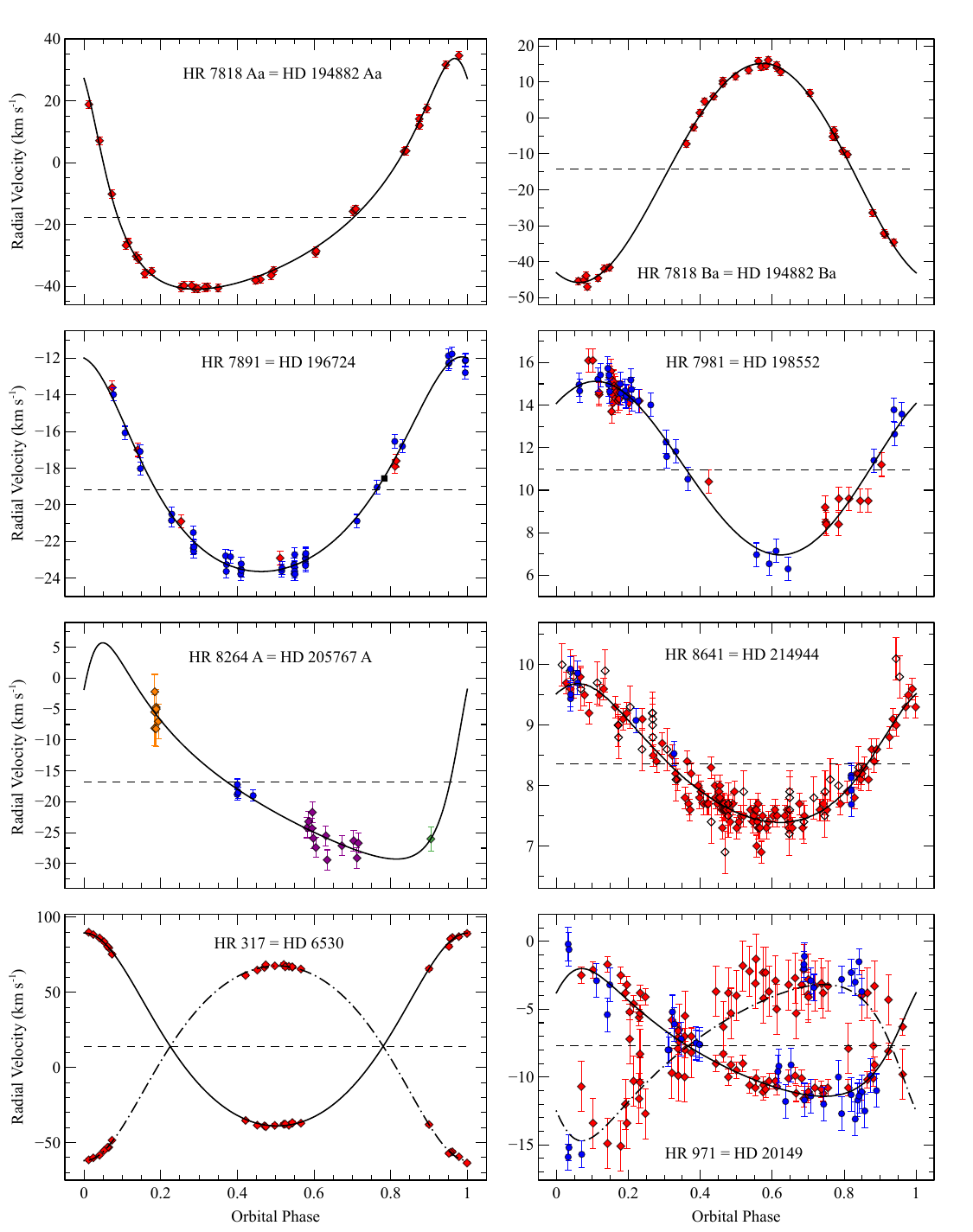}
\figcaption{Calculated orbital velocity curves and measured RVs.
Error bars are the $\sigma_{\rm {RV}}$ of the ($O - C$) values for each
star. The horizontal dashed line indicates the systemic velocity.
For double lined systems, the dash-dot curve is the calculated secondary
RV curve.
The filled red diamonds are RVs from Fairborn Observatory; the
filled blue circles are from Kitt Peak telescopes. For HR 7891, the
black square is from \citet{trif20}. For HR 8264 A, 
the orange diamonds are RVs from \citet{abt65}; the dark magenta diamonds 
are from \citet{abt74}; the light green diamond is from \citet{har37};
the blue circles are from this study and Kitt Peak telescopes by DWW.
For HR 8641, open diamonds are from KPNO CF telescope by FCF.
\label{orbplots5}}
\end{center}
\end{figure}

\begin{figure}
\begin{center}
\hspace{2em}
\includegraphics[width=6in,keepaspectratio]{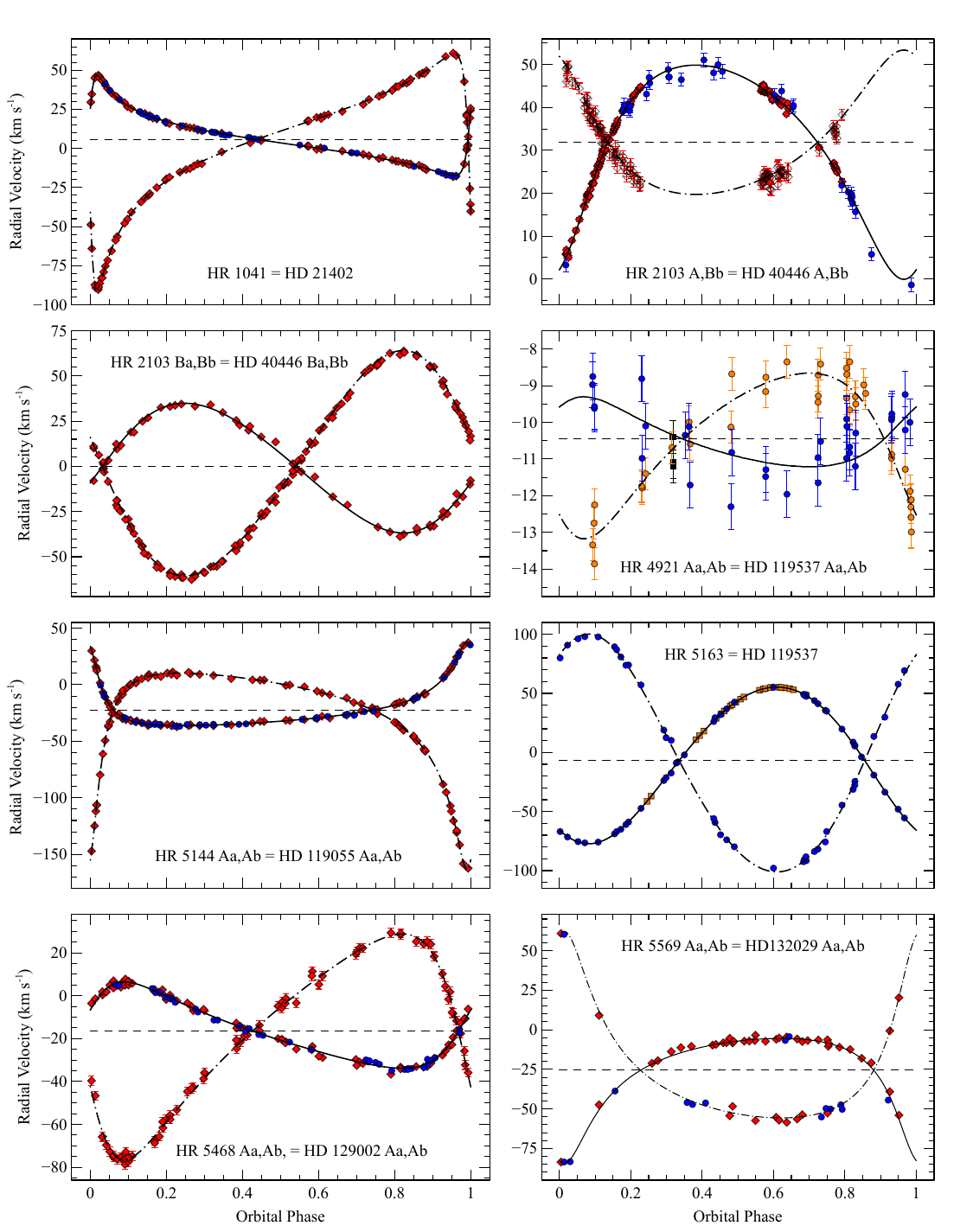}
\figcaption{Calculated orbital velocity curves and measured RVs.
Error bars are the $\sigma_{\rm {RV}}$ of the ($O - C$) values for each
star. The horizontal dashed line indicates the systemic velocity.
The dash-dot curve is the calculated secondary RV curve.
The filled red diamonds are RVs from Fairborn Observatory; the
filled blue circles are from Kitt Peak telescopes.
For HR 4921, the orange circles are
from the KPNO CF telescope by HAA and DWW; the 3
black squares are from \citet{gren99}. For HR 5163, the orange squares
are from \citet{trif20}.
\label{orbplots6}}
\end{center}
\end{figure}

\begin{figure}
\begin{center}
\hspace{2em}
\includegraphics[width=3.2in,keepaspectratio]{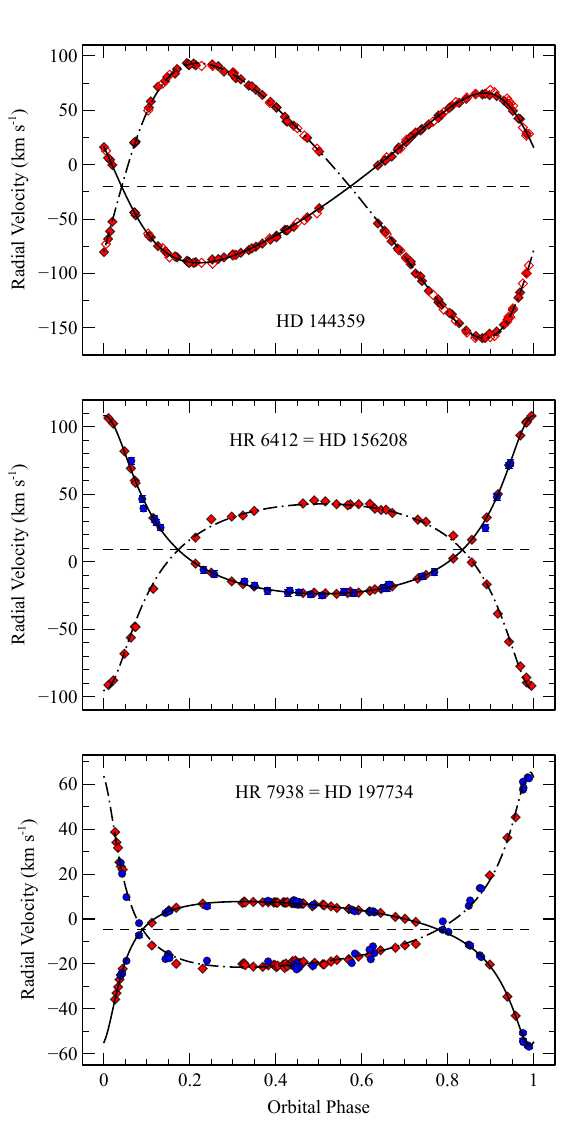}
\figcaption{Calculated orbital velocity curves and measured RVs.
Error bars are the $\sigma_{\rm {RV}}$ of the ($O - C$) values for each
star. The horizontal dashed line indicates the systemic velocity.
The dash-odot curve is the calculated secondary RV curve.
The filled red diamonds are RVs from Fairborn Observatory; the
filled blue circles are from Kitt Peak telescopes. 
For HD 144359, the open red diamonds are from the KPNO CF telescope by FCF.
\label{orbplots7}}
\end{center}
\end{figure}

Table~\ref{Orbel_sb1} is a listing of the orbital elements for the
single-lined SBs. A description of the parameters in the table columns
is in the notes immediately following the table.
Similarly, Table~\ref{Orbel_sb2} contains the orbital
elements for the double-lined and higher multiplicity SBs.
Figures~\ref{orbplots1} through Figure~\ref{orbplots7} illustrate the
computed orbital RV curves with the associated data points and
their errors.

Partial orbital elements and other properties for the 17 visual 
binaries that have orbits (VBO)
given in the \textit{Sixth Catalog of Orbits of Visual Binary Stars}
(ORB6) \citep{hart01} are listed in Table~\ref{VBwithOrbel} along with findings from this study.
Nine other stars have detections of a companion by speckle interferometry as noted in the
\textit{Fourth Catalog of Interferometric Measurements of Binary Stars} (INT4)
\citep{hart01}, or have a close companion listed in the \textit{Washington Double Star Catalog} (WDS).
Their data are also given in the table.

For the group of 23 high $v \sin i$ stars, only a single  
preliminary spectroscopic orbit (for HR~8264) has been determined, and 
only HR 5895 displayed a composite spectrum. 
VB orbits have been determined for both systems with periods of 25.5 
\citep{gon02} and 64.8 yr \citep{tok18}, respectively. 
Table~\ref{VBwithOrbel} lists additional 
information for these and other astrometric binaries, while Section 
5.1 discusses the frequency of binaries in this subset of stars.

\begin{deluxetable}{c c c D D c c}
\tablecolumns{7}
\tablewidth{0pt}
\tabletypesize{\scriptsize}
\tablecaption{Individual Radial Velocity Measurements -- Low $v \sin i$ Stars
With Orbits\label{RVs_lo_vsini}}
\tablehead{
\colhead{HR} & \colhead{HJD -- 2,400,000} & \colhead{Orb. Phase} & \multicolumn2c{RV} &
  \multicolumn2c{$(O-C)$} & \colhead{Weight} & \colhead{Source\tablenotemark{a}} \\
\colhead{} & \colhead{(days)} & \colhead{} & \multicolumn2c{(km s$^{-1}$)} &
  \multicolumn2c{(km s$^{-1}$)} &  \colhead{} & \colhead{}}
\decimalcolnumbers 
\startdata
5569A & 49455.9680 & 0.6441 &  -4.32 &  1.18 & 1.00 & KPNO \\
5569A & 50215.7496 & 0.1558 & -38.70 & -0.47 & 1.00 & KPNO \\
5569A & 56792.8688 & 0.0300 & -83.37 & -1.43 & 1.00 & KPNO \\
5569A & 56856.7068 & 0.9215 & -44.46 & -4.32 & 1.00 & KPNO \\
5569A & 57534.8130 & 0.0131 & -83.53 &  0.83 & 1.00 & KPNO \\
5569A & 58507.9738 & 0.8470 & -18.00 &  0.23 & 1.00 & FO \\
5569A & 58562.9936 & 0.4772 &  -5.40 &  2.40 & 1.00 & FO \\
5569A & 58589.9429 & 0.2757 & -19.50 &  0.43 & 1.00 & FO \\
5569A & 58611.8876 & 0.9259 & -39.10 &  3.14 & 1.00 & FO \\
5569A & 58639.8205 & 0.7535 & -10.40 & -1.89 & 1.00 & FO \\
\enddata
\tablecomments{The table columns list the following parameters:
(1) HR number \citep{hof82};
(2) heliocentric Julian Day -- 2,400,000;
(3) fractional orbital phase;
(4) the RV of the primary or secondary star; 
(5) the observed -- calculated ($O-C$) RV difference of the observation;
(6) the weight of the observation used in the orbit calculation;
(7) the source of the observation.}
\tablenotetext{a}{KPNO: Kitt Peak observation by HAA and DWW; FO: 
Fairborn Observatory observation by FCF.
\\
\\
(This table is published in its entirety in machine-readable form 
in the online article.)}
\end{deluxetable}

\begin{deluxetable}{c c D c}
\tablecolumns{4}
\tablewidth{0pt}
\tabletypesize{\scriptsize}
\tablecaption{Individual Radial Velocity Measurements -- Low $v \sin i$ Stars
That May Have Variable Radial Velocity\label{RVs_lo_vsini_var}}
\tablehead{
\colhead{HR} & \colhead{HJD -- 2,400,000} & \multicolumn2c{RV} & \colhead{Source\tablenotemark{a}} \\
\colhead{} & \colhead{(days)} & \multicolumn2c{km s$^{-1}$} & \colhead{}}
\decimalcolnumbers
\startdata
1133 & 49749.6325 & 26.13 & KPNO \\
1133 & 50724.9353 & 26.21 & KPNO \\
1133 & 50727.0061 & 26.48 & KPNO \\
1133 & 50728.0067 & 26.43 & KPNO \\
1133 & 53649.8460 & 24.82 & KPNO \\
1133 & 58488.5796 & 24.4  & FO \\
1133 & 58546.6989 & 24.4  & FO \\
1133 & 58761.7409 & 24.2  & FO \\
1133 & 58805.8752 & 24.2  & FO \\
1133 & 58872.7186 & 24.3  & FO \\
\enddata
\tablecomments{The table columns list the following parameters:
(1) HR number \citep{hof82};
(2) heliocentric Julian day -- 2,400,000;
(3) the RV of the primary or secondary star;
(4) the source of the observation.}
\tablenotetext{a}{KPNO: Kitt Peak observation by HAA and DWW; FO:
Fairborn Observatory observation by FCF.
\\
\\
(This table is published in its entirety in machine-readable form
in the online article.)}
\end{deluxetable}

\begin{deluxetable}{c c c c}
\tablecolumns{4}
\tablewidth{0pt}
\tabletypesize{\scriptsize}
\tablecaption{Individual Radial Velocity Measurements -- High $v \sin i$ Stars\label{RVs_hi_vsini}}
\tablehead{
\colhead{HR} & \colhead{HJD--2400000} & \colhead{RV} &
  \colhead{Source\tablenotemark{a}} \\
\colhead{} & \colhead{(days)} & \colhead{(km s$^{-1}$)} & \colhead{}}
\colnumbers
\startdata
5085 & 56832.7246  & $-$6.7  & KPNO \\
5085 & 56836.6719  & $-$9.8  & KPNO \\
5085 & 56856.6551  & $-$9.4  & KPNO \\
5085 & 57209.7750  & $-$8.1  & KPNO \\
5085 & 57534.8232  & $-$6.0  & KPNO \\
5085 & 60357.9380  & $-$4.6  & FO \\
5085 & 60359.9380  & $-$6.6  & FO \\
5085 & 60360.9368  & $-$6.8  & FO \\
5085 & 60361.9368  & $-$5.2  & FO \\
5085 & 60363.9368  & $-$6.9  & FO \\
\enddata
\tablecomments{The table columns list the following parameters:
(1) HR number \citep{hof82};
(2) heliocentric Julian Day--2400000;
(3) measured RV;
(4) source of the observation.}
\tablenotetext{a}{KPNO: Kitt Peak observation by DWW and HAA; FO:
Fairborn Observatory observation by FCF.
\\
\\
(This table is available in its entirety in machine-readable form in
the online article.)}
\end{deluxetable}

\clearpage
\startlongtable
\begin{deluxetable}{c c c c c r r}
\tablecolumns{7}
\tablewidth{0pt}
\tablecaption{Radial Velocity Summary -- Low $v \sin i$ Stars\label{RVsum_lo_vsini}}
\tablehead{
\colhead{HR} & \colhead{$<$RV$>$} & \colhead{$\sigma$} & \colhead{$\frac{\sigma}{\sqrt{N}}$}
       & \colhead{$N_{\rm {obs}}$} & \colhead{$\Delta \, t$} & \colhead{Status\scriptsize\tablenotemark{a}} \\
\colhead{} & \colhead{(km s$^{-1}$)} & \colhead{(km s$^{-1}$)} & \colhead{(km s$^{-1}$)}
  & \colhead{} & \colhead{(days)} & \colhead{}}
\decimalcolnumbers
\tabletypesize{\scriptsize}
\startdata
114 &   $-$9.88 & 0.80 & 0.15 & 29 & 9432 & C,CPM,$\delta$ Sct \\
196 &   $-$9.95 & 0.50 & 0.08 & 39 & 9432 & C,VB,CPM \\
317 & \nodata & \nodata & \nodata & 25 & 1051 & SB2O,NSS \\
526 &   \nodata & \nodata & \nodata & 33 & 10918 & SB1O,VBO,CPM \\
723 &     21.82 & 0.18 & 0.03 & 40 & 6144 & C,CPM \\
892 &  \nodata  & \nodata & \nodata & 36 & 10884 & SB1O,CPM? \\
971 &   \nodata & \nodata & \nodata & 81 & 11345 & SB2O \\
1018 &    14.82 & 0.35 & 0.07 & 22 & 4632 & C \\
1041 &  \nodata & \nodata & \nodata & 102 & 10885 & SB2O \\
1086 &    26.94 & 0.27 & 0.06 & 23 & 4633 & C,VBO \\ 
1133 &    25.47 & 0.91 & 0.15 & 37 & 4633 & Var?,PMA \\
1181 &    29.11 & 0.23 & 0.05 & 23 & 4633 & C \\
1675 &  $-$4.99 & 0.19 & 0.03 & 35 & 1711 & C \\
1849 &  \nodata & \nodata & \nodata & 29 & 10885 & SB1O \\
2103 &  \nodata & \nodata & \nodata & 144 & 10540 & SB3O,CPM \\
2163 &  \nodata & \nodata & \nodata & 76 & 10872 & SB1O,CPM,PMA \\
2818 &    27.13 & 0.30 & 0.05 & 31 & 7406 & C \\
3383 & \nodata & \nodata & \nodata & 194 & 12379 & SB1O,NSS \\
3657 &  $-$8.76 & 0.89 & 0.12 & 55 & 10156 & Var? \\
3689 &  $-$6.95 & 0.33 & 0.06 & 29 & 10647 & C \\
3702 &    30.36 & 0.18 & 0.03 & 43 & 10160 & C \\
3822 &    20.34 & 1.81 & 0.30 & 36 & 10884 & Var?,VB,CPM \\
3861 &    12.43 & 0.30 & 0.05 & 40 & 10169 & C \\
3894 & $-$14.11 & 1.15 & 0.51 & 5 & 10381 & Var,VBO \\
4033 &  \nodata & \nodata & \nodata & 159 & 11343 & SB1O,PMA \\
4331 & $-$11.56 & 0.17 & 0.03 & 26 & 5942 & C \\
4343 &  \nodata & \nodata & \nodata & 168 & 11047 & SB1O,PMA \\
4359 &     7.74 & 1.20 & 0.10 & 143 & 10884 & C \\
4378 & $-$37.78 & 0.32 & 0.04 & 51 & 10613 & C \\
4585 &  $-$0.74 & 0.90 & 0.15 & 35 & 4494 & Var?,CPM? \\
4670 &  $-$9.05 & 0.51 & 0.08 & 38 & 10564 & C \\
4717 &  $-$0.61 & 0.50 & 0.10 & 28 & 9604 & C,VBO,PMA \\
4738 &     0.23 & 0.67 & 0.13 & 26 & 9472 & C \\
4760 & \nodata & \nodata & \nodata & 49 & 33313 & SB1O,PMA \\
4789 & $-$12.35 & 7.40 & 1.42 & 27 & 7634 & Var,VBO \\
4805 & \nodata & \nodata & \nodata & 39 & 9472 & SB1O \\
4833 &  $-$6.21 & 1.18 & 0.20 & 35 & 5942 & Var? \\
4859 & \nodata & \nodata & \nodata & 41 & 7707 & SB1O \\
4866 & \nodata & \nodata & \nodata & 100 & 10342 & SB1O \\
4921 & \nodata & \nodata & \nodata & 31 & 8431 & SB2O,VBO \\
4948 &     3.59 & 0.44 & 0.08 & 27 & 7683 & C,CPM \\
4963 &  $-$3.89 & 0.35 & 0.06 & 30 & 7417 & C,VBO,CPM \\
4974 & $-$14.78 & 0.70 & 0.13 & 29 & 5083 & C,CPM \\
5021 & \nodata & \nodata & \nodata & 35 & 5083 & SB1O,VB \\
5057 & $-$0.26 & 0.50 & 0.08 & 40 & 1036 & C \\
5076 & $-$7.54 & 3.35 & 0.81 & 17 & 7643 & Var? \\
5079 & \nodata & \nodata & \nodata & 43 & 7707 & SB1O,VB,PMA \\
5129 &    0.00 & 1.25 & 0.22 & 32 & 7683 & Var?,$\delta$ Sct \\
5144 & \nodata & \nodata & \nodata & 111 & 11035 & SB2O,VB,CPM \\
5146 & $-$17.55 & 41.95 & 7.93 & 28 & 4712 & Var,VB,SB,CPM \\
5163 & \nodata & \nodata & \nodata & 64 & 9516 & SB2O \\
5216 & \nodata & \nodata & \nodata & 29 & 4364 & SB1O \\
5220 & $-$11.20 & 0.54 & 0.12 & 20 & 9472 & Var? \\
5229 & $-$16.20 & 0.67 & 0.12 & 30 & 8431 & C,CPM \\
5255 & \nodata & \nodata & \nodata & 58 & 10224 & SB1O,PMA \\
5280 &  $-$4.92 & 0.68 & 0.15 & 21 & 1079 & C,PMA \\
5333 & \nodata & \nodata & \nodata & 128 & 11255 & SB1O \\
5360 &  $-$6.23 & 0.55 & 0.13 & 19 & 1068 & C,PMA \\
5374 & \nodata & \nodata & \nodata & 69 & 10372 & SB1O,PMA \\
5397 &     0.68 & 1.43 & 0.25 & 34 & 10665 & C,CPM \\
5406 & $-$18.10 & 1.14 & 0.22 & 27 & 7687 & Var?,PMA \\
5415 &  $-$3.18 & 38.08 & 6.63 & 33 & 2472 & Var,SB,VB \\
5468 & \nodata & \nodata & \nodata & 108 & 10924 & SB2O,VB,PMA \\
5531 & \nodata & \nodata & \nodata & 32 & 8386 & SB1O,CPM \\
5569 & \nodata & \nodata & \nodata & 41 & 10240 & SB2O,CPM,PMA \\
5574 & \nodata & \nodata & \nodata & 95 & 35767 & SB1O,PMA \\
5665 & \nodata & \nodata & \nodata & 37 & 10285 & SB1O \\
5676 & $-$16.85 & 0.97 & 0.22 & 20 & 1121 & C \\
5748 & $-$4.60 & 0.56 & 0.09 & 37 & 10252 & C,PMA \\
5754 & $-$4.65 & 0.97 & 0.20 & 24 & 4406 & C \\
5759 & $-$9.99 & 0.33 & 0.07 & 24 & 7749 & C \\
5760 & \nodata & \nodata & \nodata & 31 & 8428 & SB2,SB1O \\
5789 & \nodata & \nodata & \nodata & 73 & 10981 & SB1O,VBO,PMA,$\delta$ Sct \\
5791 & \nodata & \nodata & \nodata & 39 & 10268 & SB1O \\
5859 & $-$10.57 & 0.44 & 0.10 & 19 & 1124 & C,PMA \\
5875 & $-$10.90 & 0.65 & 0.12 & 28 & 1126 & C \\
5886 & $-$7.57 & 0.99 & 0.20 & 25 & 1126 & C? \\
144359\tablenotemark{b} & \nodata & \nodata & \nodata & 101 & 2621 & SB2O,NSS \\
6041 & \nodata & \nodata & \nodata & 97 & 6300 & SB1O,NSS \\
6074 & \nodata & \nodata & \nodata & 76 & 10629 & SB1O,CPM \\
6170 & $-$1.10 & 0.78 & 0.17 & 20 & 1124 & Var?,PMA \\
6246 & $-$22.22 & 0.58 & 0.12 & 22 & 1125 & C,VB,CPM,PMA \\
6341 & $-$30.04 & 0.45 & 0.10 & 20 & 2152 & C,CPM \\
6362 & $-$10.43 & 0.34 & 0.05 & 39 & 10230 & C \\
6367 & $-$24.52 & 2.66 & 0.57 & 22 & 6344 & Var?,VBO,PMA \\
6378 & 0.14 & 0.46 & 0.10 & 21 & 2731 & C,VBO \\
6399 & $-$14.16 & 1.00 & 0.22 & 21 & 1126 & C,VB,CPM \\
6412 & \nodata & \nodata & \nodata & 53 & 10666 & SB2O,EB \\
6421 & $-$5.67 & 0.49 & 0.11 & 20 & 1125 & C \\
6432 & \nodata & \nodata & \nodata & 37 & 10254 & SB1O,PMA \\ 
6455 & \nodata & \nodata & \nodata & 37 & 8456 & SB1O,PMA \\
6457 & $-$19.44 & 0.71 & 0.16 & 20 & 1125 & C,CPM \\
6480 & $-$20.82 & 0.79 & 0.18 & 19 & 4405 & C \\
6499 & $-$27.63 & 0.45 & 0.09 & 24 & 7753 & C \\
6521 & $-$25.44 & 0.28 & 0.06 & 24 & 6371 & C \\
6548 & $-$13.88 & 0.82 & 0.18 & 21 & 4433 & C,VB,CPM,PMA \\
6559 & \nodata & \nodata & \nodata & 79 & 10750 & SB1O,CPM,PMA \\
6581 & $-$24.11 & 0.82 & 0.17 & 24 & 7730 & Var?,$\delta$ Sct \\
6633 & \nodata & \nodata & \nodata & 67 & 10658 & SB1O \\
6681 & \nodata & \nodata & \nodata & 27 & 4434 & SB1O,VBO,CPM,PMA \\
6753 & $-$12.11 & 0.62 & 0.13 & 23 & 7750 & C,PMA \\
6794 & $-$15.05 & 0.33 & 0.05 & 38 & 10248 & C \\
6883 & $-$39.37 & 1.30 & 0.25 & 27 & 6344 & Var? \\
7001 & $-$13.13 & 1.00 & 0.21 & 22 & 1648 & C \\
7003 & 7.56 & 76.15 & 15.88 & 23 & 4500 & Var,SB,PMA,EB,$\delta$ Sct \\
7019 & \nodata & \nodata & \nodata & 42 & 10965 & SB1O,VBO,PMA \\
7060 & $-$1.44 & 0.33 & 0.08 & 19 & 4543 & C \\
7250 & $-$22.71 & 0.28 & 0.04 & 43 & 10298 & C,PMA \\
7357 & 10.73 & 0.94 & 0.17 & 29 & 6371 & Var,PMA \\
7362 & $-$50.80 & 23.37 & 4.50 & 27 & 6371 & Var,SB,VBO \\
7371 & $-$28.43 & 0.36 & 0.06 & 41 & 10095 & C,CPM \\
7519 & $-$29.47 & 0.81 & 0.13 & 37 & 10759 & C \\
7695 & $-$24.38 & 0.93 & 0.18 & 28 & 6371 & C,CPM \\
7781 & \nodata & \nodata & \nodata & 72 & 10821 & SB1O,VB,CPM \\
7818 & \nodata & \nodata & \nodata & 33 & 552 & SB2O,VBO \\
7891 & \nodata & \nodata & \nodata & 51 & 11843 & SB1O,PMA \\
7938 & \nodata & \nodata & \nodata & 84 & 10980 & SB2O,PMA \\
7981 & \nodata & \nodata & \nodata & 60 & 10808 & SB1O,PMA \\
8272 & 3.90 & 0.35 & 0.05 & 53 & 10070 & C \\
8459 & 12.68 & 0.52 & 0.08 & 39 & 10563 & C \\
8605 & $-$7.89 & 0.35 & 0.06 & 37 & 10025 & C \\
8624 & $-$12.72 & 0.29 & 0.06 & 23 & 8385 & C \\
8641 & \nodata & \nodata & \nodata & 150 & 12833 & SB1O,NSS \\
8837 & $-$16.90 & 0.84 & 0.14 & 36 & 10679 & C,PMA \\
\enddata
\tablecomments{The table columns list the following parameters:
(1) HR number \citep{hof82};
(2) mean RV;
(3) standard deviation of the measures;
(4) standard deviation of the mean RV;
(5) number of RVs;
(6) time interval in days between the first and last observations;
(7) status of the star regarding its possible companions. Footnotes 
indicate more details.}
\tablenotetext{a}{C = constant RV; Var = possibly variable RV; CPM = Common
Proper Motion companion(s); SB1(2)(O): Single(double)-lined spectroscopic
binary, without or with (orbit). Tables 8 and 9 list the orbital elements;
VB(O) = visual or speckle binary, without or with (orbit); PMA = Proper Motion
Anomaly identified by \citet{ker19} as significant differences between the 
Hipparcos and Gaia proper motions; EB = Eclipsing Binary; $\delta$ Sct = known
or suspected delta Scuti variable; NSS = non-survey star.}
\tablenotetext{b}{HD number}
\end{deluxetable}

\newpage
\begin{deluxetable}{l D D D r r r}
\tablecolumns{7}
\tablewidth{0pt}
\tablecaption{Radial Velocity Summary  -- High $v \sin i$ Stars\label{RVsum_hi_vsini}}
\tablehead{ \colhead{HR} & \multicolumn2c{$<$RV$>$} & \multicolumn2c{$\sigma$} 
& \multicolumn2c{$\frac{\sigma}{\sqrt{N}}$} & \colhead{$N_{\rm{obs}}$} 
& \colhead{$\Delta t$} & \colhead{Status\tablenotemark{a}} \\ 
\colhead{} & \multicolumn2c{(km s$^{-1}$)} & \multicolumn2c{(km s$^{-1}$)} 
& \multicolumn2c{(km s$^{-1}$)} & \colhead{} & \colhead{(days)} & \colhead{}}
\decimalcolnumbers
\tabletypesize{\scriptsize}
\startdata
5085 & $-$6.64 &  1.55  & 0.41 & 14 & 3613  & C,CPM \\ 
5107 & $-$10.51 &  2.07  & 0.78 & 7 & 423  & VB,CPM \\ 
5112 & $-$14.55 &  1.04  & 0.23 & 21 & 3611  & C \\ 
5116 & $-$19.12 &  0.86  & 0.25 & 12 & 3604  & C \\ 
5127 & $-$3.45 &  1.80  & 0.64 & 8 & 748  & VBO,CPM \\ 
5162 & $-$7.65 &  1.52  & 0.41 & 14 & 3618  & C \\ 
5244 & $-$12.05 &  1.26  & 0.40 & 10 & 3612  & C \\ 
5290 & $-$23.59 &  0.94  & 0.17 & 32 & 3612  & C \\ 
5392 & $-$12.38 &  1.21  & 0.54 & 5 & 423  & C,CPM \\ 
5532 & 4.52 &  2.14  & 0.87 & 6 & 423  & C,CPM \\ 
5715 & 5.60 &  2.11  & 0.64 & 11 & 3604  & Var,VB \\ 
5818 & $-$23.65 &  2.37  & 0.97 & 6 & 422  & PMA \\ 
5867 & 1.00 &  0.62  & 0.28 & 5 & 422  & PMA,CPM \\ 
5895 Aa & $-$9.74 &  3.11  & 1.39 & 5 & 422 & VBO \\ 
5895 Ab & $-$15.25 &  2.47  & 0.69 & 13 & 3604 & Var,VBO \\ 
5919 & $-$26.51 &  2.49  & 0.94 & 7 & 423  & C,$\delta$ Sct \\ 
5960 & $-$13.43 &  1.79  & 0.54 & 11 & 3611  & C,$\delta$ Sct \\ 
6004 & $-$20.67 &  5.65  & 2.31 & 6 & 423  & Var?,$\delta$ Sct \\ 
6033 & $-$0.46 &  1.88  & 0.84 & 5 & 423  & C \\ 
6277 & $-$17.82 &  2.48  & 1.11 & 5 & 423  & $\delta$ Sct \\ 
6317 & $-$5.43 &  1.48  & 0.61 & 6 & 423  & VBO \\ 
6507 & $-$31.43 &  1.11  & 0.45 & 6 & 422  & C \\ 
6877 & $-$25.95 &  1.37  & 0.56 & 6 & 748  & C \\ 
8264 & $-$18.38 &  0.81  & 0.40 & 4 & 377  & VBO,SB1O \\
\enddata
\tablecomments{The table columns list the following parameters:
(1) HR number \citep{hof82};
(2) mean RV;
(3) standard deviation of the measures;
(4) standard deviation of the mean RV;
(5) number of RVs;
(6) time interval in days between the first and last observations;
(7) status of the star regarding its possible companions. 
Footnotes indicate the latter in more detail.}
\tablenotetext{a}{C = constant RV; CPM = Common Proper Motion companion(s);
SB1(2)(O) = single(double)-lined spectroscopic binary, without or with orbit;
VB(O) = visual or speckle binary, without or with orbit; $\delta$ Sct = known
or suspected $\delta$ Sct variable; PMA = proper motion anomaly identified by
\citet{ker19} as significant differences betweend the Hipparcos and Gaia
proper motions.; Var = suspected variable radial velocity.}
\end{deluxetable}

\startlongtable
\begin{deluxetable}{c c c c c c c c c D r}
\tabletypesize{\scriptsize}
\setlength\extrarowheight{-0.3pt}
\tablecolumns{11}
\tablewidth{0pt}
\tablecaption{Orbital Elements for Single-lined Spectroscopic Binaries\label{Orbel_sb1}}
\tablehead{
\colhead{HR} & \colhead{$P$} & \colhead{$T$ or $T_{0}$} & \colhead{$e$} & \colhead{$\omega$}
  & \colhead{$K$} & \colhead{$\gamma$} & \colhead{$\sigma$($O-C$)} & \colhead{$a \sin i$}
  & \multicolumn2c{$f(m)$} & \colhead{$N_{\rm{obs}}$} \\
\colhead{} & \colhead{(days)} & \colhead{(HJD$-$2,400,000)} & \colhead{} & \colhead{(deg)}
  & \colhead{(km s$^{-1}$)} & \colhead{(km s$^{-1}$)} & \colhead{(km s$^{-1}$)}
  & \colhead{($10^{6}$ km)} & \multicolumn2c{($M_\odot$)} & \colhead{}}
\decimalcolnumbers
\startdata
526     & 10990  & 60114    & 0.26  & 235    & 2.30  & $-$5.12  & 0.57    & 336    & 0.0125  & 33 \\
\nodata & 436    & 685      & 0.10  & 21     & 0.23  & 0.20   & \nodata & 37     & 0.0039  & \nodata \\
\\
892     & 2418.6 & 54236    & 0.556 & 191.7  & 5.05  & $-$7.894 & 0.35    & 139.55 & 0.0185   & 36 \\
\nodata & 7.9    & 19       & 0.013 & 2.5    & 0.11  & 0.079  & \nodata & 0.033  & 0.0013   & \nodata \\
\\
1849    & 15.1920 & 58418.70\tablenotemark{b} & 0.0\tablenotemark{c} & \nodata    & 1.59  & $-$8.57  & 0.51    & 0.33  & 0.0000064 & 29 \\
\nodata & 0.0011  & 0.56                      & \nodata              & \nodata    & 0.15  & 0.11     & \nodata & 0.032 & 0.0000018 & \nodata \\
\\
2163    & 1242.3  & 59930   & 0.318 & 304.1  & 2.658 & $-$7.759 & 0.39    & 43.050 & 0.00206   & 76 \\
\nodata & 2.4     & 18      & 0.024 & 6.0    & 0.062 & 0.060  & \nodata & 0.011  & 0.00015   & \nodata \\
\\
3383\tablenotemark{a} & 14965 & 49589 & 0.204 & 171 & 1.910 & 4.776 & 0.28 & 385  & 0.0101 & 194 \\
\nodata &                 630 &   412 & 0.029 &  13 & 0.062 & 0.036 & \nodata & 20 & 0.0011 & \nodata \\
\\
4033    & 1041.87 & 59048.8 & 0.222 & 206.8  & 5.173 & 15.348 & 0.48    & 72.276 & 0.01386   & 159 \\
\nodata & 0.37    & 8.0     & 0.011 & 3.04   & 0.069 & 0.041  & \nodata & 0.98   & 0.00056   & \nodata \\
\\
4343    & 2328.1  & 57822   & 0.417 & 183.2  & 2.824 & 6.197  & 0.41    & 82.158 & 0.00408  & 168 \\
\nodata & 5.9     & 13      & 0.016 & 2.7    & 0.043 & 0.038  & \nodata & 1.435  & 0.00021  & \nodata \\
\\
4760    & 1216.28 & 58189   & 0.220 & 187.5  & 6.17  & 15.255 & 0.38    & 100.60 & 0.0274   & 50  \\
\nodata & 0.57    & 16      & 0.014 & 4.5    & 0.13  & 0.066  & \nodata & 2.14   & 0.0018   & \nodata \\
\\
4805    & 6.82853 & 58621.47 & 0.136 & 12    & 2.508 & $-$1.624 & 0.37    & 0.2333 & 0.0000108 & 38 \\
\nodata & 0.00013 & 0.37     & 0.038 & 17    & 0.097 & 0.066  & \nodata & 0.0091 & 0.0000013 & \nodata \\
\\
4859    & 387.15  & 56634    & 0.416 & 212.1 & 2.75  & $-$8.88  & 0.47    & 13.3   & 0.00063   & 41 \\
\nodata & 0.95    & 11       & 0.083 & 8.1   & 0.32  & 0.11   & \nodata & 1.6    & 0.00023   & \nodata \\
\\
4866   & 50.18765 & 56046.876 & 0.6270 & 235.34 & 21.185 & $-$3.415 & 0.38 & 11.39  & 0.02338  & 100 \\
\nodata & 0.00024 & 0.025     & 0.0023 & 0.35   & 0.081  & 0.041  & \nodata & 0.51 & 0.00032 & \nodata \\
\\
5021 Aa & 1.7380758 & 54234.3614\tablenotemark{b} & 0.0\tablenotemark{c} & \nodata  & 23.55 & 3.38  & 0.84    & 0.5629 & 0.002352  & 34 \\
\nodata & 0.0000027 & 0.0064                     & \nodata              & \nodata  & 0.21  & 0.14  & \nodata & 0.0051 & 0.000064  & \nodata \\
\\
5079 Aa  & 919.5    & 51491     & 0.254  & 313  & 5.75  & $-$12.47 & 0.58   & 70.4   & 0.0164    & 43 \\
\nodata & 1.5     & 31        & 0.032  & 10   & 0.28  & 0.22  & \nodata & 3.5    & 0.0025    & \nodata \\
\\
5216   & 2.178597 & 53514.24  & 0.0091 & 200  & 39.51 & $-$7.55 & 0.49    & 1.1835 & 0.01392   & 29 \\
\nodata & 0.0000032 & 0.15    & 0.0036 & 25   & 0.17  & 0.11  & \nodata & 0.0051 & 0.00018   & \nodata \\
\\
5255   & 1530.7   & 59746     & 0.500  & 278.9 & 3.40 & 11.14 & 0.55    & 62.0   & 0.0040    & 58 \\
\nodata & 4.9     & 67        & 0.083  & 6.8   & 0.33 & 0.36  & \nodata & 7.0    & 0.0014    & \nodata \\
\\
5333   & 25.09473 & 59446.890 & 0.4438 & 269.2 & 8.316 & $-$0.894 & 0.53  & 2.571 & 0.001075  & 128 \\
\nodata & 0.00016 & 0.062     & 0.0081 & 1.2   & 0.077 & 0.049  & \nodata & 0.026 & 0.000033 & \nodata \\
\\
5374   & 265.992  & 59285.38  & 0.709  & 121.0 & 8.49  & $-$1.206 & 0.35  & 21.892 & 0.00591  & 69 \\
\nodata & 0.022   & 0.46      & 0.011  & 1.4   & 0.21  & 0.069  & \nodata & 0.63 & 0.00051  & \nodata \\
\\
5531 Ab & 70.3483  & 51002.92 & 0.392   & 46.2    & 43.56 & $-$23.57  & 1.18    & 38.77   & 0.469  & 32 \\
        & 0.0050   & 0.29     & 0.011   & 2.3     & 0.60  & 0.42    & \nodata & 0.57    & 0.021 & \nodata \\
\\
5574   & 421.849  & 50150.4   & 0.681  & 133.8 & 2.72  & $-$11.764 & 0.43  & 11.548 & 0.000345 & 95 \\
\nodata & 0.081   & 2.0       & 0.022  & 3.0   & 0.10  & 0.050   & \nodata & 0.53 & 0.000048 & \nodata \\
\\
5665   & 393.00   & 59558 & 0.130  & 299   & 3.54  & $-$7.61 & 0.69  & 19.0  & 0.00176  & 37  \\
\nodata & 0.39    & 26    & 0.063  & 25    & 0.20  & 0.16  & \nodata & 1.1 & 0.00030  & \nodata \\
\\
5760  & 364.649  & 57509.1  & 0.0998  & 47.1    & 20.37 & $-$17.82  & 0.47    & 101.65  & 0.3147 & 31 \\
        & 0.041    & 6.6      & 0.0060  & 7.1     & 0.18  & 0.13    & \nodata & 0.88    & 0.0082 & \nodata \\
\\
5789 Aa  & 461.92   & 54554     & 0.071  & 180   & 11.57 & $-$38.28 & 1.36  & 73.3   & 0.0735   & 73  \\
\nodata & 0.16    & 23        & 0.023  & 18    & 0.26  & 0.21   & \nodata & 1.6  & 0.0049   & \nodata \\
\\
5791   & 94.333   & 58565.2   & 0.499  & 130.2 & 3.39  & $-$20.67 & 0.50  & 3.81   & 0.000247 & 39  \\
\nodata & 0.018   & 1.2       & 0.072  & 6.5   & 0.35  & 0.11   & \nodata & 0.44 & 0.000085 & \nodata \\
\\
6041\tablenotemark{a} & 2122  & 58583  & 0.298  & 179.2 & 1.613 & $-$13.312 & 0.32 & 44.9   & 0.080302 & 97  \\
\nodata               & 16    & 41     & 0.029  & 7.6   & 0.055 & 0.040   & \nodata & 1.6 & 0.000085 & \nodata \\
\\
6074   & 47.0267  & 57554.5   & 0.073  & 317   & 7.30  & 4.247   & 0.54 & 4.706  & 0.001877  & 76  \\
\nodata & 0.0010  & 1.5       & 0.014  & 12    & 0.12  & 0.084   & \nodata & 0.077 & 0.000092 & \nodata \\
\\
6432   & 1260.2   & 57173     & 0.215  & 312.4 & 3.013 & 4.937   & 0.33 & 51.0   & 0.00333   & 37  \\
\nodata & 3.7     & 37        & 0.038  & 8.9   & 0.090 & 0.085   & \nodata & 1.6 & 0.00031   & \nodata \\
\\
6455   & 2633.7   & 56880     & 0.378  & 233.6 & 6.329 & $-$2.202  & 0.28 & 212.2  & 0.0548   & 37  \\
\nodata & 3.5     & 14        & 0.013  & 2.1   & 0.087 & 0.062   & \nodata & 3.2 & 0.0024   & \nodata \\
\\
6559 Aa  & 648.033  & 59860.19  & 0.966  & 212.5 & 54    & $-$18.33  & 0.54 & 126    & 0.19     & 79  \\
\nodata & 0.045   & 0.37      & 0.018  & 8.6   & 27    & 0.12    & \nodata & 71  & 0.32     & \nodata \\
\\
6633   & 41.0052  & 57518.88\tablenotemark{b} & 0.0\tablenotemark{c} & \nodata  & 1.294 & $-$44.449 & 0.44    & 0.729  & 0.0000092 & 67 \\
\nodata & 0.0035  & 0.43                      & \nodata              & \nodata  & 0.076 & 0.056     & \nodata & 0.043  & 0.0000016 & \nodata \\
\\
6681 Aa1  & 13.418762 & 51002.870 & 0.3658 & 79.48 & 34.32 & $-$4.443  & 0.42 & 5.894  & 0.045303  & 27  \\
\nodata & 0.000041 & 0.020    & 0.0034 & 0.71  & 0.14  & 0.098   & \nodata & 0.026 & 0.00059 & \nodata \\
\\
7019 Aa & 9064      & 61072     & 0.185  & 355   & 2.40  & 10.062  & 0.39 & 294.3  & 0.0124    & 42  \\
\nodata & 98      & 470       & 0.048  & 21    & 0.20  & 0.098   & \nodata & 25  & 0.0031    & \nodata \\
\\
7781 Aa & 260.328   & 59771.2   & 0.256  & 40.2  & 1.869 & $-$1.960  & 0.33 & 6.47   & 0.000159  & 72  \\
\nodata & 0.063   & 4.9       & 0.029  & 8.5   & 0.055 & 0.051   & \nodata & 0.20 & 0.000015 & \nodata \\
\\
7818 Aa  & 28.9409  & 59747.738 & 0.4562  & 34.3   & 37.31  & $-$17.70  & 1.40    & 13.21   & 0.1098 & 33 \\
        & 0.0058   & 0.092    & 0.0088  & 1.4    & 0.49   & 0.25    & \nodata & 0.18    & 0.0046 & \nodata \\
\\
7818 Ba  & 4.83250  & 59744.14 & 0.036   & 156    & 30.49  & $-$14.32  & 1.06    & 2.025   & 0.01427 & 33 \\
        & 0.00021  & 0.19     & 0.011   & 14     & 0.25   & 0.21    & \nodata & 0.017   & 0.00035 & \nodata \\
\\
7891  & 1277.4    & 55545     & 0.240  & 8.6   & 5.854 & $-$19.173 & 0.36 & 99.8   & 0.0243    & 51 \\
\nodata & 2.6     & 19        & 0.014  & 3.8   & 0.081 & 0.059   & \nodata & 1.4 & 0.0010    & \nodata \\
\\
7981    & 571.14  & 59326.7\tablenotemark{b} & 0.0\tablenotemark{c}  & \nodata  & 4.15  & 10.95 & 0.58    & 32.0 & 0.00423   & 60  \\
\nodata & 0.44    & 4.4                      & \nodata               & \nodata  & 0.13  & 0.11  & \nodata & 1.1  & 0.00041   & \nodata \\
\\
8264 A\tablenotemark{d} & 9057      & 53214     & 0.5    & 305   & 18    & -17     & 0.89  & 1882  & 3.2  & 25 \\
\nodata & 663     & \nodata   & \nodata & \nodata & \nodata & \nodata & \nodata & \nodata & \nodata & \nodata \\
\\
8641\tablenotemark{a} & 3773  & 58123  & 0.187 & 328.5 & 1.149   & 8.357 & 0.18  & 58.6  & 0.000562  & 150 \\
\nodata               & 21    & 100    & 0.028 & 9.7   & 0.036   & 0.024 & \nodata & 1.9 & 0.000054  & \nodata \\
\enddata
\tablecomments{The table columns list the following parameters:
(1) HR number; 
(2) orbital period in days, $P$; 
(3) time of periastron passage, $T$, or the time of maximum RV, $T_{0}$; 
(4) eccentricity, $e$; 
(5) longitude of periastron, $\omega$; 
(6) semiamplitude, $K$;
(7) systemic velocity, $\gamma$; 
(8) standard deviation of the RVs, $\sigma(O-C)$; 
(9) projected semimajor axis, $a \sin i$; 
(10) mass function, $f(m)$; 
(11) number of RVs, $N_{\rm{obs}}$.  
The second line of each entry contains the error estimate of the element.} 
\tablenotetext{a}{not part of survey}
\tablenotetext{b}{$T_{0}$}
\tablenotetext{c}{$e=0$ from statistical criterion of \citet{luc71}}
\tablenotetext{d}{high $v \sin i$ star}
\end{deluxetable}

\clearpage
\startlongtable
\begin{deluxetable}{c c c c c c c c c c c}
\tabletypesize{\scriptsize}
\setlength\extrarowheight{-0.3pt}
\tablecolumns{11}
\tablewidth{0pt}
\tablecaption{Orbital Elements for Double-lined and Higher Multiplicity Systems\label{Orbel_sb2}}
\tablehead{
\colhead{HR} & \colhead{$P$} & \colhead{$T$ or $T_{0}$} & \colhead{$e$} & \colhead{$\omega$}
  & \colhead{$K_{1,2}$} & \colhead{$\gamma$} & \colhead{$\sigma (O-C)$}
  & \colhead{$a_{1,2} \sin i$} & \colhead{$m_{1,2} \sin^{3} i$} 
	& \colhead{$N_{\rm {obs}}$} \\
\colhead{} & \colhead{(days)} & \colhead{(HJD$-$2,400,000)} & \colhead{} & \colhead{(deg)}
  & \colhead{(km s$^{-1}$)} & \colhead{(km s$^{-1}$)} & \colhead{(km s$^{-1}$)}
  & \colhead{(10$^{6}$ km)} & \colhead{($M_\odot$)} & \colhead{}}
\colnumbers
\startdata
317\tablenotemark{a} & 3.680672  & 58942.349 & 0.1733 & 358.4 & 64.14 & 13.98 & 1.08    & 3.200  & 0.3935 & 25 \\
                        & 0.000046  & 0.018     & 0.0030 & 2.3   & 0.23  & 0.16  & \nodata & 0.012  & 0.0038 & \nodata \\
317 & \nodata   & \nodata  & \nodata & \nodata & 64.88 & \nodata & 0.93 & 3.234 & 0.3890 & 25 \\
       & \nodata   & \nodata  & \nodata & \nodata & 0.40 & \nodata  & \nodata & 0.020  & 0.0032 & \nodata \\
\\
971  & 54.8261 & 59947.8  & 0.343   & 308.3   & 4.71  & $-$7.71  & 0.65    & 3.33   & 0.0030 & 81 \\
        & 0.0059  & 1.1      & 0.038   & 7.4     & 0.27  & 0.13    & \nodata & 0.20   & 0.0005 & \nodata \\
971  & \nodata  & \nodata  & \nodata & \nodata & 5.76  & \nodata & 1.77    & 4.08   & 0.0024 & 69 \\
        & \nodata  & \nodata  & \nodata & \nodata & 0.60  & \nodata & \nodata & 0.43   & 0.0004 & \nodata \\ 
\\
1041  & 96.2049  & 59258.411 & 0.79717 & 289.89  & 32.06 & 5.972   & 0.47    & 25.606 & 1.894 & 102 \\
        & 0.0014   & 0.012    & 0.00082 & 0.21    & 0.11  & 0.056   & \nodata & 0.099  & 0.021 & \nodata \\
1041 & \nodata  & \nodata  & \nodata & \nodata & 75.14 & \nodata & 0.71    & 60.02  & 0.8082 & 69 \\
        & \nodata  & \nodata  & \nodata & \nodata & 0.40  & \nodata & \nodata & 0.34   & 0.0081 & \nodata \\
\\
2103 A  & 710.55   & 554850.4 & 0.3059  & 204.4   & 24.98 & 31.87   & 0.59    & 232.3  & 1.871 & 144 \\
        & 0.20     & 2.8      & 0.0065  & 1.4     & 0.29  & 0.15    & \nodata & 2.7    & 0.064 & \nodata \\
2103 Bb & \nodata & \nodata  & \nodata & \nodata & 16.84 & \nodata & 0.59    & 156.6  & 2.775 & 114 \\
        & \nodata  & \nodata  & \nodata & \nodata & 0.42  & \nodata & \nodata & 3.9    & 0.064 & \nodata \\
2103 Ba & 28.55495 & 59550.46 & 0.1220  & 256.9   & 35.85 & 0.00    & 1.50    & 13.97   & 1.739 & 46 \\
        & 0.0019   & 0.10     & 0.0029  & 1.4     & 0.31  & 0.11    & \nodata & 0.12    & 0.037 & \nodata \\
2103 Bb & \nodata  & \nodata  & \nodata & \nodata & 62.35 & \nodata & 1.50    & 24.30   & 1.000 & 114 \\
        & \nodata  & \nodata  & \nodata & \nodata & 0.78  & \nodata & \nodata & 0.30    & 0.019 & \nodata \\
\\
4921 Aa & 2894     & 55735    & 0.290   & 315.2   & 0.95  & $-$10.452 & 0.63    & 36.3    & 0.0061 & 31 \\
        & 29       & 92       & 0.060   & 9.8     & 0.20  & 0.075   & \nodata & 7.6     & 0.0033 & \nodata \\
4921 Ab & \nodata  & \nodata  & \nodata & \nodata & 2.26  & \nodata & 0.44    & 86      & 0.0026 & 40 \\
        & \nodata  & \nodata  & \nodata & \nodata & 0.67  & \nodata & \nodata & 26      & 0.0013 & \nodata \\
\\
5144 Aa & 14.7306416 & 58550.2269 & 0.7018 & 26.87 & 36.95 & $-$22.433 & 0.65    & 5.33    & 0.7427 & 111 \\
        & 0.0000095  & 0.0025    & 0.0012 & 0.14  & 0.15  & 0.066   & \nodata & 0.023   & 0.0088 & \nodata \\
5144 Ab & \nodata  & \nodata  & \nodata & \nodata & 87.25 & \nodata & 1.02    & 12.590  & 0.3143 & 74 \\
        & \nodata  & \nodata  & \nodata & \nodata & 0.53  & \nodata & \nodata & 0.079   & 0.0035 & \nodata \\
\\
5163  & 7.1995999 & 58622.214 & 0.0763 & 145.73  & 66.29 & $-$6.812  & 0.40    & 6.5435  & 2.072  & 64 \\
        & 0.0000034 & 0.018    & 0.0012 & 0.84    & 0.093 & 0.063   & \nodata & 0.0092  & 0.014  & \nodata \\
5163  & \nodata  & \nodata  & \nodata & \nodata & 100.61 & \nodata & 2.19   & 9.932   & 1.3654 & 36 \\
        & \nodata  & \nodata  & \nodata & \nodata & 0.43   & \nodata & \nodata & 0.043  & 0.0075 & \nodata \\
\\
5468 Aa & 102.0223 & 58548.42  & 0.3962 & 290.6   & 20.15 & $-$16.49  & 1.13    & 25.96   & 2.284 & 108 \\
        & 0.0045   & 0.20     & 0.0052  & 1.0     & 0.19  & 0.12    & \nodata & 0.26    & 0.067 & \nodata \\
5468 Ab & \nodata  & \nodata  & \nodata & \nodata & 52.65 & \nodata & 2.12    & 67.8    & 0.874 & 71 \\
        & \nodata  & \nodata  & \nodata & \nodata & 0.84  & \nodata & \nodata & 1.1     & 0.023 & \nodata \\
\\
5569 Aa & 33.75058 & 58951.894 & 0.5113  & 163.1  & 39.46 & $-$25.59  & 1.63    & 15.74   & 1.267 & 41 \\
        & 0.00080  & 0.082     & 0.0065  & 1.2    & 0.56  & 0.25    & \nodata & 0.24    & 0.050 & \nodata \\
5569 Ab & \nodata  & \nodata  & \nodata & \nodata & 58.9  & \nodata & 1.90    & 23.51   & 0.848 & 22 \\
        & \nodata  & \nodata  & \nodata & \nodata & 1.3   & \nodata & \nodata & 0.54    & 0.030 & \nodata \\
\\
144359\tablenotemark{b} & 4.3843700 & 54431.4730 & 0.2817 & 68.67 & 78.02 & $-$20.22 & 0.91 & 4.514 & 2.120 & 101 \\
               & 0.0000050 & 0.0037 & 0.0014 & 0.36 & 0.19 & 0.11 & \nodata & 0.011 & 0.014 & \nodata \\
144359  & \nodata  & \nodata  & \nodata & \nodata & 126.41 & \nodata & 1.26 & 7.313 & 1.3087 & 101 \\
          & \nodata  & \nodata  & \nodata & \nodata & 0.48   & \nodata & \nodata & 0.028 & 0.0075 & \nodata \\
\\
6412  & 38.32712 & 59695.126 & 0.5076 & 358.84  & 66.08 & 8.80    & 0.42    & 30.005 & 3.219 & 62 \\
        & 0.00017  & 0.020    & 0.0018 & 0.32    & 0.16  & 0.11    & \nodata & 0.081  & 0.058 & \nodata \\
6412  & \nodata & \nodata & \nodata & \nodata & 69.22 & \nodata & 2.37    & 31.44  & 3.072 & 35 \\
        & \nodata & \nodata & \nodata & \nodata & 0.84  & \nodata & \nodata & 0.38   & 0.041 & \nodata \\
\\
7938  & 315.656  & 58422.94 & 0.6470  & 198.50 & 32.17  & $-$4.666  & 0.31    & 106.5   & 3.61 & 84 \\
        & 0.021    & 0.19     & 0.0048  & 0.50   & 0.40   & 0.072   & \nodata & 1.4     & 0.12 & \nodata \\
7938  & \nodata  & \nodata  & \nodata & \nodata & 43.50 & \nodata & 1.23    & 144.0 & 2.671 & 83 \\
        & \nodata  & \nodata  & \nodata & \nodata & 0.76  & \nodata & \nodata & 2.6   & 0.082 & \nodata \\
\enddata
\tablecomments{The table columns list the following parameters:
(1) HR number and component designation; 
(2) orbital period in days, $P$; 
(3) time of periastron passage, $T$, or the time of maximum RV, $T_{o}$; 
(4) eccentricity, $e$; 
(5) longitude of periastron, $\omega$;
(6) semiamplitude, $K_{1,2}$;
(7) systemic velocity, $\gamma$; 
(8) standard deviation of the RVs, $\sigma (O-C)$; 
(9) projected semimajor axis, $a_{1,2} \sin i$; 
(10) minimum mass, $m_{1,2} \sin^{3} i$;
(11) number of RVs, $N_{\rm {obs}}$. The second line of each 
component entry contains the error estimate of the element.}
\tablenotetext{a}{Not part of survey}
\tablenotetext{b}{HD number; not part of survey}
\tablenotetext{c}{Mass function}
\end{deluxetable}

\startlongtable
\begin{deluxetable}{l c c c c c l}
\centering
\tablewidth{0pt}
\tablecolumns{7}
\setlength\extrarowheight{-.5pt}
\tablecaption{Survey Stars with Orbits in the \textit{ORB6}\tablenotemark{a}, Speckle Measurements in the
\textit{INT4}\tablenotemark{b}, or Entries in the \textit{WDS}\tablenotemark{c}\label{VBwithOrbel}}
\tablehead{
\colhead{HR} & \colhead{WDS} & \colhead{$P$} & \colhead{$a$ or $\rho$} 
& \colhead{$V_{\rm A}$} & \colhead{$V_{\rm B}$} & \colhead{Notes} \\
\colhead{} & \colhead{identifier} & \colhead{(yr)} & \colhead{(")}
  & \colhead{(mag)} & \colhead{(mag)} & \colhead{}}
\colnumbers
\tabletypesize{\scriptsize}
\startdata
196   & 00453+5513 & 385\tablenotemark{d} & 0.7 & 5.4 & 9.9 & WDS: 4 additional CPM components; \\
\nodata   & \nodata    & \nodata & \nodata & \nodata & \nodata & this study: constant RV $= -9.95 \pm0.08$ km $s^{-1}$ \\
\\
526 Aa,Ab & 01493+4754 & 36.9 & 0.122 & 6.47 & \nodata & VBO \citep{zirm14}; speckle pair could \\
\nodata   & \nodata    & \nodata & \nodata & \nodata & \nodata & also be Ba,Bb; this study: SB1 orbit, $P = 30.1$ yr \\
\\
1086    & 03344+2428 & 522.16 & 0.625 & 6.60 & 6.86 & VBO \citep{scar02} \\
\nodata   & \nodata    & \nodata & \nodata & \nodata & \nodata & this study: constant RV $=26.94 \pm0.06$ km $s^{-1}$ \\
\\
3822 Aa,Ab & 09356-1935 & 56\tablenotemark{d} & 0.227 & 6.33 & 11.1 & resolved by speckle \citep{tok24}; \\
\nodata   & \nodata    & \nodata & \nodata & \nodata & \nodata & this study: var? RV $= 20.34 \pm0.30$ km $s^{-1}$ \\
\\
3894 & 09521+5404 & 106.35 & 0.337 & 5.28 & 5.39    & VBO \citep{jos21} \\
\nodata & \nodata    & 0.68 & 0.002 & \nodata & \nodata & this study: RV increased by $\sim$3 km $s^{-1}$ starting $\sim$JD 2450500 \\
\\
4717    & 12243+2606 & 27.25 & 0.170 & 5.2  & 9.6  & VBO \citep{tok24} \\
\nodata & \nodata    & 1.76  & 0.012 & \nodata & \nodata & this study: constant RV$=-0.61 \pm0.09$ km s$^{-1}$ \\
\\
4789    & 12349+2238 & 36.00 & 0.192 & 4.96 & 6.90 &  VBO \citep{hart14} \\
\nodata & \nodata    & \nodata & \nodata & \nodata &  \nodata & this study: sharp rise in RV of $\sim$20 km s$^{-1} \sim$JD 2449700 \\
\\
4921  & 12597$-$0349 & 7.906 & 0.078 & 5.9 & 8.3 & VBO \citep{doco15}; this study: composite \\
\nodata & \nodata    & \nodata & \nodata & \nodata & \nodata & spectrum; Ab: SB1O, $P = 7.92 \pm0.08$ yr \\
\\
4963  & 13099$-$0532   & 695.  & 1.243 & 4.49 & 6.83 & VBO \citep{zirm15}; this study: constant \\
\nodata & \nodata    & \nodata & \nodata & \nodata & \nodata & RV $ = -3.86 \pm0.07$ km s$^{-1}$; 1977 SB orbit likely spurious \\
\\
5021    & 13189+0341 & 1077\tablenotemark{d} & 1.135--1.60  & 6.62 & 9.55 & AB: speckle obs. by \citet{hor11,hor12} \\
\nodata & \nodata    & \nodata & \nodata & \nodata & \nodata & this study: Aa: SB1O, $P = 1.738^{d}$; EB: \citep{prsa22} \\
\\
5079    & 13282+5035 & 104\tablenotemark{d} & 0.173 & 7.0 & 11.6 & resolved by speckle (E. Horch, INT4) \\
\nodata    & \nodata & \nodata & \nodata & \nodata & \nodata  & this study: SB1, $P = 919.5^{d}$ \\
\\
5107\tablenotemark{e}  & 13347$-$0036 & 158\tablenotemark{d}   & 1.8   & 3.40    & 10. & WDS: VB, B is M4-7V \\
\nodata & \nodata    & \nodata & \nodata & \nodata & \nodata  & this study: constant RV $= -10.5 \pm0.8$ km s$^{-1}$ \\
\\
5127\tablenotemark{e}    & 13375+3618 & 244.98  & 1.0058 & 4.98   & 6.95    & VBO \citep{izm19} \\
\nodata & \nodata    & 7.60    & 0.0033 & \nodata & \nodata & this study: constant RV $= -3.4 \pm0.6$ km s$^{-1}$  \\
\\
5146  & 13415$-$2327 & 164\tablenotemark{d} & 0.184--0.240 & 6.93 & 8.13  & INT4: many visual and interferometric measures since 1959 \\
\nodata & \nodata    & \nodata & \nodata & \nodata & \nodata &  this study: rapid and large line profile variations \\
\\
5715\tablenotemark{e}    & 15201+5158 & 31.2\tablenotemark{c} & 0.166--0.26 & 5.52 & 11.5 & INT4: rapid change in $\theta$ indicating short $P$ \\
\nodata & \nodata    & \nodata & \nodata & \nodata & \nodata  & this study: variable RV \\
\\
5789    & 15348+1032 & 1150.1  & 6.38  & 4.17 & 5.16 & VBO \citep{izm19}; notes to ORB6 indicate \\
\nodata & \nodata    & 451.4   & 3.41  & \nodata & \nodata & orbit is premature; this study: Aa, SB1O, $P = 461.92^{d}$ \\
\\
5895\tablenotemark{e}  & 15513$-$0305 & 64.842  & 0.562 & 5.2  & 7.8  &  VBO \citep{tok18} \\
\nodata & \nodata    & \nodata & \nodata & \nodata & \nodata &  this study: variable RV \\
\\
6317\tablenotemark{e}    & 17005+0635 & 13.111  & 0.1063 & 6.6 & 9.8 & VBO \citep{toko24} \\
\nodata & \nodata    & 0.056   & 0.0025 & \nodata & \nodata & this study: RV $= -5.4 \pm0.6$ km s$^{-1}$ over 423$^{d}$  \\
\\
6367    & 17082$-$0105 & 203.6 & 0.5716 & 6.32  & 7.75 & VBO \citep{mas06} \\
\nodata   & \nodata    & \nodata & \nodata & \nodata & \nodata & this study: variable RV \\
\\
6378    & 17104$-$1544 & 87.76 & 1.307  & 3.05  & 3.27 & VBO \citep{doco23}; this study: composite \\
\nodata   & \nodata  & 1.10  & 0.105  & \nodata & \nodata & spectrum, RV$_{sharp-lined} = 0.14 \pm0.14$ km s$^{-1}$  \\
\\
6548    & 17346+0935 & 197\tablenotemark{d} & 0.37 & 5.8 & 10.1 & resolved by speckle (Horch, INT4) \\
\nodata   & \nodata    & \nodata & \nodata & \nodata & \nodata & this study: constant RV $ = -13.88 \pm0.18$ km s$^{-1}$ \\
\\
6681    & 17563$-$1549 & 69.5  & 0.338  & 6.45  & 7.50 & Aa,Ab VBO \citep{wais23a} \\
\nodata   & \nodata  & 7.3   & 0.024  & \nodata & \nodata & this study: Aa, SB1O \\
\\
7019    & 18402+3822 & 23.53 & 0.100  & 6.87  & 8.06 & VBO \citep{doco21} \\
\nodata   & \nodata  & 0.27  & 0.005  & \nodata & \nodata & this study: SB1O, $P = 24.82 \pm0.27$ yr. \\
\\
7362    & 19253$-$2431 & 5.6985 & 0.0664 & 5.8  & 5.8  &  VBO \citep{doco23} \\
\nodata   & \nodata  & 0.0015 & 0.0003 & \nodata & \nodata & this study: blended triple or double-lined spectra \\
\\
7818    & 20251+5936 & 90.42  & 0.1378 & 6.94 & 7.63 &  VBO \citep{dav24}; \\
\nodata   & \nodata  & 0.51   & 0.0055 & \nodata   & \nodata & this study: FCF determined separate SB1O for Aa, Ba \\
\\
8264\tablenotemark{e}    & 21378$-$0751 & 25.5   & 0.28   & 4.8  & 7.0  & VBO \citep{gon02} \\
\nodata   & \nodata  & 3.     & 0.03  & \nodata   & \nodata  & this study: preliminary SB1O, $P = 24.8 \pm1.8$ yr. \\
\enddata
\tablecomments{The table columns list the following parameters:
(1) HR number and component designation
(2) WDS (Washington Double Star Catalog) identifier
(3) $P$ in years from ORB6 if available, otherwise estimated
(4) the semi-major axis $a$ if listed in ORB6, or the angular speparation $\rho$ from the WDS
(5) visual magnitude $V$ of component A
(6) visual magnitude $V$ of component B
(7) notes concerning the orbit, if available, and the results from this study.}
\tablenotetext{a}{Sixth Catalog of Orbits of Visual Binary Stars}
\tablenotetext{b}{Fourth Catalog of Interferometric Measurements of Binary Stars}
\tablenotetext{c}{The Washington Double Star Catalog}
\tablenotetext{d}{$P$ estimated from Kepler's third law}
\tablenotetext{e}{high $v \sin i$ star}
\end{deluxetable}

\clearpage

\clearpage

\clearpage

\clearpage
\vspace{2.5in}

\clearpage

\clearpage

\clearpage
\section{Discussion}
\subsection{Multiplicity Statistics}
It is of course well known that a magnitude limited
sample such as the one discussed here, may be affected by the Branch
Bias \citep{bra76}, whereby some binary or multiple stars will be
included in the sample that might otherwise be fainter than the
magnitude limit cutoff. It is worth noting that at the time this survey
commenced, many of the A-type stars had no parallax measurements or their
parallaxes were of low accuracy compared to those from HIP2
and now from GEDR3. We have attempted to mitigate the Branch Bias by 
estimating whether the secondary member (or members)
would add sufficient luminosity to create a combined magnitude brighter than
$V = 6.5$ mag, the nominal faint limit of The Bright Star Catalog 
\citep{hof82}, herafter BSC. The BSC is particularly
problematic because the magnitude limit is ``hazy at best.'' 
Nevertheless, we have evaluated the systems and concluded that 8 of the SBs
would not have been included in the BSC if based on the primary 
star's magnitude alone. There are also 3 VBs, 2 with common proper motion (CPM)
and proper motion anomaly \citep[PMA,][]{ker19}, and 
2 non-binary stars in the sample with $V$ fainter than  6.5 mag that 
we deducted 
from the sample total.  These 15 stars are listed in Table~\ref{starsV>6.5}.

Among the 119 survey stars with $v \sin i \leq 120$ km s$^{-1}$, we have 
determined single-lined orbits (SB1O) for 33 systems and double-lined 
or higher multiplicity orbits (SB2O) for 10. The former have 1 quadruple
system (HR 7818) in which both visual components have short period
SB1 orbits. The latter have orbits for the components of 1 spectroscopic
triple system, HR 2103A,Ba,Bb.
We have also included 4 stars, HRs 5146, 5415, 7003, and 7362, as SB1 systems
even though we were not able to derive orbits for them. They all have large
and rapid CCF variations and are significantly brighter than the 
canonical $M_{V}$ values associated with their
spectral types. HR 7003 is also an eclipsing system. Some or all of them may
eventually be measured as SB2 or higher multiplicity systems, but our value
for the MF would remain unchanged. Appendix A contains more 
details of these particular stars. Tables~\ref{Orbel_sb1} 
and~\ref{Orbel_sb2} list the orbital elements for the SB1 and SB2/SB3 
systems respectively. Together these indicate a spectroscopic multiplicity
fraction (MF) = 47/119 = 0.40.  After deducting the systems with 
$V_{A} > 6.5$, we find a MF of 37/104 = 0.36.
For 2 $M_{\sun}$ primaries with semi-major axes, $a$, less than 10 au, 
\citet{moe21}, after
correcting for missing M dwarf and white dwarf (WD) companions, determined 
a bias-corrected MF of 0.40 $\pm$ 0.06 (their Table A1).
All of our SBs except HR 8264 have $a \sin i$ values less than 2.25 au, so 
assuming no very low values of $i$ ($<13$\arcdeg), $a$ would also be less than
10 au. HR 8264 has a VBO, is a high $v \sin i$ star with a tentative SB1 orbit,
and is not included in our multiplicity statistics. A recent LBI survey by
\citet{defur25} detected 7 companions in a sample of 54 chemically normal 
A-type stars within 80 pc. From these detections they derived a 
completeness-corrected companion frequency of 
0.25$^{+0.17}_{-0.11}$ over their projected 
separation sensitivity of 0.01--27.54 au. Five of the stars in their 
survey are also included in our two samples, HRs 3861, 5229, and 5875, 
which have low $v \sin i$, and HRs 5112 and 6507, which have high 
$v \sin i$.  Neither \citet{defur25} nor our study detected any companions 
for these 5 stars. The relatively large error limits on the
\citet{defur25} MF do allow an overlap with our value of 0.36, and perhaps 
a larger sample or longer observing interval would lessen the difference. 
Indeed, their companion frequency in this recent work is an increase from
their smaller sample of 26 A-type stars in \citet{defur22} where they derived 
a companion frequency of 0.19$^{+0.11}_{-0.06}$.

Additionally, 26 visual binaries were found to have orbits in  
ORB6, speckle observations listed in INT4, or have entries in 
the WDS catalog.  Three of the VBs have
spectroscopic orbits corresponding to visual orbits (HRs 526, 4921, 7019),
9 have spectroscopic orbits for sub-systems (HRs 526, 5021, 5079, 5144, 5468, 
5789, 6681, 7781, 7818), and 2 (HRs 5146, 5415) presently have no SB orbits
but present strong evidence for being close SBs as discussed above. Altogether,
we then have a total of 11 VBs as unique systems. For the VBs without 
orbits, we estimated periods from the available $\rho$ values, 
parallaxes, and system mass estimates based on spectral types, as 
discussed in \citet{toko24}. All of these are listed in 
Table~\ref{VBwithOrbel}. 
Finally, at least 28 stars have suspected CPM companions and 35 have 
a PMA. Eleven of the stars with CPM companions and 11 with PMAs are 
unique without other indications of companions. All together, the 
fractions of VB, CPM, and PMA systems total 33/119 = 0.28, and after 
the magnitude limit corrections discussed above, we find
a MF of 30/104 = 0.29. We note that the assignment of a ``G''
flag in the original Hipparcos Catalog (ESA), hereafter
HIP and/or a PMA in Gaia (GDR2; \citet{ker19}) is a useful indicator 
of multiplicity. Thirty-five of our 119 sample stars have either or both HIP
and \emph{GDR2} flags for PMA and 24 of these have SB, VB, or CPM companions, or 68\%.
Table~\ref{MFtable} summarizes the MF of each of these types of
systems as well as the total of all types. Adding the corrected fractions 
of VBs (0.096), CPM systems (0.096), and PMA systems (0.096) to the SB 
fraction (0.356) described above, yields a total MF = 0.64. There are few
other modern spectroscopic studies of intermediate mass stars, but one
that is similar in precision is the analysis of the Pleiades cluster
by \citet{tor20}. He finds a lower limit to the MF of $\sim$37\%, including 
astrometric systems. He also notes
that the survey was conducted over $\sim$1 -- 2 yr, so additional longer
period systems may be present. Our study found 60\% of just the SBs had
periods longer than 1 yr.

\begin{figure}[ht]
\begin{center}
\includegraphics[width=4in,keepaspectratio]{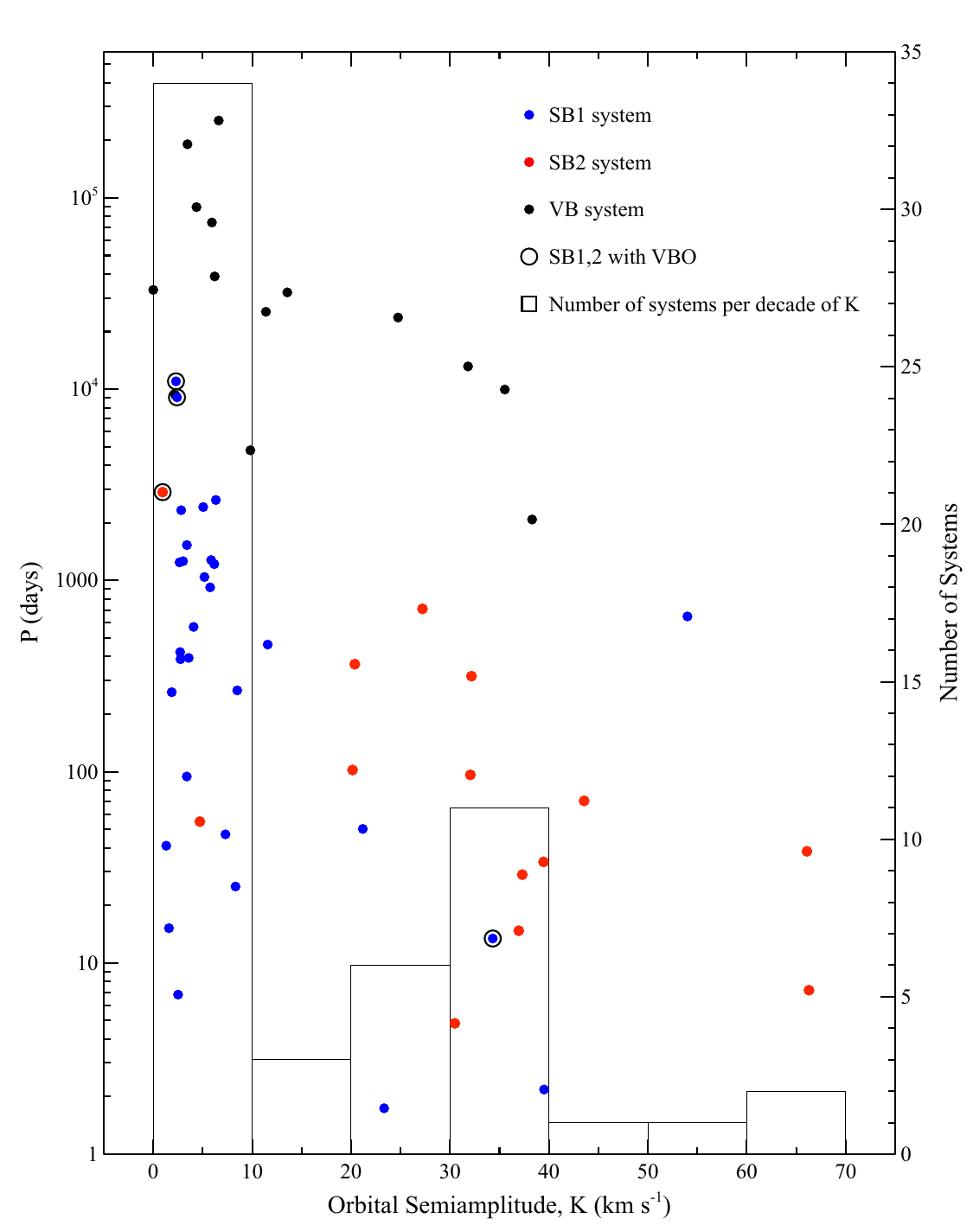}
\figcaption{The frequency of orbital periods with respect to the orbital
semiamplitude, $K$. For visual systems, some of the relatively large values of $K$ are due to 
the semi-major axis $a$ being equal to the relative orbit as discussed in Section 5.1.
\label{P_vs_K}}
\end{center}
\end{figure}

Referring again to the work of \citet{moe21}, their Table A1 lists a 
compilation MF of 0.57 $\pm$ 0.08 for 2.0 $M_{\sun}$ systems with $a <$ 100 au.
While all of our systems including the VBs, except 2 (HRs 1086 and 4963), 
have $a$ values less than 100 au, most of the stars with CPM companions 
and perhaps some with PMA would have separations larger than 100 au. 
Table A1 of \citet{moe21} also lists a total MF = 0.72 $\pm$ 0.09,
corrected for missing M and WD companions. It is difficult to estimate 
an incompleteness value for our sample, but it is safe to say that any 
MF is a lower limit. In Figure~\ref{P_vs_K} it is obvious that systems 
with the lowest values of $K$ are by far the most numerous in our sample. 
It can also be seen in Figure~\ref{vsini_P} that there does
not seem to be any preference for low amplitude binaries to have lower 
values of $v \sin i$ since 15/22 = 68\% of the systems with $P \leq 100$ 
days have $v \sin i \geq 50$ km s$^{-1}$. It would therefore not be 
surprising that some low amplitude systems could have been missed.

We have identified a total of 149 companions to our 119 sample stars 
including all of the known SBs, VBs, CPM, and PMA components, 
for a companion frequency (CF) of 149/119 = 1.25. After correcting for 
the Branch Bias, we find 128/104 = 1.23. As reported in \citet{off23}, 
\citet{moe21} derived a value of 0.99 $\pm$ 0.13 correcting for 
incompleteness and removing systems with WD companions.

Our relatively few observations of the 23 high $v \sin i$ ($>$120 km~s$^{-1}$)
stars yielded one preliminary spectroscopic orbit, which is for 
HR~8264. There is a previous spectroscopic orbit for this star by 
\citet{abt65} who used literature values and his new RVs
and also subsequently obtained RVs \citep{abt74}. With our 4 more recent RVs,
the 2 sets of RVs from \citet{abt65} and \citet{abt74}, and one RV from 
\citet{har37}, we were able to determine an orbit with $P = 24.8 \pm 1.8$ yr.
Our spectroscopic period is in reasonable agreement with the 
astrometric result of $P$ = 25.5 $\pm$ 3 yr that was determined by
\citet{gon02}.

While there are earlier measures of many of these stars, in most cases 
the listed RV values are from observatory surveys done during the 
first half of the 20th century or an assessment of each star's average 
velocity based on those surveys and found in the General Catalogue 
of Stellar Radial Velocities \citep[hereafter GCRV]{wil53} or its 
successors. Such 
RVs are of relatively poor quality compared to those in this paper. 
It is of interest to again mention the work of \citet{tor20} in 
regard to RV measurements of stars with high $v \sin i$ in the 
Pleiades Cluster. He determined RVs for 33 rapidly rotating B, A, 
and early F stars by fitting a Gaussian curve to 5 strong Balmer 
lines in high S/N spectra. His average internal uncertainty is 
1.7 km s$^{-1}$ while our average SD for the 23 high $v \sin i$ stars 
reported here is 1.67 km s$^{-1}$.

There are also 4 VBs with orbits in ORB6 and speckle observations 
of one other (HR 5715) in the INT4 that enabled an estimate of the 
period to be made. Information pertaining to these 5 systems is 
also listed in Table~\ref{VBwithOrbel}. Without further spectroscopic
RV measurements, it isn't as yet possible to make any definitive 
statements concerning the frequency of binaries among this sample of
high $v \sin i$ stars, and we do not include them in our statistics. 
The accuracy of our measurements does seem to encourage further
work however, especially if high S/N spectra can be obtained over a 
period of a few years. 

\begin{figure}[ht]
\begin{center}
\includegraphics[width=3.5in,keepaspectratio]{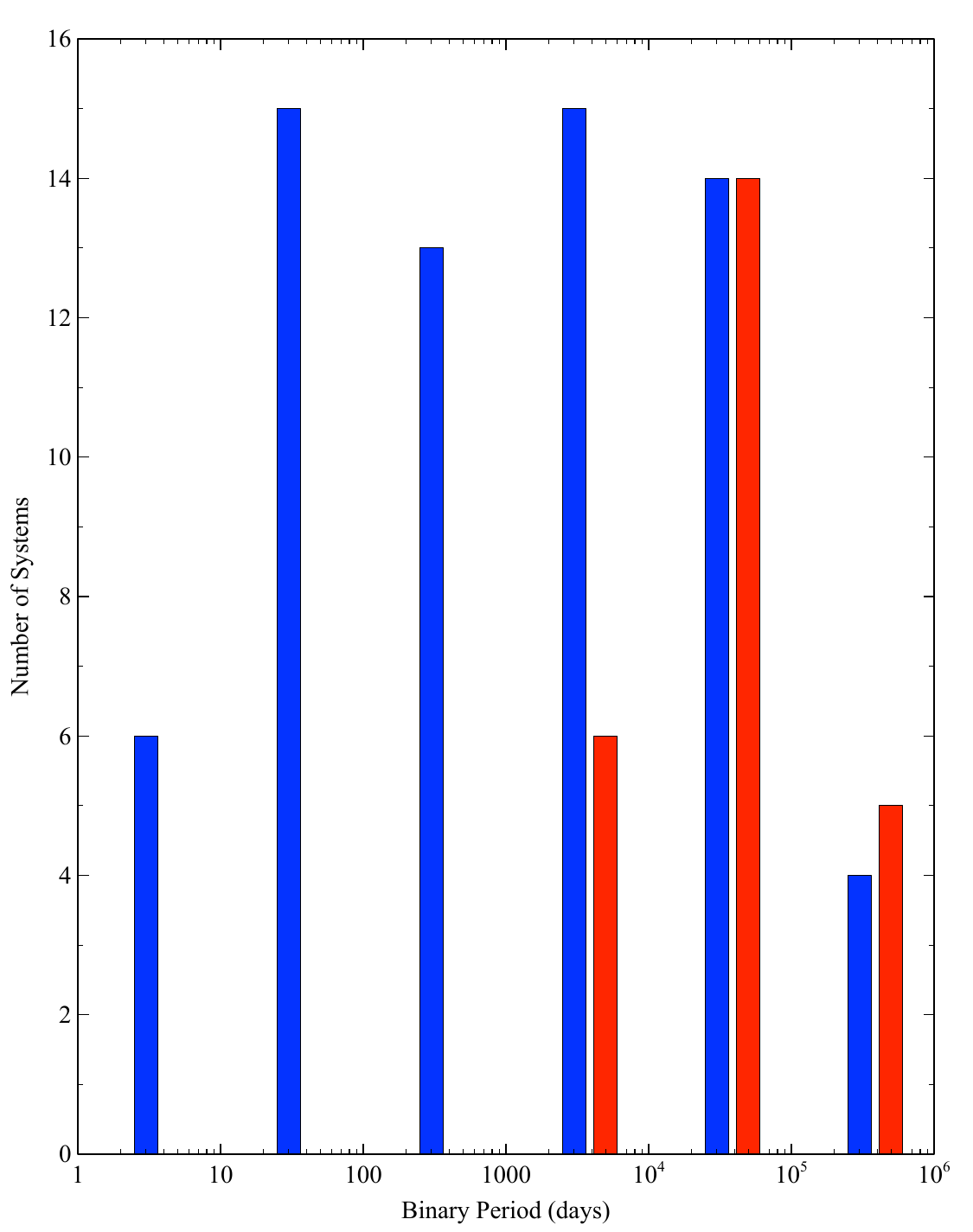}
\figcaption{Derived spectroscopic and known or estimated VBO periods for the sample stars.
The periods of VB systems without orbits were estimated from Kepler's 
third law, as noted in the footnote to Table ~\ref{VBwithOrbel} comments: 
$P = (\rho / \varpi)^{3/2}M^{-1/2}$. The blue bars are the number of 
systems per decade of $P$ and the red bars indicate the number of 
visual orbits or estimated periods.  \label{sb_periods}}
\end{center}
\end{figure}

A graph of the number of systems plotted against $P$ per decade of 
$P$ in days is shown in Figure~\ref{sb_periods} for the systems in 
our study, including the high $v \sin i$ stars. At least five of these systems 
with periods $>$10$^{4}$ days have shorter $P$ sub-systems as part of their
hierarchical architectures (HRs 5021 Aa, 5079 Aa, 5789 Aa, 
6681 Aa1, 7818 Aa,Ba).
A few others show evidence of short term variations but no orbits have been
determined. These are discussed individually in more detail in Appendix A.1.

It may be of some significance that Figure~\ref{sb_periods} indicates 
relatively low numbers of short $P$ ($<10$ days) binaries while those 
are generally the easiest to detect with their often larger values of 
$K$. In Figure~\ref{P_vs_K} where we show the relation of orbital $K$ 
values and $P$, we see that systems with low values of $K$ 
are the most common and more short period systems might be found with 
higher precision RV measurements. It is noteworthy though that of the 
approximately 163 A0 -- A9 IV,V systems listed in the 
\emph{Ninth Catalogue of Spectroscopic
Binary Orbits} (SB9) \citep{pour04}, 82 have a value of $P < 10$ days.
Most of these were studied before the year 2000, suggesting that many 
of the most readily detected sytems are already known.

It should be noted that the rather large values
of $K$ for some of the visual binaries is from our computation of 
\( K = 2\varpi (a \sin i)/P(1 - e^{2})^{1/2} \), where $a$ is either 
the semi-major axis of the relative orbit, or the observed angular 
separation of the components. In either case, \( a = a_{1} + a_{2} \).

\startlongtable
\begin{deluxetable}{l c c l}
\centering
\tablewidth{0pt}
\tablecolumns{4}
\tablecaption{Survey Stars with $V_{A} > 6.5$ mag}\label{starsV>6.5}
\tablehead{
	\colhead{HR} & \colhead{$V_{\rm {BSC}}$} & \colhead{Status} & \colhead{Comments} \\
\colhead{} & \colhead{(mag)} & \colhead{} & \colhead{}}
\colnumbers
\tabletypesize{\scriptsize}
\startdata
1086 A,B     & 5.92 & VBO       & $V_{A} = 6.60, V_{B} = 6.86$ \\
3702         & 6.62 & single    & constant RV \\
4378         & 6.65 & single    & constant RV \\
5021 Aa,B & 5.99 & SB1O,VBO & $V_{A} = 6.62, V_{B} = 9.55$ \\
5079 Aa   & 6.80 & SB1O,VB      & $V_{A} = 6.96, V_{B} = 11.58$ \\
5146 A,B     & 6.59 & VB, CPM   & $V_{A} = 6.93, V_{B} = 8.13$ \\
5163 Aa,Ab   & 6.51 & SB2O      & \nodata \\
5397 A,B     & 6.47 & CPM & $V_{A} = 6.61, V_{B} = 7.16$ \\
5415 A,B     & 6.95 & VB \& SB? & $V_{A} = 7.11, V_{B} = 7.56$ \\
5760 Aa,Ab   & 6.46 & SB2      & SB1O \\
5791 Aa      & 6.56 & SB1O      & \nodata \\
6412 Aa,Ab   & 6.17 & SB2O      & nearly equal components \\
7019 A,B     & 6.45 & SB1O      & $V_{A} = 6.87$ \\
7357 Aa,Ab?  & 6.64 & PMA and RV change & \nodata \\
7818 Aa,Ba & 6.44 & 2SB1O  & quadruple; $V_{Aa} =6.94, V_{Ba} = 7.63$ \\
\enddata
\tablecomments{The table columns list the following parameters: 
(1) HR number of the star and the component designations;
(2) $V$ mag from the BSC; 
(3) The system status or if the star is single, as determined from 
the data and/or references;
(4) Further information on the status from our work and the component magnitudes 
from the WDS.}
\end{deluxetable}

\startlongtable
\begin{deluxetable}{c c c c c c c c}
\tablenum{12}
\centering
\tablewidth{0pt}
\tablecolumns{8}
\tablecaption{Multiplicity Fraction By Companion Type}\label{MFtable}
\tablehead{
\colhead{} & \multicolumn{3}{c}{Number of Systems} & \colhead{} & \multicolumn{3}{c}{Multiplicity Fraction} \\
   \cline{2-4} \cline{6-8} \\
\colhead{Companion} & \colhead{Full} & \colhead{Unique} & \colhead{Branch Bias} & \colhead{}
   & \colhead{Full} & \colhead{Unique} & \colhead{Branch Bias} \\
\colhead{Type} & \colhead{Sample} & \colhead{Systems} & \colhead{Corrected} & \colhead{}
   & \colhead{Sample} & \colhead{Systems} & \colhead{Corrected} \\
\colhead{(1)} & \colhead{(2)} & \colhead{(3)} & \colhead{(4)} & \colhead{} & \colhead{(5)} & \colhead{(6)}
   & \colhead{(7)}}
\tabletypesize{\scriptsize}
\startdata
SB1 & 37 & 37 & 29 & & 0.311 & 0.311 & 0.279 \\
SB2 & 10 & 10 &  8 & & 0.084 & 0.084 & 0.077 \\
VB  & 25 & 11 & 10 & & 0.210 & 0.092 & 0.096 \\
CPM & 28 & 11 & 10 & & 0.235 & 0.092 & 0.096 \\
PMA & 35 & 11 & 10 & & 0.294 & 0.092 & 0.096 \\
\cline{1-8}
Total & 135 & 80 & 67 &  & 1.134 & 0.671 & 0.644 \\
\enddata
\tablecomments{The table columns list the following quantities:
  (1) type of companion detected;
  (2) number of survey stars with each companion type in full sample;
  (3) the number of unique systems with a particular companion type;
  (4) the number of unique systems after Branch Bias correction;
  (5) the MF of the full sample of each companion type;
  (6) the MF of the number of unique systems of each companion type;
  (7) the Branch Bias corrected MF of each companion type.}
\end{deluxetable}

\subsection{Apparently Constant Velocity Stars}
 
 A number of our A stars with apparently constant velocities have RVs 
listed in Gaia Data Release 3 \citep[hereafter GDR3]{gaia23}. 
However, as \citet{blo23} explained, the wavelength range of the Gaia 
spectra, 8460 -- 8700~\AA, is small and is optimized for cool stars.
For A stars, this spectral region contains five hydrogen Paschen lines,
three of which are blended with lines of the calcium infrared triplet.
This serious blending makes the determination of Gaia A star RVs
problematic. \citet{blo23} went on to extensively discuss the RV reduction
pipeline, which compares each program A star spectrum to a grid of synthetic
spectra to determine the best fit RV of the star. This enabled
the determination of RVs that are within $\pm$3 km~s$^{-1}$ of their
reference star values. However, \citet{blo23} noted that there remains a 
magnitude dependent effect in the Gaia hot star RVs. While they provided 
an equation to approximately correct for that effect, which ranged 
from 0.5 to 2 km~s$^{-1}$ for the brightest stars, they also noted 
that individual stars can behave differently. Thus, we have not applied
any RV corrections.

\begin{figure}[h]
\begin{center}
\includegraphics[width=5in,keepaspectratio]{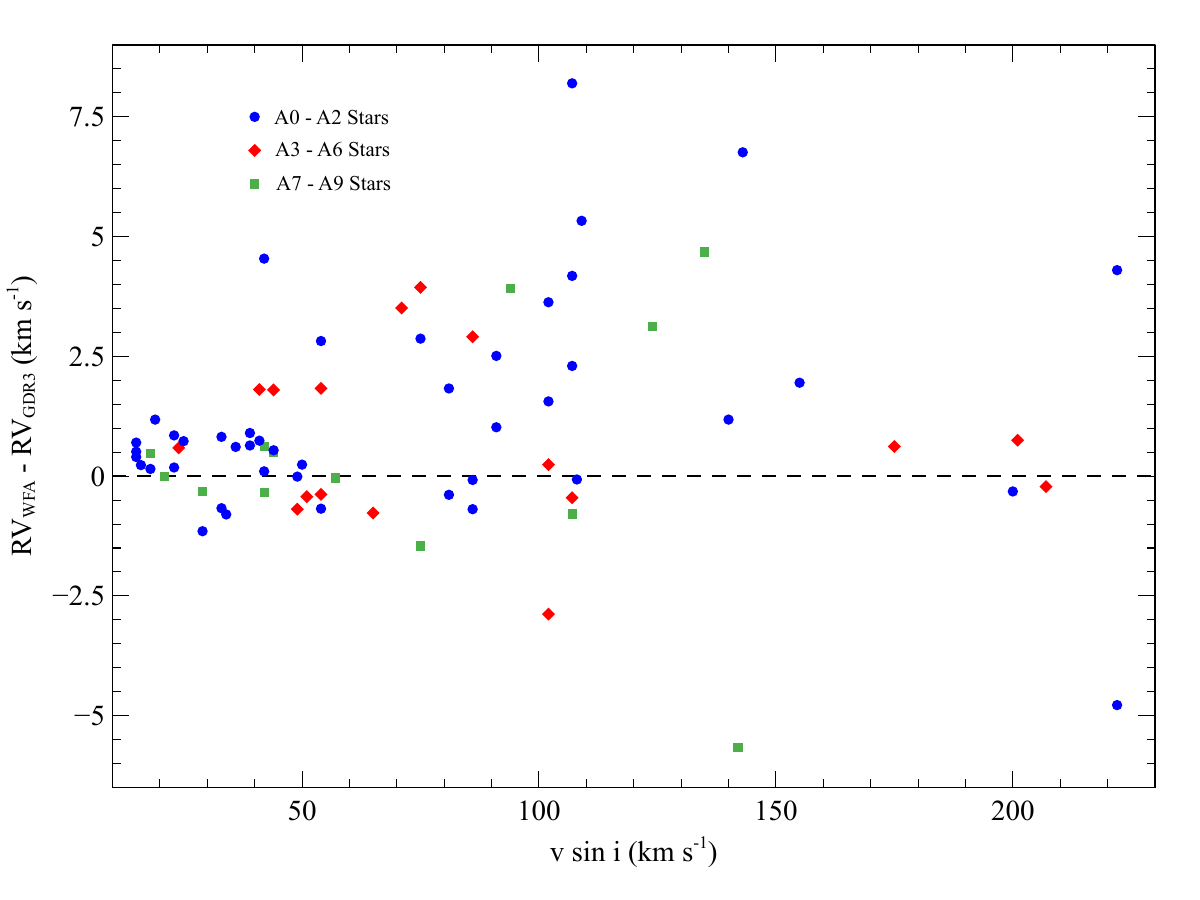}
\figcaption{The difference between the average RVs from this study and those of
GDR3 for stars not known to be binaries. For reference, the dashed line indicates
zero difference.\label{RV-RVgdr3}}
\end{center}
\end{figure}

For our survey stars without orbits,
there are 62 low $v \sin i$ and 12 high $v \sin i$ stars that
have GDR3 RVs. We have divided those stars into three spectral class
groups, A0--A2 (44 stars), A3--A6 (17 stars) and A7--A9 (13 stars). In 
Figure~\ref{RV-RVgdr3} we plot the RV difference between
our average RV and the uncorrected GDR3 value for those 74 stars versus
$v \sin i$. Only two of the 39 stars with $v \sin i$ $\leq$ 
70 km~s$^{-1}$ have RV differences $\geq$ 2 km~s$^{-1}$ regardless 
of spectral subclass. Stars with $v \sin i$ values greater than $\sim$70
km~s$^{-1}$ have a much larger scatter with the RV difference indicating
that our RVs are generally significantly more positive than those of GDR3.
     
In the individual star notes of the Appendices, we compare our RVs   
for the low $v \sin i$ stars with the uncorrected GDR3 RV values 
if the latter differ by $>$3 km~s$^{-1}$ from our values or if the 
GDR3 RVs have a range greater than 10 km~s$^{-1}$. 
It is certainly possible that at least some of these
stars with the large differences are low amplitude, long period
binaries and/or have large amplitude $\delta$~Sct or $\gamma$ Dor
RV changes. For the 23 stars in our high $v \sin i$ sample, we 
mention the GDR3 RV for each star that has one.

\vspace{-0.2in}
\subsection{Rotational Velocities of Narrow-lined Stars}

It may not have been as clear in 1992 when this survey commenced, 
that stellar rotation rates in binary systems are most likely 
determined in the formative stages of close systems. The evidence we 
present here, that few stars in this sample have orbital periods short
enough (i.e. $<10$ days) to be significant in the reduction of 
rotation, reinforces the conclusions of the recent work of 
\citet{he23} and \citet{smi24} on early-type stars as discussed in 
our Introduction.

\begin{figure}[h!]
\begin{center}
\includegraphics[width=5in,keepaspectratio]{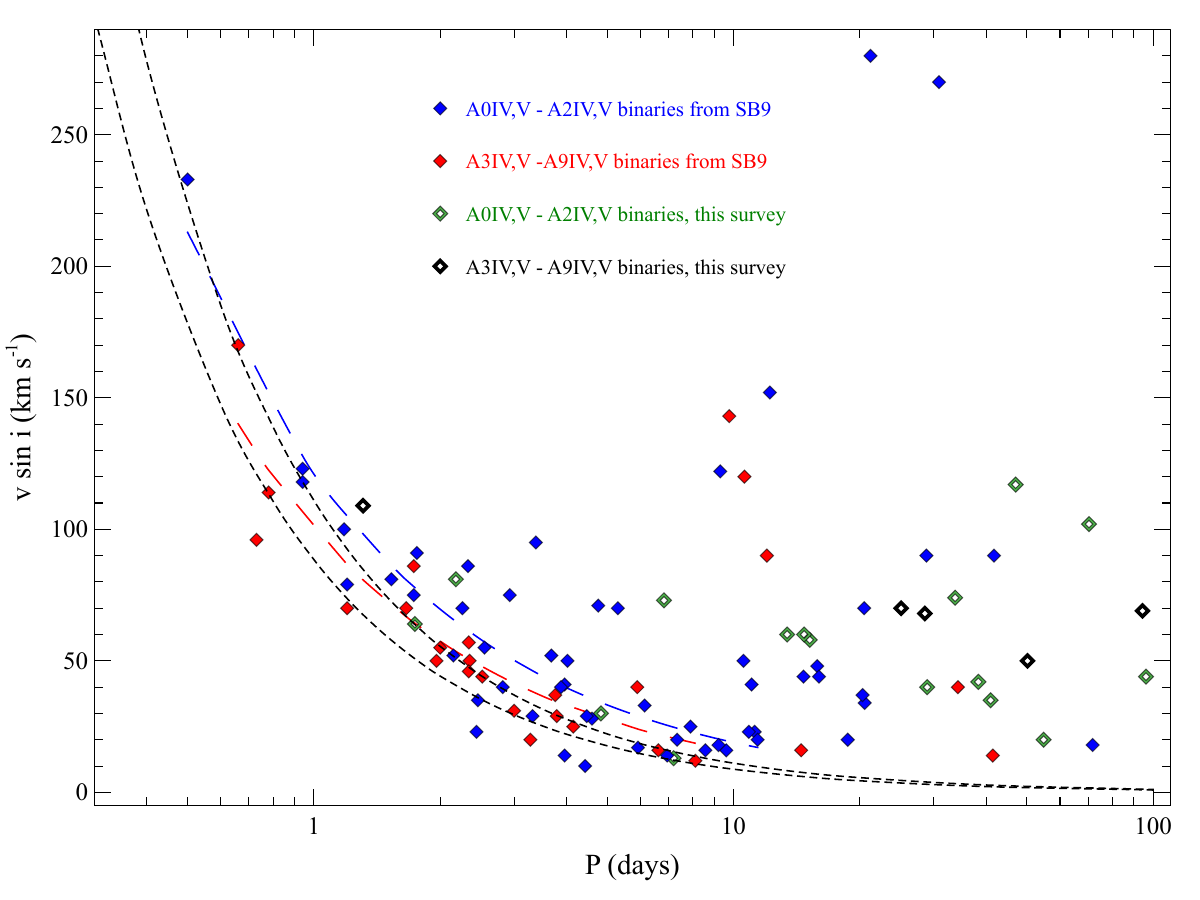}
\figcaption{Measured $v \sin i$ versus orbital period from the SB9 
catalog as well as the values for orbits determined in this survey. 
The two curves with short dashes are the calculated equatorial 
rotation values for A0V (upper curve) and A9V stars as described 
in the text. The two curves with the long dashes are power law fits 
to the A0 -- A2 values (blue) and the A3 -- A9 (red) values from 
0.5 -- 10.0 days. \label{vsini_P}}
\end{center}
\end{figure}

Another way to examine the relationship between $v \sin i$ and 
orbital period is by plotting $v \sin i$ versus $P$ as shown in 
Figure~\ref{vsini_P}. Here we have plotted two curves (short black dashes) 
showing the theoretical equatorial rotation velocity if the stellar 
rotation period and the orbital period are synchronous. The values of 
rotational velocity $v$ for the curves are given by 
$v = 2\pi R_{*} R_{\sun} /P_{\rm sec}$ where $R_*$ is the radius of the 
star in solar units and $P_{\rm sec}$ is the orbital period in seconds.
The value of $R_{\sun} = 6.9634 \times 10^5$ km. For $R_*$ we used the value 
$2.193 R_{\sun}$ for the A0V curve, and $1.747 R_{\sun}$ for the A9V curve 
from \citet{pec13}, but provided more recently on a web site by E.
Mamajek\footnote{\url{https://www.pas.rochester.edu/~emamajek/EEM\_dwarf\_UBVIJHK\_colors\_Teff.txt}}.

The binaries listed in the SB9 catalog that are fainter than about the 
cutoff of the BSC ($V \geq 6.5$ mag) do not for the most part have measured 
values of $v \sin i$, so the number 
of points in Figure~\ref{vsini_P} is small, especially for the 
shortest periods. Nevertheless, for periods $<10$ days, there does 
appear to be evidence of synchronization or partial synchronization, 
especially for the A3 -- A9 stars. 
For a pair of A6V stars rotating synchronously and using the value of 
$a/R = 6.78$ 
from Table 1 of \citet{hur02}, we find a period of approximately 2.04 days. 
Since the late A-type stars have a main sequence (MS) lifetime approximately 5 times longer
than early A stars, many of them may be considerably older and have had 
more time for tidal forces to act. 

\clearpage
\section{Summary}
We have completed the first, extensive, modern spectroscopic survey 
of normal, low $v \sin i$ A-type dwarfs and sub-giants. From our RVs
of the 119 stars in our sample that span more than three decades, we 
have determined spectroscopic orbits for 43 separate stars
including several multiple systems. Our resulting multiplicity 
values are consistent with earlier estimates and indicate that a large 
number of systems await detection among the brightest stars. Just 
considering our detected SB frequency of 36\%, the number of undetected 
normal A star binaries with $v \sin i$ $ \leq 120$ km s$^{-1}$ may 
be more than 75 in the BSC.

We have also shown that there is little evidence for rotational braking 
to have caused the significant numbers of low $v \sin i$ stars, although 
data from SB9 may show some synchronization between orbital and rotational 
periods in late A stars. This result reinforces recent investigations
concerning star formation and evolution.

Our initial radial velocity results for stars with $v \sin i > 120$ 
km s$^{-1}$ indicate sufficient precision to detect low amplitude 
binaries among these more numerous stars.

\begin{acknowledgments}
This research has made use of the Washington Double Star Catalog maintained 
at the U.S. Naval Observatory. We particularly thank Dr. Douglas Gies 
(Georgia State University) for help in maintaining access to the Catalog. 
The SIMBAD database, operated at CDS, Strasbourg, France has been
invaluable in identifying numerous papers that we reference.
This research has made use of the Astrophysics Data 
System, funded by NASA under Cooperative Agreement 80NSSC21M0056. 

This work used the \emph{Standard Deviation Calculator} at 
\url{https://www.calculatorsoup.com/calculators/statistics/standard-deviation-calculator.php}. 
We have used the curve fitting routines at \url{https://curve.fit/}.
Data visualization and plotting were performed using Veusz \url{https://veusz.github.io/i}.
This work has made use of data from the European Space Agency (ESA) mission
\emph{Gaia}
(\url{https://www.cosmos.esa.int/gaia}), processed by the \emph{Gaia}
Data Processing and Analysis Consortium (DPAC,
\url{https://www.cosmos.esa.int/web/gaia/dpac/consortium}). Funding for the DPAC
has been provided by national institutions, in particular the institutions
participating in the \emph{Gaia Multilateral Agreement}.
Astronomy at Tennessee State University was supported by the state of
Tennessee through its Centers of Excellence program. 
It is with great appreciation that we thank the many telescope 
operators on Kitt Peak that have worked to enable the observations 
and shared their companionship, good will and humor over the years.
\begin{verse}
\emph{...Souls that have toil'd, and wrought, and thought with me-- \\
That ever with a frolic welcome took \\
The thunder and the sunshine, and opposed \\
Free hearts, free foreheads...} \\
\vspace{0.5em}
from \emph{Ulysses} by Alfred, Lord Tennyson
\end{verse}
\end{acknowledgments}


\appendix
\section{Notes on Individual Stars}

In the following Appendices, one for the low $v \sin i$ stars 
and one for the high $v \sin i$ stars, we provide brief comments on 
selected stars. 
Much of the information on the VB components comes from  HIP, WDS, INT4, 
ORB6, and the BSC. References to previously published RV measurements 
are typically found in the SIMBAD \emph{Collections of Measurements} 
section for each star. Section 5.1 discusses the issues related to 
the use of RVs from older, mostly photographic spectra. A number of 
our stars have RVs listed in Gaia DR3 (\citet{gaia23}, 
hereafter GDR3). The low $v \sin i$ stars with GDR3 mean RVs that 
differ by more than 3 km~s$^{-1}$ from ours are noted in Appendix A.1.
A more detailed discussion of GDR3 mean RVs in relation to ours is given
in Section 5.2.  In Appendix A.2, when our high (>120 km~s$^{-1}$) $v \sin i$ 
stars have GDR3 mean RVs, they are listed for comparison.

We mention that many of our program stars are listed in HIP with 
a ``G'' multiplicity flag. That flag signifies that the star has an 
accelerated (non-linear) proper motion solution, which likely results 
from an unseen astrometric companion. In addition, we note that several 
of our stars have been assigned an ``S'' quality flag in HIP, which 
indicates a suspected binary but for which no definitive non-single 
star model was obtained. As mentioned previously in the notes to 
Table~\ref{RVsum_lo_vsini}, \citet{ker19} also
identified a number of our stars as having possible companions by
finding significant differences between the Hipparcos and Gaia proper
motions, and their PMA.


\subsection{Low $v \sin i$ A stars}

\emph{HR 114 = HD 2628}:  The star has CPM with ADS 409B, which has 
$V$ = 13.1 mag, $\rho$ = 2\farcs4 (WDS). There are also more distant 
CPM companions at 136\farcs6, $V$ = 12.27 mag and at 192\farcs1, 
$V$ = 14.71 mag. 
Our RV of $-$9.88 $\pm$ 0.15 km s$^{-1}$ is consistent with the value
of $-$10.2 km s$^{-1}$ listed in the GCRV. The star is identified 
as a $\delta$ Sct variable in the BSC. \\

\emph{HR 196 = HD 4222}: The WDS lists the star as the primary 
of the VB, BU~492AB, with $\rho$ = 1\farcs9 and $\Delta$m = 6.1 mag plus 
three other fainter and more distant CPM components. The WDS also 
lists newly identified components Aa and Ab that have $\rho$ = 0\farcs7 
and $\Delta$m = 4.5 mag. The present RV measurements appear constant at
$-$9.95 $\pm$ 0.08 km s$^{-1}$. Older RV averages range from 
$-$5.0 km~s$^{-1}$ from three spectra \citep{cam28} to $-$11.2 km~s$^{-1}$
from two spectra \citep{har37}. \\ 

\emph{HR 317 = HD 6530}: \citet{roy14} identified it as an 
SB2 from changes in its cross-correlation function (CCF). We have 
derived an SB2O for it with $P = 3.68$ days and $e = 0.1733$. The 
non-zero $e$ value is likely due to the system being young. The
star is not part of our survey, and so it is not included in our
multiplicity analyses. \\
 
\emph{HR 526 = HD 11031}: The star is the primary of a relatively 
close VB with $\Delta$$V$ = 0.75 and $\rho$ = 2\farcs0 (WDS). Besides 
this pair,  
two other much more widely separated visual components are also listed
in the WDS.
Speckle measurements (INT4) have detected a close pair with the
separation and position angle varying between 0\farcs147 and 
0\farcs121, and 193\fdg5 and 251\fdg2, respectively, between 1984 and 1994.
No indication of the relative magnitudes is given. An orbit for this 
close pair was determined by \citet{zirm14} with $P$ = 36.9 yr. In ORB6
that orbit was given a grade of 4 or preliminary.
With our RVs, we determined an SB1O with $P = 30.1$ yr. Since both 
stars of the VB would be more or less always on the slit,
it is not yet known whether the close pair is the A or B component of
the VB.
The system appears to be quintuple (INT4, WDS).  \\

\emph{HR 723 = HD 15385}: The WDS reports components A and B with a 
$\rho$ = 217\farcs5 and $\Delta$$V$ = 3.5 mag. The CPM component, B, is 
itself a close VB that has $\rho$ = 0\farcs5, and component magnitudes of 
$V$ = 10.2 and 10.5. For component A we found a constant RV = 21.82 
$\pm$ 0.03 km s$^{-1}$. \\

\emph{HR 892 = HD 18543}: Our RVs resulted in an orbit with $P$ = 2418.6
days. A companion with  $\Delta$$V$ = 7.3 mag and $\rho$ = 3\farcs0 is 
listed in WDS, but there is no indication as to whether it is physically 
associated. The parallaxes for A and B are 8.94 $\pm$ 0.19 and 8.12 
$\pm$ 0.21 mas, respectively. \\

\emph{HR 971 = HD 20149}: \citet{roy14} discovered this star to have an 
asymmetric CCF.  Our RVs produced an SB2O with $P$ = 54.8261 days. \\

\emph{HR 1041 = HD 21402}: Our RVs yielded an SB2O with $P$ = 
96.2049 days. \\

\emph{HR 1086 = HD 22091}: This star is a VB consisting of a nearly 
equal pair of stars. 
The WDS lists $V_{A}$ = 6.60 mag, A3V, $V_{B}$ = 6.86 mag, and 
$\rho$ = 0\farcs8. \citet{scar02} determined a visual orbit with 
$P$ = 522.16 yr, which was given a grade of 3 or moderately reliable  
in ORB6. Our spectra include both stars on the slit, 
and no velocity changes were detected. \\

\emph{HR 1133 = HD 23193}: In HIP this star has a ``G'' flag indicating
an acceleration solution, and \citet{ker19} flagged it as having a PMA. 
While our average RV is 25.47 $\pm$ 0.15 km s$^{-1}$, GDR3 determined
a mean RV of 20.93 $\pm$ 0.16 km~s$^{-1}$.  From five Mount Wilson 
observations, \citet{wil50} found an RV of 21.5 km s$^{-1}$, which is
similar to that of GDR3.  \\
 
\emph{HR 1849 = HD 36473}:  Our measures yielded a low amplitude SB1O
with $P$ = 15.1920 days, but several spectra have the
appearance of very blended double lined (DBL) spectra. The $M_{V}$ is 
consistent with its spectral type. \\

\emph{HR 2103 = HD 40446}: Long suspected of being variable in
velocity \citep{dwor74} and noted as an SB2 by \citet{roy14}, the 
spectrum of this system appears as a blended DBL in most of our 
spectra. However, the system is actually triple. 

As a result, the analysis of this system required additional steps.
We first determined the long period orbit of the strongest component 
with SB1. Using the GLS program of \citet{dan66}, we next analyzed 
the weak, narrow lined component, which is the secondary of the short 
period system. To obtain that solution, all the long-period elements 
were fixed at values of the above SB1 solution. The very weak, broad 
lines of the short-period primary were the last to be examined. For 
that GLS solution, all elements were fixed at the previously determined 
values except for the short period semi-amplitude and the center of mass
velocity.  Finally, to represent the secondary RVs of the long period orbit,
we combined its predicted RV plus the observed residual RV for the
short period secondary that were computed in its GLS solution.

The A component yielded an orbit with $P$ = 710.55 days. The RVs from 
the Bb component were used to show the long-period secondary velocities
in an SB2 solution.  The B component was resolved into components Ba and Bb,
whose velocities were in effect the residuals of the fit to the long-period
orbit. They produced an SB2O with $P$ = 28.5495 days. A companion with 
$V$ = 11.8 mag and $\rho$ = 19\farcs1 is listed in the WDS. \\

\emph{HR 2163 = HD 41841}: A ``G'' multiplicity flag in HIP
indicates an acceleration of its proper motion. 
Our RVs yielded an SB1O with $P$ = 1242.3 days. A companion with 
$\Delta m$ = 5.1 mag and $\rho$ = 44\farcs3 is listed in the BSC. \\

\emph{HR 3383 = HD 72660}:  A narrow--lined A star previously observed
as a possible RV standard \citep{fek99}, HR 3383 is not part of our 
survey. We began observations in 1990
April and continued acquiring them through 2024 March for a
total of 12378 days or 33.9 yr. A Julian date plot revealed that the
star has low amplitude RV variability and that the orbital period is
somewhat longer than the range of our observations. We determined
orbits with periods between 13000 and 17000 days in steps of 1000 days.
Then with all elements varied we obtained a best final orbit with a
period of 14965 days or 40.97 yr and a semi-amplitude of 1.9 km~s$^{-1}$.
Because the final period is 21\% greater than our range of RVs, we
consider the orbit to be a preliminary result. \\

\emph{HR 3657 = HD 79248}: A relatively slowly rotating star with
$v \sin i$ of 42 km~s$^{-1}$, its RV standard deviation 
is about twice that expected for its rotational velocity. Our average 
RV of $-$8.76 $\pm$ 0.12 km~s$^{-1}$ differs by about $-$17 
km s$^{-1}$ compared to the average of $+$7.8 km~s$^{-1}$ from 
four RVs of \citet{you45}. If the average RV of \citet{you45} has a sign 
error, the two RVs would be compatible. \\

\emph{HR 3822 = HD 83104}: This star appears to have a very 
blended DBL spectrum in about one half of our data. \citet{gren99} also 
noted it as a blended DBL. Measured as a single-lined spectrum,
its RV SD of 1.81 km s$^{-1}$ is significantly higher than expected 
for its $v \sin i$ of 57 km~s$^{-1}$. With speckle \citet{tok24} recently
detected a close companion, nearly 5 mag fainter, at a separation of 
0\farcs227. The primary seems under luminous for its spectral type. 
Component B with $\Delta V$ = 3.10 mag and $\rho$ = 51\farcs8 is listed 
in the WDS. \\

\emph{HR 3894 = HD 85235}: This star has a composite spectrum 
consisting of a narrow lined and a broad lined component. We were 
only able to measure the RV of the sharp component, which shows a 
clear positive trend of about 3 km s$^{-1}$ in 9600 days (26.3 yr). 
The star has a VBO with $P$ = 106.35 yr  and $a$ = 0\farcs337 
\citep{jos21}, which has been given a quality rating of 2 or reliable
in ORB6. \\

\emph{HR 4033 = HD 89021}: HIP lists a ``G'' multiplicity flag 
indicating a possible companion. Our data yielded an orbit with 
$P$ = 1041.87 days. \\

\emph{HR 4331 = HD 96723}: We determined a constant RV = $-$11.56 
$\pm$ 0.03 km s$^{-1}$. From 27 HARPS RVs, \citet{trif20} determined
an average RV of $-$11.85 $\pm$ 0.03 km s$^{-1}$. \\
 
\emph{HR 4343 = HD 97277}: \citet{fle91} noted that this star was 
found to be a source in the ROSAT all-sky X-ray survey, which along 
with EUV and existing optical data led them to suggest that the 
source of x-rays is not the A star but a hitherto unknown WD companion. 
HIP lists a ``G'' multiplicity flag for the system.  Other groups
including \citet{sma97} and \citet{due97} searched
for radial velocity changes in HR 4343, while \citet{gren99} obtained
RVs as part of a survey of southern B8--F2 stars observed by HIP.
The first two groups concluded HR 4343 might be a binary of a few
years, while \citet{bur01} estimated an upper limit of 9.5
yr based on not resolving the binary with HST, its parallax, and
assumed masses. Indeed, RVs from these sources combined with our
own yielded an orbit with $P$ = 2328.1 days (6.37 yr). Most of the data
obtained by \citet{sma97} and \citet{due97} was acquired during
phases when little velocity change occured. \\

\emph{HR 4359 = HD 97633}:  Considered for status as an early type RV 
standard by \citet{fek99}, our average velocity is consistent with
that work. \\

\emph{HR 4585 = HD 104181}: \citet{ker19} listed an RV = $-$3.390 
km s$^{-1}$ from \citet{and12} for this star and also flagged it as 
having a PMA. \citet{derosa11} detected a possible CPM companion at 
3\farcs2 with $\Delta H_{\rm p}$ = 9.2 mag. 
The mean of our RVs = $-$0.74 with a $\sigma$ of 0.90 km s$^{-1}$, 
but more observations are needed to confirm a possible 72 day period. \\

\emph{HR 4670 = HD 106819}: This star has a moderately large 
$v$~sin~$i$ of 75 km~s$^{-1}$. We determined a mean RV of $-$9.05 
$\pm$ 0.08 km~s$^{-1}$ while the GDR3 value is $-$12.99 $\pm$ 0.22 km~s$^{-1}$. 
Thus, the two average velocities differ by 3.94 km~s$^{-1}$. \\
 
\emph{HR 4717 = HD 107966}: HIP lists a ``G'' multiplicity flag 
indicating a possible companion. \citet{hor11} detected a secondary
with speckle having $\rho$ = 0\farcs24 and $\Delta$m = 4.40 and 3.87 mag
at 5620 and 6920 \AA, respectively. A VBO was determined by \citet{tok24} with
a $P$ = 27.25 $\pm 1.76$ yr, which was given a grade of of 4 or preliminary
in ORB6. Our present RVs do not show a significant velocity change. \\

\emph{HR 4760 = HD 108844}: HIP lists a ``G'' multiplicity flag, and 
\citet{ker19} detected a PMA. Our RVs yielded an SB1 orbit with 
$P$ = 1215.1 days. \\

\emph{HR 4789 = HD 109485}: The system has a VBO with $P$ = 33.04 yr, 
$a$ = 0\farcs219, and e = 0.898 \citep{hart14}, and ORB6 gave it 
a grade of 3 or moderately reliable. The WDS lists $\Delta$$V$  
= 1.94 mag.  Slightly asymmetric 
CCFs were seen about 3 yr before periastron, although the CCF was
not unusual the year just before periastron. However, a significant 
change in RV started about 1.5 yr before periastron, increasing by 
about 20 km s$^{-1}$.
Our RVs do not span a sufficient fraction of the period to produce 
an orbit, and other available velocities do not have enough
precision to be useful.  \\

\emph{HR 4805 = HD 109860}: Our RVs resulted in an SB1O with 
$P$ = 6.82860 days. \\

\emph{HR 4833 = HD 110462}: Although $\sigma$ is equal to three 
times that expected from its $v \sin i$, which is 50 km~s$^{-1}$, 
no reasonable orbit has been found. \\

\emph{HR 4859 = HD 111270}: Our RVs yielded an SB1O with 
$P$ = 387.15 days. \\

\emph{HR 4866 = HD 111421}: Our RVs produced an SB1O with 
$P$ = 50.18765 days. \\

\emph{HR 4921 = HD 112846}:  This star is the primary of a wide 
VB with $\rho$ = 20\farcs6 and $\Delta V$ = 4.5 mag. \citet{mca87b}
observed that the primary is also a close VB, designated 
CHR~39Aa,Ab. From speckle observations \citet{doco15} computed an 
orbit for this close pair that resulted in a $P$ = 7.98 yr and 
$a$ = 0\farcs078, and in ORB6 it was assigned a grade of 2 or 
reliable. We found that the primary has a composite spectrum 
consisting of narrow lined and broad lined components, and so the 
large $v \sin i$ of 114 km~s$^{-1}$ refers to the broad lined
component.  Our RVs produced an SB2O with $P$ = 7.92 yr (2894 days), 
but we regard it as preliminary due to the difficulty of measuring 
the broad-lined component. \\

\emph{HR 4948 = HD 113865}  This star is the brightest component
of five. According to the WDS, the A component is separated by 
6\farcs2 from a much fainter BC pair that has a CPM with A. The WDS
lists two more visual components that have separations of 40 and 
139\arcsec. Our RVs show no significant change. \\

\emph{HR 4963 = HD 114330}: The brighter star has a long period VBO 
\citep{zirm15} with $P$ = 695 yr and $a$ = 1\farcs243 that was given
a grade of 4 or preliminary in ORB6, and $\Delta$m = 2.34 mag.
The B component has a CPM at $\rho$ = 6\farcs9 and $\Delta$m = 4.99 mag.
There is also a C component at 71\farcs6. \citet{bea77} determined a
provisional SB1O for the primary, HR 4963, with $P$ = 17.8 yr, but 
our data obtained over 20.3 years show no significant change in RV,
and we conclude the orbit is likely spurious. Our average RV of $-$3.86 
$\pm$ 0.06 km s$^{-1}$ is similar to the center-of-mass RV of 
$-$4.19 $\pm$ 0.45 km s$^{-1}$ proposed by \citet{bea77}. \\

\emph{HR 4974 = HD 114504}: The WDS lists a companion with $\Delta$$V$ =
4.05 mag and $\rho$ = 107\farcs8. Our RVs show no significant changes 
with the average RV = $-$14.78 $\pm$ 0.13 km s$^{-1}$. \\

\emph{HR 5021 = HD 115709}: This star was discovered to be a
VB by HIP with $\Delta H_{\rm p}$ = 2.91 mag and has a 
separation of $\rho$ = 1\farcs1 \citep{hor11}.
In about half of our spectra, the CCF resembles a very blended 
DBL spectrum. This may be caused by the VB primary's RV (the SB1) shifting 
relative to the VB secondary velocity. Our spectra yielded an SB1O with
$P$ = 1.7380758 days.\\

\emph{HR 5076 = HD 117242}: With a $v \sin i$ of 107 km~s$^{-1}$, this star
has one of the largest rotational velocities in our low $v$~sin~$i$ sample.
The CCFs of a number of its spectra have the appearance of very blended 
DBL CCFs. \\

\emph{HR 5079 = HD 117281}: HIP lists a ``G'' acceleration solution, and 
\citet{ker19} found a PMA. One speckle detection was reported by E. Horch
(INT4) of a faint secondary at $\rho$ = 0\farcs173 and $\Delta m$ = 4.62.
Our RVs produced an SB1O with $P$ = 919.5 days. The CCFs of a 
number of spectra have the appearance of very blended DBL CCFs. \\

\emph{HR 5129 = HD 118660}: This star, which has a moderately large 
$v \sin i$ of 94 km~s$^{-1}$, is another star for which the CCFs of 
many of our spectra have the appearance of very blended DBL CCFs. 
\citet{joshi06} reported $\delta$~Sct pulsation periods of about 
1 and 2.5 hours, while \citet{sark24} determined it to be a mild Am 
star with additional $\delta$ Sct pulsations. Thus, it is possible that 
the observed line profile asymmetries result from the 
$\delta$~Sct pulsations. Our mean RV is 0.00 $\pm$ 0.22 km~s$^{-1}$
while that of GDR3, $-$3.92 $\pm$ 0.21 km~s$^{-1}$, is somewhat 
different. \\

\emph{HR 5144 = HD 119055}: The star is the primary of a VB 
with component B at $\rho$ = 4\farcs4 and $\Delta$$V$  = 3.84 mag (WDS). 
Two other CPM companions are listed in the WDS.  Our RVs 
for the VB primary yielded an SB2O with P = 14.7306416 days. 
The RV of the B component = $-$21.996 km s$^{-1}$ \citep{and12}, 
which is very similar to the systemic velocity of $-$22.433 km s$^{-1}$ 
for our SB2O. The visual components A and B also have similar 
parallaxes (GDR3), reinforcing the CPM status. \\

\emph{HR 5146 = HD 119086}: \citet{fin59} discovered this star to 
be a close VB, and HIP determined $\Delta H_{\rm p}$ = 1.20 mag. 
According to the WDS, it currently has a separation of 0\farcs3, and 
its position angle has changed by 124$^\circ$ since its discovery. 
Component C at 31\farcs0 separation has a CPM. 
Our spectra reveal up to 4 broad-lined components all blended into one 
rapidly changing CCF. The nature of the rapid changes suggests a 
hierarchical pair of SB2s. However, no orbital solution has been 
determined so far although periods around a day might exist. \\

\emph{HR 5163 = HD 119537}: \citet{dwor74} listed the star as an 
SB, and the CCF of \citet{roy14} showed two unequal components. Our 
spectra likewise show a DBL system, and our RVs produced an SB2O 
with $P$ = 7.199676 days and a mass ratio of 0.66. \\

\emph{HR 5216 = HD 120874}: Our RVs resulted in an SB1O with 
$P$ = 2.178597 days. \\

\emph{HR 5220 = HD 120934}: Our mean RV of $-$11.20 km s$^{-1}$ seems
a bit discrepant with the value of $-$15.5 km s$^{-1}$ from the GCRV. \\

\emph{HR 5229 = HD 121164}: Two CPM companions are listed in the WDS.
We found the star to have a constant RV. \\

\emph{HR 5255 = HD 121996}: HIP lists a ``G'' multiplicity flag, 
and \citet{ker19} found a PMA. Our RVs produced an SB1O with $P$ = 
1530.7 days. \\

\emph{HR 5280 = HD 122866}: While \citet{ker19} concluded that this 
star has a PMA, from our data, its RV appears to be constant. \\
 
\emph{HR 5333 = HD 124713}: Our RVs produced an SB1O with 
$P$ = 25.09473 days. \\

\emph{HR 5360 = HD 125349}: An ``S'' quality flag is given by HIP
indicating a suspected non-single star. \citet{ker19} also flagged it as 
having a PMA. The star has a moderate $v \sin i$ of 86 km~s$^{-1}$. Our 
data show no significant velocity change with an average 
RV = $-$6.23 $\pm$ 0.13 km~s$^{-1}$. This value is in good agreement with 
the $-$6 km~s$^{-1}$ average of \citet{pal68} from five RVs and also with
the GDR3 value of $-$5.54 $\pm$ 0.72 km~s$^{-1}$. However, The GDR3 RVs 
have a large range of 11.79 km~s$^{-1}$. Seven much older RVs from 
\citet{plas21} also have a significant range, $-$16.2 to $-$4.8 km s$^{-1}$, 
and a rather different average RV of $-$12.3 $\pm$ 3.4 km s$^{-1}$. \\

\emph{HR 5374 = 125658}: \citet{ker19} detected a PMA while our 
RVs produced an SB1O with $P$ = 265.992 days. \citet{osa59} classified
the star as Am, but \citet{cow69} and AM95 have classified it as a 
normal A star. Possible evidence of a secondary spectrum 
has been detected in its CCF, but higher resolution spectra would be 
useful. \\

\emph{HR 5397 = HD 126367}:  Although a companion approximately
1 mag fainter was detected during a 1934 lunar occultation (INT4, WDS), a
1988 speckle observation with a 3.6 m telescope detected no companion 
with a separation $\geq$0\farcs038 \citep{mca93}. The WDS lists a more distant 
visual companion, B, with $\Delta$$V$ = 0.55 mag and $\rho$ = 35\farcs0. 
This component B has a close visual companion, C, 
\citep{hei56} with $\rho$ = 1\farcs20. Our data show no significant RV 
changes for the bright A component. \\

\emph{HR 5406 = HD 126722}: Although flagged in HIP2 with an acceleration
solution and by \citet{ker19} as having a PMA, our data indicate a 
constant RV of $-$18.10 $\pm$ 0.22 km s$^{-1}$. However, 
\citet{ker19} list an RV = $-$14.784 km s$^{-1}$ from \citet{and12}. 
As part of our low rotational velocity sample, this star has a 
relatively high $v \sin i$ of 102 km~s$^{-1}$. \\

\emph{HR 5415 = 127067}: The star is a wide VB with HR~5414. Their 
separation is 25\farcs4, and $\Delta V$ is 0.5 mag.  Our spectra of 
the brighter star, HR 5415, show a very blended, rapidly changing 
CCF similar to that of HR 5146 with possibly up to 4 components. 
No orbital solution has been determined so far, but periods around a 
day might exist. \\

\emph{HR 5468 = HD 129002}: Noted by \citet{ker19} as having a PMA, 
the recent study of \citet{elm25} 
detected a much fainter visual component with a separation of 
$\sim$0\farcs3. While their atmospheric modeling favored a late M 
dwarf spectral type for this component, their dynamic modeling, 
assuming a mass of 2.05 $M_{\sun}$ for the A star primary, produced a 
brown dwarf with a mass of $\sim65 M_{\rm Jup}$. 
A preliminary orbit by \citet{elm25} resulted in a period of 26 yr and 
an extremely eccentric orbit of 0.9.  We find that the visual 
primary of this system is an SB2, and our RVs produced an orbit with 
a $P$ = 102.0223 days. The components of our SB2 have 
a mass ratio of 0.38 and minimum masses of 2.28 and 0.87 $M_{\sun}$. 
Thus, the total minimum mass of component A is 3.15 $M_{\sun}$ or 
at least 50\% larger than that adopted by \citet{elm25}.
\\

\emph{HR 5531 = HD 130841}: \citet{sliph04} first detected the 
DBL nature of the spectrum of this very bright composite system, which 
has a very broad lined primary and a narrower lined secondary. 
\citet{fuhr14} provided the history of the system, discussed the 
great difficulty in the RV measurement of the primary, and provided 
an orbit for the secondary that has $P$ = 70.34 days. \citet{wais23b} 
measured the fluxes of the 2 stars interferometrically and 
derived $M_{Aa} = 1.95 M_{\sun}$ and $M_{Ab} = 1.79 M_{\sun}$ resulting
in a mass ratio $q$ of 0.92. Like those of \citet{fuhr14}, our 
spectra show a very
blended DBL system, and we determine a $P$ for the narrow lined 
secondary, Ab, of 70.3483 days.
Our orbit with the very difficult to measure primary velocities 
yielded a similar orbital period but a DBL solution did not give 
reasonable masses.
HR~5531 has a CPM with HR~5530, which is 4\arcmin   away, and the 
latter is itself a VBO with $P$ = 16.229 yr. Three other CPM 
members are listed in the WDS as part of the larger system. \\

\emph{HR 5569 = HD 132029}: The WDS lists a CPM companion at
5\farcs1 separation and $V$ = 11.1 mag, although this would probably
not account for the asymmetric CCF seen in over half of our
spectra. HIP lists an ``S'' quality rating. We determined the
primary to be an SB2 with $P$ = 33.75058 days. \\

\emph{HR 5574 = HD 132145}: \citet{ram89} suspected this star to be 
an SB, and it was also flagged by \citet{ker19} as having a PMA. 
Our RVs yielded an SB1O with $P$ = 421.849 days. \\

\emph{HR 5665 = HD 135263}: We determined an SB1O with 
$P$ = 393.00 days. \\

\emph{HR 5676 = HD 135502}: Our average RV appears to be constant 
at $-$16.85 $\pm$ 0.22 km s$^{-1}$. This value is similar to those 
from earlier investigations. \\

\emph{HR 5748 = HD 137928}: \citet{ker19} indicated a PMA for the 
star, but our average RV appears to be constant at $-$4.60 $\pm$ 0.09 
km s$^{-1}$. This value is in reasonable agreement with the
value of $-$5.9 km~s$^{-1}$ of \citet{har37} from three observations. \\

\emph{HR 5760 = HD 138341}: Our spectra show very blended CCFs 
with the secondary lines nearly twice as broad as those of the 
primary, so the star is an SB2. An SB1O for the primary produced 
$P$ = 364.429 days. An attempt to obtain an orbit for the secondary 
resulted in a similar period, but other elements were not consistent 
with those of the primary presumably due to the blending problems. \\

\emph{HR 5789 = HD 138918}: HR 5789 is a VB with HR 5788, the latter 
being 1.1 mag fainter and 4\farcs0 distant.  HR 5789 was noted by 
\citet{ker19} as having a PMA.  Many of our spectra of it appear to 
show very blended CCFs, however, the star is a $\delta$~Sct variable 
and this pulsation may cause the observed line asymmetries. 
\citet{wais23c} discovered a 1.8 $M_{\sun}$ companion from 
interferometric data and deduced a $P$ of $\sim$265 days.
We obtained an SB1 orbit for HR 5789 with $P$ = 461.92 $\pm$ 0.16 days.

The GCRV lists an average RV of $-$37.9 km~s$^{-1}$ that is quite close to 
the system RV of $-$38.28 km s$^{-1}$ from our orbit for HR 5789. \\

\emph{HR 5791 = HD 138963}: \citet{mend25} obtained a non-detection 
of duplicity with a speckle camera on the 8.1 m Gemini South telescope. 
Our RVs yielded an orbit with a $P$ = 94.333 days. \\

\emph{HR 5859 = HD 140775}: \citet{mca87a} reported not resolving 
the system with speckle interferometry. However, \citet{ker19}  
flagged the star as having a PMA. Our RVs show it to be constant with 
$-$10.57 $\pm$ 0.10 km~s$^{-1}$. 
\\
 
\emph{HR 5886 = HD 141653}: Many of the CCFs appear to be due to 
two very broad-lined blended components. Our average RV measured as a 
single star is $-$7.57 $\pm 0.20$ km s$^{-1}$.  The value of $M_{V} = 
0.476$ mag seems anomalously high compared to the nominal value of
1.16 mag for a A1 V star. \\

\emph{HD 144359}: \citet{alb47} found this star, classified as A0~IV
\citep{bid88}, to have a very large RV range. The GDR3 determined an  
initial SB2O with $P$ = 4.385 days, $e$ = 0.289, and a 
secondary to primary mass ratio of 0.67. While our improved solution
has similar values of $P$ (4.384370 days) and $e$ (0.282), our mass
ratio of 0.617 is somewhat smaller and produces minimum masses of 
2.1 and 1.3 $M_{\sun}$. G. Henry (private communication) found that 
the binary has low amplitude variability associated with orbital
phase. This star is not part of our survey.  \\

\emph{HR 6041 = HD 145788}: A narrow lined A star \citep{roy02}, it
was investigated as a possible early-type RV standard. 
Observations were begun in 2004 January and continued through 
2021 April. The star has a period of 2122 days or 5.8 yr and a 
semi-amplitude of just 1.6 km~s$^{-1}$. It is not part of our 
survey. \\
 
\emph{HR 6074 = HD 146738}: Our RVs produced an orbit with 
$P$ = 47.0267 days. The star has four much fainter companions ranging
from $\rho$ = 53\farcs7 to 122\farcs9 as noted in the WDS. \\

\emph{HR 6170 = HD 149650}: The PMA found by \citet{ker19} indicates 
a possible companion. Its $v \sin i$ value of 102 km~s$^{-1}$ is 
one of the largest in our low $v \sin i$ sample. Our mean RV appears 
to be constant at $-$1.10 $\pm$ 
0.17 km s$^{-1}$, but the GDR3 average RV is somewhat different at 
$-$4.73 $\pm$ 0.45 km~s$^{-1}$. A single RV of $-$24.1 km s$^{-1}$ 
from \citet{har37} is very different. \\

\emph{HR 6246 = HD 151862}:  HIP lists this star with an ``S''
quality solution indicating a possible companion. It has a visual secondary
with $\Delta$$V$ = 4.07 mag and a separation of 4\farcs9 (WDS). Two 
other faint and more distant CPM companions are listed in the WDS. Our 
data indicate no significant RV changes. \\

\emph{HR 6341 = HD154228}: A companion noted in the WDS with 
$\Delta$$V$ = 0.27 mag and $\rho$ = 300\farcs7 has been judged to be 
non-physical. Four other fainter and distant CPM companions are
listed in the WDS. \\

\emph{HR 6367 = HD 154895}: This star is a VB with $P$ = 203.6 yr 
and $a$ = 0\farcs572 \citep{mas06} and is given a grade of 3 or
moderately reliable in ORB6. Its components have $\Delta$m  
= 2.10 mag. The results of HIP2 and \citet{ker19} both suggest a 
possible shorter period binary companion. Historical RVs  
for this system are very inconsistent. From Mount Wilson Observatory 
spectra, \citet{ada12} found a velocity range of $-$8 to $-$39 km~s$^{-1}$. 
The notes section of INT4 reports that the primary is an SB with an 
observed range of 55 km~s$^{-1}$. Is this mistakenly the sum of two 
other values, $-$17.7 km s$^{-1}$ \citep{shaj32} and a single RV of 
$-$37.3 km s$^{-1}$ \citep{har37}?
Our RVs show a modest range of 7 km s$^{-1}$ with an average
value of $-$24.52 $\pm 0.57$ km s$^{-1}$. 
The CCFs are often asymmetric in this rather broad-lined star
($v \sin i$ = 76 km~s$^{-1}$). 
Various low amplitude orbits can be found, but more RVs are needed 
to confirm variability and identify the correct period. \\

\emph{HR 6378 = HD 155125}:  \citet{doco23} computed a VBO with 
$P$ = 87.76 yr and $a$ = 1\farcs307 that was given the grade of
2 or reliable with good coverage in ORB6., The components have a
$\Delta$m = 0.51 mag. 
Our spectra show a very sharp-lined component plus a broader, weaker 
component. With both stars always on the slit, no significant RV 
changes were measured for the sharp-lined star. \\

\emph{HR 6399 = HD 155860}: The star is a VB with $\rho$ = 4\farcs9 and 
$\Delta$$V$ 3.17 mag (WDS).  Its A and B components have a CPM 
according to the BSC. In the WDS another CPM component is listed at 
26\farcs7 and $V$ = 14.7 mag. Our RVs refer to the A component 
and show no significant RV changes. That component has a $v \sin i$
of 102 km~s$^{-1}$, one of the largest in our low rotational velocity
sample.\\

\emph{HR 6412 = HD 156208}: HIP discovered this star to be a partial
eclipsing binary of the Algol type with an eclipse depth of 0.13 mag. 
They also determined a period of 7.70 days. It was later assigned the 
variable star name V2368 Oph. \citet{harm11} used photometry and RVs 
from several observatories to determine the correct orbital period of 
38.327 days and obtained a preliminary eclipsing binary solution. 
They discussed at some length the difficulties in determining RVs from the 
very blended lines seen in their spectra and the effect on the derived 
physical properties of the system. They adopted a solution with a mass ratio
of 1.0 that produced an orbital inclination of 86.1\arcdeg and
masses of 2.68 $M_{\odot}$. 

    We used our RVs to obtain a new SB2O. We adopted the 
period of \citet{harm11} and first determined the orbit of the 
narrower lined primary. The orbit of the broader lined secondary 
was determined with only the semi-amplitude and center-of-mass 
velocity allowed to vary. The center-of-mass velocities of the two
solutions differed by 6 km~s$^{-1}$, so that amount was added to
the much more uncertain secondary RVs, and after appropriately
weighting the RVs, an SB2 solution was obtained.
While our orbit is an improvement compared to that of \citet{harm11},
throughout much of the orbit, the lines of the two components are
significantly blended.

Our SB2O produces minimum masses of 3.219 and 3.072 $M_{\odot}$ for 
the primary and secondary, respectively. Adopting the orbital 
inclination of \citet{harm11}, our masses are 15 to 20\% larger than
theirs. \\

\emph{HR 6421 = HD 156295}:  The CCF is sometimes assymetric 
resembling a very blended DBL spectrum, but measured as a single 
spectrum no significant RV variations were noted. \\

\emph{HR 6432 = HD 156653}: HIP lists a ``G'' multiplicity flag, 
and \citet{ker19} flagged it as having a PMA.  From six observations 
\citet{you39} found a mean RV of 7.3 km~s$^{-1}$ and a range of 
24 km~s$^{-1}$ causing him to conclude that the RV of this star 
is probably variable.  Our RVs produced an SB1O with $P$ = 
1260.2 days, a center-of-mass velocity of 4.9 km~s$^{-1}$, but 
a total amplitude of just 6 km~s$^{-1}$. \\

\emph{HR 6455 = HD 157087}: HIP lists a ``G'' possible multiplicity flag
and \citet{ker19} indicated that it has a PMA. Our RVs yielded an SB1 
orbit with $P$ = 2633.7 days.  \\

\emph{HR 6457 = HD 157198}: The WDS lists a very wide companion 
with $\Delta V$ = 4.21 mag and $\rho$ = 223\farcs3.  Three other CPM
companions are listed in the WDS.  The CCF is sometimes asymmetric 
resembling a very blended DBL spectrum, but measured as a single 
spectrum, no significant RV variations were noted. \\


\emph{HR 6480 = HD 157728}: The CCF is sometimes assymetric 
resembling a very blended DBL spectrum, but measured as a single 
spectrum no significant RV variations were noted. Five measures from
\citet{plas21} have an averge RV = $-$20.7 km~s$^{-1}$ and our average 
RV = $-$20.82 $\pm 0.18$ km~s$^{-1}$. \\

\emph{HR 6499 = HD 158067}: The WDS lists 2 companions with 
$V \sim11$ mag and $\sim$50\arcsec distant, but their changes in position 
seem to be too rapid to be physical. They also have smaller parallaxes. \\

\emph{HR 6521 = HD 158716}: \citet{ram89} determined the projected
rotational velocities of 41 A0 stars including HR 6521. From the 
measurement of a single line profile, they suggested HR 6521 to be a 
possible SB2.
However, we find its RV to be constant at $-$25.44 $\pm 0.06$ 
km s$^{-1}$. \\

\emph{HR 6548 = HD 159480}: \citet{ker19} flagged the star 
as having a PMA. \citet{hor11} have detected a close secondary with 
$\Delta$m = 3.97 mag at 5620~\AA\ and 0\farcs369 separation. With a
mean RV of $-$13.88 $\pm$ 0.18 km~s$^{-1}$, our resultant velocity 
shows no significant variations and agrees well with an earlier 
mean RV of $-$14.9 km~s$^{-1}$ from \citet{plas21}. Four other very 
wide CPM companions are listed in the WDS. \\

\emph{HR 6559 = HD 159834}: An entry in the notes to the WDS states 
that component A in the AB pair, is an SB. \citet{ker19} flagged it 
as having a PMA. Our RVs yielded an SB1 orbit with $P$ = 648.033 days. 
The eccentricity appears to be as high as 0.97, but data near 
periastron are needed to define the orbit more precisely. The WDS 
lists a CPM companion (B) with $\Delta$$V$ = 3.35 mag and $\rho$ = 10\farcs4.
Our one measurement of the RV of this much fainter B component is 
$-$17.2 km s$^{-1}$, which is close to the system RV of 
$-$18.3 km s$^{-1}$ determined from our SB1O. \\

\emph{HR 6581 = HD 160613}: This star is a known $\delta$ Sct star and 
has a relatively high $v \sin i$ value of 107 km~s$^{-1}$. Our average 
RV of $-$24.11 $\pm$ 0.17 km~s$^{-1}$ is $\sim$6 km s$^{-1}$ more 
positive than the GCRV average RV. The GDR3 mean value of $-$32.3 $\pm$ 
0.24 km~s$^{-1}$ differs substantially from our RV, and is more in 
line with that of the GCRV. 
\\

\emph{HR 6633 = HD 161941}: Our RVs produced a low amplitude SB1O 
with $P$ = 41.0052 days. \\

\emph{HR 6681 = HD 163336}: This star is the primary, ADS 10891A, 
of a CPM binary, AB, with $\Delta$m = 3.2 mag and $\rho$ = 20\farcs6 (WDS). 
\citet{ker19} flagged it as having a PMA. \citet{na85} 
found component A to have a variable RV. \citet{wais23a} discussed the total 
AB system and showed that it is quintuple. Component A is a triple 
system and component B, BD$-$15~4723, is an SB2 with a $P$ of 729 days that
was previously determined by \citet{tok19}. \citet{wais23a} used RVs 
from several different sources to compute an SB1O of $P$ = 13.4191 days 
for the Aa pair. Then the RVs and astrometry were combined to determine 
masses for those two components, which they identify as Aa1 and 
Aa2. For Aa1 we computed a substantially improved orbit with 
$P$ = 13.4188 days. Using adaptive optics images, \citet{wais23a} also 
discovered a new close companion, Ab, and calculated a $P$ of $\sim69.5$ yr 
for the Aa and Ab pair. However, in ORB6 this orbit was assigned a grade
of 5 or indeterminate. \citet{tok19} estimated an approximate period 
of 30,000 yr for the wide AB pair. \\
 
\emph{HR 6753 = HD 165358}: 
\citet{ker19} flagged it as having a PMA. Two faint companions are 
listed in the WDS, but both are identified as optical in the BSC,
and their parallaxes appear discrepant. This star has a rather large
$v \sin i$ of 109 km~s$^{-1}$. Our RVs indicate 
a constant RV of $-$12.11 $\pm 0.13$ km s$^{-1}$. However, the GDR3
mean RV is $-$17.44 $\pm$ 0.28 km~s$^{-1}$, which differs from ours by  
5.33 km~s$^{-1}$. \\

\emph{HR 6883 = HD 169111}: This star has a relatively large $v \sin i$ 
of 107 km~s$^{-1}$. Our mean RV is $-$39.37 $\pm$ 0.25 km~s$^{-1}$ while
that of GDR3 is $-$43.55 $\pm$ 0.36 km~s$^{-1}$, and so differs by
4.18 km~s$^{-1}$. An RV = $-$56.3 km~s$^{-1}$ was listed by 
\citet{pet28} from 5 observations. \\

\emph{HR 7001 = HD 172167}: Famously known as $\alpha$ Lyrae or Vega, this
star may have more references than almost any other star. Our measured RV = 
$-$13.13 $\pm$ 0.21 km s$^{-1}$ is similar to the other 19 RVs listed 
on SIMBAD that have a range from $-$12.6 to $-$20.6 km s$^{-1}$. The 
GCRV RV = $-$13.9 km~s$^{-1}$. The scatter of our RVs results in a
$\sigma$ of 1.00 km~s$^{-1}$. This is larger than would be expected 
from its $v \sin i$ of 24 km s$^{-1}$, but this larger $\sigma$ is 
most likely due to the very short exposure
times needed to avoid saturation. \\

\emph{HR 7003 = HD 172187}: Yet another star that \citet{ker19} 
flagged as having a PMA. From TESS photometric observations, 
\citet{prsa22} determined
it to be an eclipsing binary, later named V533 Lyr, with a $P$ of  
1.31283 days, The TESS photometry also caused \citet{chen22} to identify
it as a $\delta$ Sct variable. Our spectra show very rapid and 
large CCF changes from night to night that are consistent with the 
short orbital period. The measured $M_{V}$ is 0.797 mag while the 
nominal value for an A6 V star is 2.12 mag. No spectroscopic orbit has 
been determined due to difficulties measuring the very broad and 
blended line profiles. \\

\emph{HR 7019 = HD 172741}: HIP discovered this star to be a 
close VB, designated HDS~2644. It was also flagged by \citet{ker19} as 
having a PMA.  From 5 observations \citet{mas26} determined a visual orbit 
with a period of $P$ = 23.55 yr or 8601.6 days and $e$ = 0.18. While 
\citet{mas26} indicate that this is a first orbit, a similar orbit was 
computed by \citet{doco21} and given the grade of 3 or moderately 
reliable in ORB6.  
Both orbits are in good agreement with our SB1O of 24.8 yr or 9064 days 
and $e$ = 0.18. 
The WDS lists a $\Delta$$V$ of 1.19 mag, and we measure RVs 
for a single component with symmetric, moderately narrow lines.  
All our KPNO CCF profiles have a very broad base or pedestal from the 
secondary star under the sharp component, which we were not able to measure. \\

\emph{HR 7060 = HD 173664}: This star has a moderate $v \sin i$ value
of 71 km~s$^{-1}$. Our average RV for it is $-$1.44 $\pm$ 0.08 km~s$^{-1}$
while the GDR3 mean RV is $-$4.95 $\pm$ 0.36 km~s$^{-1}$. \\

\emph{HR 7250 = HD 178187}: \citet{ker19} flagged it as 
having a PMA. Our average RV appears to be constant at $-$22.64 $\pm 0.04$ 
km s$^{-1}$. This value agrees well with earlier values listed in SIMBAD. \\
 
\emph{HR 7357 = HD 182239}: HIP listed the star with a ``G'' multiplicity
flag, and \citet{ker19} identified it as having a PMA. 
At the time our observations commenced in 1993 April (JD 2449081), the RV 
was decreasing rapidly from an intial RV $\sim$12.4 to $\sim$10.5 km s$^{-1}$ 
over an interval of $\sim$80 days. The RV remained near this lower value 
for the remainder of our observing interval that ended in 2010 
September (JD~2455451). 
The GDR3 mean RV is 10.77 $\pm 1.08$ km s$^{-1}$, very similar to
our most recent RVs. Although the Gaia time series for this star
covers 869 days, no specific epoch is given. However, during that 2.4 yr
interval, the total range of RVs is 11.1 km~s$^{-1}$. Given that the 
$v \sin i$ for this star is just 57 km~s$^{-1}$, this large range 
suggests a possible high eccentricity system. \\
  
\emph{HR 7362 = HD 182369}: A well known close VB with an orbital 
period of 5.6985 yr, $a$ = 0\farcs064, and components of nearly equal 
brightness \citep{doco23}. In ORB6 the orbit was given a grade of 2 
or reliable. Our spectra indicate that one of the VB 
components is also an SB2 with a period of a few days. The value of 
$M_{V}$ = 0.553 mag is much larger than the nominal value of 2.32 for an 
A8 V star. Due to the severe blending of all 3 components, a satisfactory 
orbit has not been obtained. \\

\emph{HR 7371 = HD 182564}: The WDS lists a very faint companion,
$\Delta$m = 8.6 mag at 13\farcs1. Our data indicate a constant RV 
of $-$28.43 $\pm 0.06$ km s$^{-1}$. Other values listed in SIMBAD 
are similar. \\
 
\emph{HR 7695 = HD 191174}: The WDS lists 2 CPM companions, 
B at 5\farcs7 and $V$ = 9.54 mag plus C at 153\farcs0 and $V$ = 11.42
mag. The average of our RVs = $-$24.38 $\pm 0.18$ km s$^{-1}$,
however, the CCF often appears to have an extended blue wing. \\

\emph{HR 7781 = HD 193592}: The WDS lists the current separation of 
the VB components A and B as $\rho$ = 3\farcs9 with a $\Delta$$V$ for 
the pair of 1.5 mag. Three other faint companions also are listed in 
the WDS. Our spectra show a slightly assymetric CCF on a few occasions, 
and as an SB1, we determined an orbit with $P$ = 260.328 days. AM95 
classify the spectral type of B as F3~V. Notes in the WDS and BSC 
claim that both components A and B are SBs. The SIMBAD references 
provide no obvious support for such a claim. \\

\emph{HR 7818 = HD 194882}:  \citet{dav24} determined a substantially
revised VBO with $P$ = 90.42 yr and $a$ = 0\farcs1378. It was given a
grade of 2 or reliable in ORB6.  The WDS lists a $\Delta$$V$ = 0.69 mag 
for the pair.
Our spectra show a very blended pair of lines from which we derive two
different SB1 orbits with $P$ = 28.9409 days for Aa and 
$P$ = 4.83250 days for Ba. \\

\emph{HR 7891 = HD 196724}:  HIP assigned the star a ``G'' flag, and 
\citet{ker19} listed the star as having a PMA.  \citet{roy14} 
suspected the star of having a variable RV.  
We determined an SB1O with $P$ = 1277.4 days. \\

\emph{HR 7938 = HD 197734}:  \citet{ker19} flagged the star 
as having a PMA. Our spectra show a DBL system with the primary and secondary 
spectra mostly blended together except when the secondary had large positive
RVs near phases 0 and 1. We derived an SB2O with $P$ = 315.656 days. \\

\emph{HR 7981 = HD 198552}: Both HIP and \citet{ker19} flagged this star 
as a possible binary. While \citet{mca87a} reported not resolving the 
system with speckle interferometry, we have detected binary motion with $P$ = 
571.14 days. \\
 
\emph{HR 8641 = HD 214994}: A narrow--lined A star previously observed 
as a possible RV standard \citep{fek85}. It is not part of our survey.
\citet{fek99} computed an SB1O with $P$ = 3955 $\pm$ 238 days or 10.83 
$\pm$ 0.65 yr. With additional RVs the present orbit has $P$ = 3773 
$\pm$ 21 days or 10.330 $\pm$ 0.058 yr. \\

\emph{HR 8837 = HD 219290}: Another star that \citet{ker19} 
flagged as having a PMA. However, our RVs appear constant with 
an average of $-$16.90 $\pm$ 0.18 km~s$^{-1}$. 

\subsection{High $v \sin i$ A stars}

\emph{HR 5085 = HD 117376}:  
There are 5 components listed in the WDS but the D and E components 
are not thought to be physially connected due to their disparate 
proper motions and parallaxes.  Curiously, it is the C component 
that is HR 5085, the brightest member. According to \citet{tok21},
the AB pair is in a 500 yr orbit that is also listed in ORB6 but
graded indeterminate.  Our average RV of $-$6.6 $\pm 0.4$
km s$^{-1}$ agrees well with the mean value of $-$7.0 km~s$^{-1}$
from the GCRV as well as the GDR3 average value of $-$8.6 $\pm$ 
0.8 km~s$^{-1}$. However, we note that the GDR3 RVs have a relatively
large range of 11.2 km~s$^{-1}$.
\\

\emph{HR 5107 = HD 118098}: This star has a $v \sin i$ of 222 km~s$^{-1}$,
the highest value in our program.
The WDS lists one companion having $V$ = 10 mag and $\rho$ = 1\farcs8. 
The notes in the WDS state that ``zet Vir = GJ 3792. Observations 
over 4.75 years indicate the pair has common proper motion and shows 
clear evidence of orbital motion.
\citet{hink10} constrain elements $a$ >24.9 au, $P$ >124 yr, and
$e$ >0.16. Mass of the companion is estimated at 0.168$^{+0.012}_{-0.016}$
$M_{\sun}$, suggesting a spectral type of M4V-M7V.'' \citet{ker19} flagged it
as having a PMA due to differences between the HIP2 and GDR2 proper motions.
Our average RV is $-$10.5 $\pm 0.8$ km s$^{-1}$ while the GDR3 mean 
value is $-$5.7 $\pm$ 1.1 km~s$^{-1}$.  The GCRV value is $-$13.2 
km~s$^{-1}$ from 15 measurements. \\

\emph{HR 5112 = HD 118232}: Our average RV of $-$14.6 $\pm$ 0.2 
km s$^{-1}$ is fairly close to the mean value of $-$12.0 km~s$^{-1}$ from 
the GCRV. However, SIMBAD lists many other RVs ranging from $-$4.9 to 
$-$23.3 km s$^{-1}$. We consider the RV to be constant. \\

\emph{HR 5116 = HD 118295}: Our mean RV is $-$19.1 $\pm$ 0.2 
km s$^{-1}$ compared to the GDR3 value of $-$22.2 $\pm$ 0.2 km~s$^{-1}$
\citet{you45} obtained a value of $-$26.1 km s$^{-1}$. \\ 

\emph{HR 5127 = HD 118623}:  \citet{izm19} determined the orbit of 
the AB pair. Given a grade of 3 or moderately reliable in ORB6, it has 
a $P$ = 245.0 $\pm 
7.6$ yr and $a$ = 1\farcs006 $\pm 0\farcs033$. Thus, both the A and B 
components with $V$ = 4.98 and 6.95 mag, respectively, would be on the 
spectrograph slit together. Four additional fainter and more distant 
CPM companions are listed in the WDS. Our average RV of $-$3.4 $\pm 0.6$ 
km s$^{-1}$ is in reasonable agreement with the GCRV mean RV = $-$6.0 
$\pm$ km~s$^{-1}$. 
\\

\emph{HR 5162 = HD 119476}: Our average RV = $-$7.6 $\pm$ 0.4 km~s$^{-1}$ 
while SIMBAD lists values ranging from 2.2 to $-$15.4 km~s$^{-1}$. \\

\emph{HR 5244 = HD 121607}: Our mean RV is $-$12.0 $\pm$ 0.4 km~s$^{-1}$. 
More recently, from seven RVs over a span of 768 days, \citet{beck15} 
measured an average RV of $-$23.84 km~s$^{-1}$ with a $\sigma$ of 1.5 
km~s$^{-1}$. The GDR3 value is $-$16.7 $\pm$ 0.3 km~s$^{-1}$. Older
RVs by \citet{you45} and \citet{pal68} are $-$27.9 and $-$19.0 km~s$^{-1}$, 
respectively. Taken altogether, these RVs strongly suggest that this 
star is a binary. \\

\emph{HR 5290 = HD 123255}: Our average RV is $-$23.6 $\pm 0.2$ km s$^{-1}$
compared with the GDR3 value of $-$17.9 $\pm$ 0.4 km~s$^{-1}$. Both
differ significantly from the four old Mount Wilson Observatory RVs that 
have an average value of $-$36.5 $\pm 4.3$ km s$^{-1}$ \citep{chris38}. 
Our study saw no significant variation over the 3612 days of observation. \\

\emph{HR 5392 = HD 126248}: The WDS lists 2 CPM companions, one 
with $\rho$ = 158\farcs5 and $V$ = 7.32 mag, and the other with 
$\rho$ = 3\farcs4 and $V$ = 10.2 mag. HIP2 has a flag
indicating a possible non-single star. 
Our average RV is $-$12.4 $\pm 0.5$ km~s$^{-1}$ and in agreement with 
the GDR3 value of $-$13.3 $\pm$ 0.4 km~s$^{-1}$.
However, SIMBAD lists a number of much earlier RVs ranging from 
15.7 to $-$20.0 km s$^{-1}$. \\

\emph{HR 5532 = HD 130917}: The WDS lists 2 CPM companions, 
B at $\rho$ = 61\farcs0 and $V$ = 16.5 mag , and C at 110\farcs3 
and $V$ = 12.94 mag. The A component is listed as an SB by the BSC. 
This star has the highest $v \sin i$ in our survey, 222 km~s$^{-1}$. 
Our average RV is 4.5 $\pm 0.9$ km~s$^{-1}$ while the GDR3 value
is 0.2 $\pm$ 0.3 km~s$^{-1}$. 
Earlier measures listed in SIMBAD
range from $-$14.7 to 6.0 km s$^{-1}$. \\

\emph{HR 5715 = HD 136729}: The WDS lists one close companion for 
which the INT4 has 5 entries including one from HIP that does not 
resolve the pair, with a range of $\rho$ from 0\farcs166 to 0\farcs26. 
The angular change appears to be fairly rapid with 
a 2\fdg 9 change between 2 consecutive years of measures by 
\citet{mca89,mca93}. \citet{derosa11} determined the magnitudes 
of components A and B to be 5.52 and 11.5 at 7980 \AA, 
respectively. Our first group of 5 RV measurements between JDs 
2455678 and 2457208 have an average RV of 3.5 $\pm 0.3$ km s$^{-1}$, 
while the second group of 6 RVs between JDs 2460379 and 2460390 yielded
an average RV = 7.4 $\pm 0.4$ km s$^{-1}$. We conclude that this 
change is due to the RV motion in the VBO. The GDR3 mean RV is $1.2$ $\pm$
0.6 km~s$^{-1}$. Earlier RV measures listed in SIMBAD fall between 
3.4 and 12.3 km s$^{-1}$. \\

\emph{HR5818 = HD 139493}: \citet{mca87a} first detected this star as 
a close VB, but after multiple failed attempts at confirmation, 
\citet{mca93} listed it as an apparently spurious VB. However, 
HIP listed it with a ``G'' flag and \citet{ker19} also flagged 
it as having a PMA. Its nature as a VB has been restored. At a  
$v \sin i$ of 200 km~s$^{-1}$, this star has a very large rotational
velocity.
Our average RV is $-$23.6 $\pm 1.0$ km s$^{-1}$, in very good agreement
with the GDR3 mean RV of $-$23.3 $\pm$ 1.5 km~s$^{-1}$. However, the
GDR3 RV range of 11.3 km~s$^{-1}$ is one of the four largest
of our stars with GDR3 RVs.
\citet{you45} determined an RV of $-$20.3 km s$^{-1}$ from observations
taken five decades earlier. 
 \\

\emph{HR 5867 = HD 141003}: The WDS lists 3 CPM companions, however 
the AC pair is now considered non-phyical due to statistically 
different parallaxes and proper motions. \citet{ker19} found a PMA. 
\citet{beck15} determined a linearly increasing RV of 0.96 
km~s$^{-1}$~yr$^{-1}$ 
from 32 observations obtained over 1082 days. Their Table 6 lists 
an RV of 3.32 km s$^{-1}$, which characterizes the range of the RV 
change seen over the approximate time period of JD 2453950 to 2455000. 
Our data cover the interval of JD 2456787 to 2457209 with an average RV = 
1.0 $\pm 0.3$ km s$^{-1}$. \citet{bori23} list an RV = $-$1.96 $\pm 
0.23$ km s$^{-1}$ from observations centered on approximately 
JD 2452320. All of these observations seem consistent with a period of
several years. \\

\emph{HR 5895 = HD 141851}: \citet{mca87a} discovered HR 5895 to be a VB.
\citet{tok18} determined the most recent VBO for the 
$V$ = 5.2 and 7.8 mag components that have $a$ = 0\farcs562 and 
$P$ = 64.842 yr. His orbit was given a grade of 3 or moderately 
reliable in ORB6. Our spectra show a very broad CCF with a sharp 
component superposed. The broad component is difficult to measure, but 
we derived an RV = $-$9.2 $\pm 1.9$ km s$^{-1}$ from 4 spectra. Our 
four early spectra of the sharp component (B) obtained during the 
interval of JD 2456786.9 to 2457208.8 yielded an RV = $-$11.8 $\pm$ 
1.1 km s$^{-1}$, while a second set of nine observations that commenced 
3151 days later during JDs 2460360.0 to 2460390.9 have an average 
RV of $-$16.6 $\pm 0.1$ km s$^{-1}$. Finally, there are 10 HARPS RVs
\citep{trif20} obtained between JDs 2453875.6 and 2453990.5 with and
average RV = $-$10.2 $\pm 0.1$ km s$^{-1}$. The approximately 6 
km s$^{-1}$ difference between the earlier HARPS RVs and our last 
measures likely indicates orbital motion. \\

\emph{HR 5919 = HD 142500}:  A known $\delta$ Sct star, it appears to 
have a constant RV.  Our average RV = $-$26.5 km s$^{-1}$ and is similar 
to the mean RV of $-$26.3 $\pm$ 0.6 km s$^{-1}$ from GDR3, as well as 
$-$24.0 km s$^{-1}$ from the GCRV. \\

\emph{HR5960 = HD 143466}: A known $\delta$ Sct star, our mean RV 
is $-$13.4 $\pm 0.5$ km s$^{-1}$. \citet{ker19} listed an RV = $-$11.0 
km s$^{-1}$ from the compilation of \citet{and12}. Earlier RVs listed 
in SIMBAD range from $-$4.0 to $-$30.0 km s$^{-1}$, but based on our 
RVs, we consider it to have a constant velocity. \\

\emph{HR 6004 = HD 144874}: Our AST spectra of this very rapidly rotating
($v \sin i$ = 142 km~s$^{-1}$) star, show changing line profile bumps, 
particularly at red wavelengths, suggesting that the star is a $\delta$ Sct 
or perhaps a $\gamma$~Dor pulsator. The CCFs of our KPNO spectra show
asymmetric line profiles, likely also the result of pulsation.
However, the star seems 
to have been considered as an SB from its earliest RV measurements 
\citep{plas18}. Its spectra were among the first obtained with the then 
new 1.8 m telescope at DAO. Based on the 4 spectra with a measured RV
range of 29 km s$^{-1}$ and an average of $-$27.0 km s$^{-1}$, it was 
considered to be an SB. A second publication 
by \citet{plas20} titled \emph{One Hundred Spectroscopic Binaries} 
repeated the previous 4 observations but the third and fourth were 
different, perhaps remeasured. Two additional observations were also 
included. A third reference to all 6 RVs was subsequently reported by 
\citet{plas21}. SIMBAD lists the same RVs under other compilations, 
but all refer to the earlier work. Our own RVs yield an average
RV = $-$20.7 $\pm 2.3$ km s$^{-1}$.  Additional observations 
are needed to determine if the RV variations result from pulsation or
are due to it being an SB. \\

\emph{HR 6033 = HD145607}: The mean of our 5 RV observations is $-$0.5
km~s$^{-1}$. The average RV of $-$1.1 $\pm$ 0.5 km s$^{-1}$ from GDR3 
is consistent with our result. 
Various decades of old RVs from $-$6.4 to 15.9 km s$^{-1}$ are listed in
SIMBAD.  We consider the RV to be constant. \\

\emph{HR 6277 = HD 152569}: A known low amplitude $\delta$ Sct star 
\citep{kaye00}, it appears to have had only a single RV measurement 
of $-$21 $\pm 7$ km~s$^{-1}$ \citep{shaj32}. Our average RV from 5 
measures is $-$17.8 $\pm 1.1$, and we consider it to have a constant 
RV from this limited data. \\

\emph{HR 6317 = HD 153653}:  \citet{toko24} determined a revised VBO 
for this close binary with $P$ = 13.1 yr, $e$ = 0.72, $a$ = 0\farcs106, 
and a rather low $i$ of 24.9\arcdeg. This orbit was assigned a grade of
3 or moderately reliable in ORB6. Earlier, \citet{rob18} measured a 
magnitude difference of 2.3 for the pair.
Five RVs from \citet{pet28} have a total range of 
44.6 km s$^{-1}$ over 425 days and yielded an average RV of $-$33.1 
$\pm 9.4$ km s$^{-1}$. This range caused him to claim that the star is 
an SB. Two other RVs listed in SIMBAD have values 
of $-$10.0 and 7.20 km s$^{-1}$. Our five RVs have an average value 
of $-$5.4 $\pm 0.6$ km s$^{-1}$ over 424 days. 
Our RVs do not support the star being a short period binary. \\

\emph{HR 6507 = HD 158352}: \citet{thomas23} included HR 6507 in 
two adaptive optics infrared imaging surveys but did not find an 
optical companion to the star. More recently using LBI, 
\citet{defur25} also detected no close optical companion to it. 
However, the star was earlier of interest for a 
different reason.  \citet{abtm73} found this broad lined star
with $v \sin i$ = 180 km~s$^{-1}$ to have a couple of narrow Ti~II lines 
probably due to absorbing shells. \citet{domsmith77} reexamined the 
star and noted that it barely displayed shell star characeteristics 
with weak narrow absorption lines of Ti~II in spectra obtained in 
1974-75. Two decades later \citet{jaschek98} also examined the star 
but noted that at blue wavelengths they found no clear evidence of 
a shell.  \\

\emph{HR 6877 = HD 168914}: \citet{mca87a} included this star
in a search for duplicity amoung bright stars but obtained a negative
result. \\ 

\emph{HR 8264 = HD 205767}: A high $v \sin i$ star with a previous   
orbit by \citet{abt65}.  With combined data as discussed in Section 5.1,
we derived a preliminary orbit with $P$ = 24.8 $\pm$ 1.8 yr. This period
is in good agreement with an astrometric orbit having $P$ = 25.5 $\pm$
3 yr determined by \citet{gon02}.  \\


\begin{thebibliography}{}


\bibitem[H. A. Abt(1965)]{abt65}
   Abt, H. A. 1965, \apjs, 11, 429
\bibitem[Abt(2009)]{abt09}
   Abt, H. A. 2009, \aj, 138, 28
\bibitem[H. A. Abt \& S. G. Levy(1974)]{abt74}
   Abt, H. A., \& Levy, S. G., 1974, \apj, 188, 291 
\bibitem[H. A, Abt \& N. Morrell(1995)]{abt95}
   Abt, H. A., \& Morrell, N. 1995, \apjs, 99, 135
\bibitem[H. A. Abt \& K. I. Moyd(1973)]{abtm73}
   Abt, H. A., \& Moyd, K. I. 1973, \apj, 182, 809
\bibitem[W. S. Adams(1912)]{ada12}
   Adams, W. S. 1912, \pasp, 24, 129
\bibitem[V. A. Albitzky(1947)]{alb47}
   Albitzky, V. A. 1947, Izv. Krym. Astrofiz. Obs., 1, 23
\bibitem[E. Anderson \& C. Francis(2012)]{and12}
   Anderson, E., \& Francis, C. 2012, AstL, 38, 331A
\bibitem[E. S. Barker et al.(1967)]{bar67}
   Barker, E. S., Evans, D. S., \& Laing, J. D. 1967, R. Obs. Bull., 130, 355
\bibitem[N. Bastian \& S. E. de Mink(2009)]{bas09}
   Bastian, N., \& de Mink, S. E. 2009, \mnras, 398, L11
\bibitem[W. R. Beardsley \& E. R. Zizka(1977)]{bea77}
   Beardsley, W. R., \& Zizka, E. R. 1977, \rmxaa, 3, 109B
\bibitem[J. C. Becker et al.(2015)]{beck15}
   Becker, J. C., Johnson, J. A., Vanderburg, A., \& Morton, T. D. 2015,
   \apjs, 217, 29
\bibitem[W. P. Bidelman et al.(1988)]{bid88}
   Bidelman, W. P., Ratcliff, S. J. \& Svolopoulos, S. 1988, \pasp. 100,828-829
\bibitem[R. Blomme et al.(2023)]{blo23}
   Blomme, R., Frem\'{a}t, Y., Sartoretti, P., et al. 2023, \aap, 674, A7 
\bibitem[B. W. Bopp et al.(1970)]{bopp70}
   Bopp, B. W., Evans, D. S., Laing, J. D., \& Deeming, T. J. 1970, \mnras,
   147, 355
\bibitem[S. V. Borisov et al.(2023)]{bori23}
   Borisov, S. V., Chilingarian, I. V., Rubtsov, E. V. et al. 2023, \apjs, 266, 11
\bibitem[D. Branch(1976)]{bra76}
   Branch, D. 1976 \apj, 210, 392
\bibitem[M. R. Burleigh et al.(2001)]{bur01} 
   Burleigh, M. R., Barstow, M. A., Schenker, K. J., et al.  2001, 
   \mnras, 327, 1158
\bibitem[W. W. Campbell(1928)]{cam28}
   Campbell, W. W. 1928, Publ. Lick Obs. 16, 1
\bibitem[J.-M. Carquillat \& J.-L. Prieur(2007)]{car07}
   Carquillat, J.-M., \& Prieur, J.-L. 2007, \mnras, 380, 1064
\bibitem[X. Chen(2022)]{chen22}
   Chen, X., Ding, X., Cheng, L., et al. 2022, \apjs, 263, 34
\bibitem[W. H. Christie \& O. C. Wilson(1938)]{chris38}
   Christie, W. H., \& Wilson, O. C. 1938, \apj, 88, 34
\bibitem[A. Cowley et al.(1969)]{cow69}
   Cowley, A., Cowley, C., Jaschek, M., \& Jaschek, C. 1969, \aj, 74, 375
\bibitem[ W. Daniels(1966)]{dan66}
   Daniels, W. 1966, Univ. of Maryland, Dept. of Physics and Astronomy 
   Tech. Rep. No. 579
\bibitem[J. W. Davidson et al.(2024)]{dav24}
   Davidson, J. W., Horch, E. P., Majewski, S. R., et al. 2024, \aj, 167, 117 
\bibitem[M. De Furio et al.(2022)]{defur22}
   De Furio, M., Gardner, T., Monnier, J., et al. 2022, \apj, 941, 118
\bibitem[M. De Furio et al.(2025)]{defur25}
   De Furio, M., Gardner, T., Monnier, J. D., et al. 2025, \apj, 990, 54
\bibitem[R. J. De Rosa et al.(2011)]{derosa11}
   De Rosa, R.J., Bulger, J., Patience, J., et al. 2011, \mnras, 415, 854
\bibitem[R. J. De Rosa et al.(2014)]{derosa14}
   De Rosa, R.J., Patience, J., Wilson, P. A., et al. 2014, \mnras, 437, 1216
\bibitem[J. A. Docobo \& P. Campo(2015)]{doco15}
   Docobo, J. A., \& Campo, P. 2015, IAU Comm. 26, Info. Circ., 187, 1
\bibitem[J. A. Docobo \& P. Campo(2021)]{doco21}
   Docobo, J. A., \& Campo, P. 2021, IAU Comm. 26, Info. Circ., 205, 1
\bibitem[J. A. Docobo et al.(2023)]{doco23}
   Docobo, J. A.,Docobo, Campo, P., Mendez, R. A., \& Costa, E. 2023, 
   IAU Comm. 26, Info. Circ., 211, 1
\bibitem[J. F. Dominy \& M. A. Smith(1977)]{domsmith77}
   Dominy, J. F., \& Smith, M. A. 1977, \apj, 217, 494
\bibitem[A. Duquennoy \& M. Mayor(1991)]{duq91}
   Duquennoy, A., \& Mayor, M. 1991, \aap, 248, 485
\bibitem[R. Duemmler et al.(1997)]{due97} 
   Duemmler, R., Pelt, J., Korhonen, H., \& Iliev, I. 1997, 
   \aap, 328, L37
\bibitem[M. M. Dworetsky(1974)]{dwor74}
   Dworetsky, M. M. 1974, \apjs, 28, 101
\bibitem[J. A. Eaton \& M. H. Williamson(2004)]{ew04}
   Eaton, J. A., \& Williamson, M. H. 2004, SPIE, 5496, 710
\bibitem[J. A. Eaton \& M. H. Williamson(2007)]{ew07}
   Eaton, J. A., \& Williamson, M. H. 2007, \pasp, 119, 886
\bibitem[M. El Morsy et al.(2025)]{elm25}
   El Morsy, M., Currie, T., Lacy, B., et al. 2025, \apj, 995, L4
\bibitem[ESA(1997)]{esa97}
   The Hipparcos and Tycho Catalogues (ESA SP-1200), (Noordwijk: ESA)
\bibitem[F. C. Fekel(1985)]{fek85}
   Fekel, F. C. 1985, in IAU Coll. 88, Stellar Radial Velocities, ed.
   A.G. David Philip \& D. W. Latham (Schenectady, NY: L. Davis Press), 335
\bibitem[F. C. Fekel(1999)]{fek99}
   Fekel, F.C. 1999, in ASP Conf. Ser. 185, Precise Stellar Radial 
   Velocities, IAU Coll. 170, ed. J. B. Hearnshaw \& C. D. Scarfe 
   (San Francisco: ASP), 378
\bibitem[F. C. Fekel \& R. F. Griffin(2011)]{fg11}
   Fekel, F. C., \& Griffin, R. F., 2011, The Observatory, 131, 283 
\bibitem[F. C. Fekel et al.(2013)]{fetal13}
   Fekel, F. C., Rajabi, S., Muterspaugh, M. W., \& Williamson, M. H.
   2013, \aj, 145, 111
\bibitem[F. C. Fekel et al.(2009)]{fetal09}
   Fekel, F. C., Tomkin, J., \& Williamson, M. H. 2009, \aj, 137, 3900
\bibitem[F. C. Fekel et al.(2015)]{fetal15}
   Fekel, F. C., Williamson, M. H., Muterspaugh, M. W., et al. 2015,
   \aj, 149, 63 
\bibitem[W. S. Finsen(1959)]{fin59}
   Finsen, W. S. 1959, Monthly Notices Astron. Soc. South Africa, 18, 56
\bibitem[M. J. Fitzpatrick(1993)]{fit93}
   Fitzpatrick, M. J. 1993, in ASP Conf. Ser. 52, Astronomical Data 
   Analysis Software and Systems II, Ed. R. J. Hanish, R. V. J. 
   Brissenden, \& J. Barnes (San Francisco: ASP), 472
\bibitem[T. A. Fleming et al.(1991)]{fle91}
   Fleming, T. A., Schmitt, J. H. M. M., Barstow, M. A., \& 
   Mittaz, J. P. D. 1991, \aap, 246, L47
\bibitem[K. Fuhrmann et al.(2014)]{fuhr14}
   Fuhrmann, K., Chini, R., Barr, A., et al. 2014, \mnras, 
   437, 2303
\bibitem[T. Gardner et al.(2022)]{getal22}
   Gardner, T., Monnier, J. D., \& Fekel, F. C. 2022, \aj, 164, 184
\bibitem[Gaia Collaboration(2021)]{gaia21}
   Gaia Collaboration, Brown, A. G. A., Vallenari, A., Prusti, T., et al. 
   2021, \aap, 649, A1
\bibitem[Gaia Collaboration et al.(2023)]{gaia23}
   Gaia Collaboration, Vallenari, A., Brown, A. G. A., Prusti, T., et al. 
   2023, \aap, 674, A1
bibitem[A. Vallenari et al.(2023)]{val23}
   Gaia Collaboration, Vallenari, A., Brown, A. G. A., et al. 2023, \aap, 
   674, A34
\bibitem[G. Gontcharov \& O. Kiyaeva(2002)]{gon02}
   Gontcharov, G., \& Kiyaeva, O. 2002, \aap, 391, 647
\bibitem[S. Grenier et al.(1999)]{gren99}
   Grenier, S., Burnage, R., Faraggiana, R., et al. 1999, \aaps, 135, 503
\bibitem[J. C. Hall et al.(1994)]{hal94}
   Hall, J. C., Fulton, E. E., Huenemoerder, D. P., Welty, A. D. \& Neff, J. E.
   1994, \pasp, 106, 315
\bibitem[J. C. Hall \& G. W. Lockwood(1995)]{hal95}
   Hall, J. C., \&  Lockwood, G. W. 1995, \apj, 438, 404
\bibitem[P. Harmanec et al.(2011)]{harm11}
   Harmanec, P., Bozic, H., Mayer, P., et al. 2011, \aap, 531, 49
\bibitem[W. E. Harper(1937)]{har37}
   Harper, W. E. 1937, Publ. Dom. Astrophys. Obs., 7, 1
\bibitem[W. I. Hartkopf \& B. D. Mason(2014)]{hart14}
   Harkopf, W. I., \& Mason, B. D. 2014, IAU Comm. 26, Info. 
   Circ., 184, 1
\bibitem[W. I. Hartkopf et al.(2001)]{hart01}
   Harkopf, W. I., Mason, B. D., \& Worley, C. E. 2001, \aj, 122, 3472; 
   see also https://crf.usno.navy.mil/wds-orb6/
\bibitem[C. He et al.(2022)]{he22}
   He, C., Sun, W., Li, C., et al. 2022, \apj, 938, 42
\bibitem[C. He et al.(2023)]{he23}
   He, C., Li, C., Sun, W., et al. 2023, \mnras, 525, 5880 
\bibitem[W. D. Heintz(1956)]{hei56}
   Heintz, W. D. 1956, \mnras, 116, 248
\bibitem[S. Hinkley et al.(2010)]{hink10}
   Hinkley, S., Oppenheimer, B. R., Brenner, D. et al. 2010, \apj, 712, 421
\bibitem[D. Hoffleit(1964)]{hof64}
   Hoffleit, D. 1964, The Bright Star Catalogue (3rd rev. ed.; New Haven: 
   Yale Univ. Observatory)
\bibitem[D. Hoffleit \& C. Jaschek(1982)]{hof82}
   Hoffleit, D., \& Jaschek, C. 1982, The Bright Star Catalogue (4th rev.
   ed.; New Haven: Yale Univ. Observatory)
\bibitem[E. H\o{}g et al.(2000)]{hog00}
   H\o{}g, E., Fabricius, C. Makarov, V. V., et al. 2000, \aap, 355, L27
\bibitem[E. P. Horch et al.(2011)]{hor11}
   Horch, E. P., Gomez, S. C., Sherry, W. H., et al. 2011, \aj, 141, 343
\bibitem[E. P. Horch et al.(2012)]{hor12}
   Horch, E. P., Bahi, L. A. P., Gaulin, J. R., et al. 2012, \aj, 143, 10
\bibitem[J. R. Hurley et al.(2002)]{hur02}
   Hurley, J. R., Tout, C. A., \& Pols, O. R. \mnras, 329, 897
\bibitem[I. S. Izmailov(2019)]{izm19}
   Izmailov, I. S. 2019, Astronomy letters, 45, 30
\bibitem[C. Jaschek \& Y. Andrillat(1998)]{jaschek98}
   Jaschek, C., \& Andrillat, Y. 1998, \aaps, 130, 597
\bibitem[Joshi et al.(2006)]{joshi06}
   Joshi, S., Mary, D. L., Martinez, P., et al. 2006, \aap, 455, 303
\bibitem[J. Josties \& B. D. Mason(2021)]{jos21}
   Josties, J., \& Mason, B. D. 2021, IAU Comm. 26, Info. Circ., 205, 1
\bibitem[A. B. Kaye et al.(2000)]{kaye00}
   Kaye, A. B., Henry, G. W. \& Rodriguez, E. 2000, IAU Inform. Bull. Var. Stars,
   4850, 1-3
\bibitem[P. Kervella et al.(2019)]{ker19}
   Kervella, P., Arenou, F., Mignard, F., Th\'{e}venin, F., 
   2019, \aap, 623A, 72
\bibitem[L. B. Lucy \& M. A. Sweeney(1971)]{luc71}
   Lucy, L. B., \& Sweeney, M. A. 1971, \aj, 76, 544
\bibitem[G. W. Marcy \& K. J. Benitz(1989)]{mar89}
   Marcy, G. W., \& Benitz, K. J. 1989, \apj, 344, 441
\bibitem[B. D. Mason et al.(2006)]{mas06}
   Mason, B. D., Hartkopf, W. I., Wycoff, G. L., \& Holdenried, E. R.  
   2006, \aj, 132, 2219
\bibitem[B. D. Mason \& W. I. Hartkopf(2026)]{mas26}
   Mason, B. D., \& Hartkopf, W. I. 2026, \aj, 171, 149
\bibitem[H. A. McAlister et al.(1987b)]{mca87b}
   McAlister, H. A., Hartkopf, W. I., Hutter, D. J., \& 
   Franz, O. G. 1987b \aj, 93, 688
\bibitem[H. A. McAlister et al.(1987a)]{mca87a}
   McAlister, H. A., Hartkopf, W. I., Hutter, D. J., Shara, M. M., 
   \& Franz, O. G. 1987a, \aj, 93, 183
\bibitem[H. A. McAlister et al.(1989)]{mca89}
   McAlister, H. A., Hartkopf, W. I., Sowell, J. R., Dombrowski, E. G., 
   \& Franz, O. G.  1989, \aj, 97, 510-531
\bibitem[H. A. McAlister et al.(1993)]{mca93}
   McAlister, H. A., Mason, B. D., Hartkopf, W. I., \& Shara, M. M. 1993, 
   \aj, 106, 1639 
\bibitem[R. A. Mendez et al.(2025)]{mend25}
   Mendez, R. A., Tokovinin, A., Costa, E., \& Dirk, M. 2025, \aj, 169, 226
\bibitem[M. Moe \& K. M. Kratter(2021)]{moe21}
	Moe, M., i\& Kratter, K. M. 2021, \mnras, 507, 3593
\bibitem[D. G. Monet(1979)]{mon79}
   Monet, D. G. 1979, \apj, 234, 275
\bibitem[B. Nordstr\"{o}m \& J. Andersen(1985)]{na85}
	Nordstr\"{o}m, B. \& Andersen J. 1985, \aaps, 61, 53
\bibitem[S. S. R. Offner et al.(2023)]{off23}
   Offner, S. S. R., Moe, M., Kratter, K. M., et al., in
   Protostars and Planets VII, Chapter 8, The Origin and Evolution 
   of Multiple Star Systems, ed. S. Inutsuka, Y. Aikawa, T. Muto, 
   K. Tomida, \& M. Tamura, (San Francisco, CA: ASP) 
\bibitem[K. Osawa(1959)]{osa59}
   Osawa, K. 1959, \apj, 130, 159
\bibitem[D. R. Palmer et al.(1968)]{pal68}
Palmer, D. R., Walker, E. N., Jones, D. H. P., \& Wallis, R. E. 1968,
   Royal Obs. Bull., 135, 385
\bibitem[M. J. Pecaut \& E. E. Mamajek(2013)]{pec13}
   Pecaut, M. J., \& Mamajek, E. E. 2013, \apjs, 208, 9
\bibitem[R. M. Petrie(1928)]{pet28}
   Petrie, R. M., Publ. Dominion Astrophys. Obs., 4, 81 
\bibitem[J. S. Plaskett (1918)]{plas18}
   Plaskett, J. S. 1918, \jrasc, 12, 460-463
\bibitem[J. S. Plaskett et al.(1920)]{plas20}
   Plaskett, J. S., Harper, W. E., Young, R. K., \& Plaskett, H. H. 1920,
   Publ. Dominion Astrophys. Obs., 1, 163-185
\bibitem[J. S. Plaskett et al.(1921)]{plas21}
   Plaskett, J. S., Harper, W. E., Young, R. K., \& Plaskett, H. H. 1921, 
   Publ. Dominion Astrophys. Obs., 2, 1
\bibitem[D. Pourbaix et al.(2004)]{pour04}
   Pourbaix, D., Tokovinin, A. A., Batten, A. H. et al. 2004, \aap, 424, 727
\bibitem[A. Prsa et al.(2022)]{prsa22}
   Prsa, A., Kochoska, A., Conroy, K. E., et al. 2022, \apjs, 258, 16
\bibitem[M. Ramella et al(1989)]{ram89}
   Ramella, M., Gerbaldi, M., Faraggiana, R., \& B$\odot{o}$hm, C. 
   1989, \aap, 209, 233
\bibitem[L. C. Roberts \& B. D. Mason(2018)]{rob18}
   Roberts, L. C., \& Mason, B. D. 2018, \mnras, 473, 4497
\bibitem[F. Royer et al.(2002)]{roy02}
   Royer, F., Grenier, S., Baylac, M.-O., Gomez, A. E., \& Zorec, J. 
   2002, \aap, 393, 897
\bibitem[F. Royer et al.(2007)]{roy07}
   Royer, F., Zorec, J., \& Gomez, A. E. 2007, \aap, 463, 671
\bibitem[F. Royer et al.(2014)]{roy14}
   Royer, F., Gebran, M., Monier, R., et al. 2014, \aap, 562, A84
\bibitem [H. N. Russell(1902)]{rus02}
   Russell, H. N. 1902, \apj, 15, 252
\bibitem[C. H. Sandberg Lacy \& F. C. Fekel(2011)]{slf11}
   Sandberg Lacy, C. H., \& Fekel, F. C. 2011, \aj, 142, 185
\bibitem[M. Sarkar et al.(2024)]{sark24}
   Sarkar, M., Joshi, S., Dupret, M.-A, et al. 2024, \mnras, 534, 3211
\bibitem[M. Scardia et al.(2002)]{scar02}
   Scardia, M., Prieur, J.-L., Koechlin, L., \& Aristidi, E. 2002, 
   IAU Comm. 26, Info. Circ., 146 
\bibitem[C. D. Scarfe(2010)]{sca10}
   Scarfe, C.  D. 2010, Obs, 130, 214
\bibitem[G. Shajn, \& V. Albitzky(1932)]{shaj32}
   Shajn, G. \& Albitzky, V. 1932, \mnras, 92, 771
\bibitem[V. M. Slipher(1904)]{sliph04}
   Slipher, V. M. 1904, \apj, 20, 146
\bibitem[B. Smalley et al.(1997)]{sma97}
   Smalley, B., Kellett, B. J., Wonnacott, D., \& Stickland, D. J. 
   1997, \mnras, 284, 457
\bibitem[C. L. Smith et al.(2024)]{smi24}
   Smith, C. L., Moe, M., \& Kratter, K. M. 2024, \apj, 975, 153 
\bibitem[S. J. Thomas et al.(2023)]{thomas23}
   Thomas, S. J., Rodgers, B., van der Bliek, N. S. et al. 2023,
   \aj, 165, 135
\bibitem[A. A. Tokovinin(2018)]{tok18}
   Tokovinin, A. A. 2018, IAU Commission 26 , Double Star Info. Circ., 
   195, 6
\bibitem[A. Tokovinin(2019)]{tok19}
   Tokovinin, A. 2019, \aj, 158, 222
\bibitem[A. Tokovinin(2021)]{tok21}
   Tokovinin, A. A. 2021, IAU Commission 26 , Double Star Info. Circ.,
   203, 5
\bibitem[A. A. Tokovinin(2024)]{toko24}
   Tokovinin, A. A. 2024, \aj, 168, 190
\bibitem[Tokovinin et al.(2024)]{tok24}
   Tokovinin, A., Mason, B. D., Mendez, R. A., \& Costa, E., 
   2024, \aj, 168, 28
\bibitem[G. Torres(2020)]{tor20}
   Torres, G. 2020, \apj, 901, 91
\bibitem[T. Trifonov et al.(2020)]{trif20}
   Trifonov, T., Tal-Or, L., Zechmeister, M., et al. 2020, \aap, 636, A74
\bibitem[S. Udry et al.(1998)]{udry98}
   Udry, S., Mayor, M., Latham, D. W., et al., 1998 in ASP Conf. Ser. 154
   Cool Stars, Stellar Systems and the Sun: Tenth Cambridge Workshop,
   ed. R. A. Donahue \& J.A. Bookbinder, (San Francisco, CA: ASP), 2148
\bibitem[F. van Leeuwen(2007)]{vanl07}
   van Leeuwen, F. 2007, \aa, 474, 653
\bibitem[I. Waisberg et al.(2023a)]{wais23a}
   Waisberg, I., Klein, Y., \& Katz, B. 2023, \mnras, 521, 5255
\bibitem[I. Waisberg, Y. Klein, \& B. Katz(2023b)]{wais23b}
   Waisberg, I., Klein, Y., \& Katz, B. 2023, Research Notes of the AAS, 7, 180 
\bibitem[I. Waisberg et al.(2023c)]{wais23c}
   Waisberg, I., Klein, Y., \& Katz, B. 2023, Research Notes of the AAS, 7, 231
\bibitem[J. Wilsing(1893)]{wil93}
   Wilsing, J. 1893, Astronomische Nachtrichten, 134, 89
\bibitem[R. E. Wilson(1953)]{wil53}
   Wilson, R. E. 1953, General Catalogue of Stellar Radial Velocities,
   (Washington, D.C.: Carnegie Inst.), publication 601
\bibitem[R. E. Wilson \& A. H. Joy(1950)]{wil50}
   Wilson, R. E., \& Joy, A. H. 1950, \apj, 111, 221
\bibitem[R. H. Wolfe et al.(1967)]{wol67}
   Wolfe, R. H., Horak, H. G., \& Storer, N. W. 1967, in Modern 
   Astrophysics, ed. M. Hack (New York: Gordon \& Breach), 251
\bibitem[S. C. Wolff \& G. W. Preston(1978)]{wol78}
   Wolff, S. C., \& Preston, G. W. 1978, \apjs, 37, 371
\bibitem[R. K. Young(1939)]{you39}
   Young, R. K. 1939, Pub. David Dunlap Obs., 1, 71 
\bibitem[R. K. Young(1945)]{you45} 
   Young, R. K. 1945, Pub. David Dunlap Obs., 1, 311
\bibitem[J.-P. Zahn(1977)]{zah77}
   Zahn, J.-P. 1977, \aap, 57, 383
\bibitem[H. Zirm \& F.M. Rica(2014)]{zirm14}
   Zirm, H., \& Rica, F. M. 2014, IAU Comm. 26, Info. Circ., 183
\bibitem[H. Zirm(2015)]{zirm15}
   Zirm, H. 2015, IAU Comm. 26, Info. Circ. 185
\bibitem[Zucker et al.(2022)]{zuc22}
   Zucker C., Goodman, A. A., Alvesm H, et al. 2022, \nat, 601, 334
\end{thebibliography}
\end{document}